\documentclass[fleqn,usenatbib]{mnras}

\usepackage{newtxtext,newtxmath}
\usepackage[OT2, T1]{fontenc}
\usepackage[russian, english]{babel}

\DeclareRobustCommand{\VAN}[3]{#2}
\let\VANthebibliography\thebibliography
\def\thebibliography{\DeclareRobustCommand{\VAN}[3]{##3}\VANthebibliography}

\usepackage{graphicx}	
\usepackage{amsmath}	
\usepackage{color}
\usepackage{booktabs}
\usepackage{tcolorbox}

\usepackage{multirow}

\newcommand{\changetext}[1]{#1}

\newcommand{\Sha}{\text{\foreignlanguage{russian}X}}

\title[Image combination for Roman I]{Simulating image coaddition with the {\textit{\textbf{Nancy Grace Roman Space Telescope}}}: I. Simulation methodology and general results}

\author[Hirata et al.]{Christopher M. Hirata$^{1,2,3}$,
Masaya Yamamoto$^4$,
Katherine Laliotis$^{1,2}$,
Emily Macbeth$^{1,2,3}$,
\newauthor{M.A.~Troxel$^4$,
Tianqing Zhang$^5$,
Kaili Cao$^{1,2}$,
Ami Choi$^6$,
Jahmour Givans$^{7,8}$,
Katrin Heitmann$^{9}$,
}
\newauthor{
Mustapha Ishak$^{10}$,
Mike Jarvis$^{11}$,
Eve Kovacs$^{9}$,
Heyang Long$^{1,2}$,
Rachel Mandelbaum$^{5,12}$,
Andy Park$^5$,
}
\newauthor{
Anna Porredon$^{13}$,
Christopher W. Walter$^4$,
W. Michael Wood-Vasey$^{14}$}\\
$^{1}$Center for Cosmology and AstroParticle Physics (CCAPP), 191 West Woodruff Ave, The Ohio State University, Columbus, OH 43210, USA\\
$^{2}$Department of Physics, The Ohio State University, 191 West Woodruff Ave, Columbus, OH 43210, USA\\
$^{3}$Department of Astronomy, The Ohio State University, 140 West 18th Avenue, Columbus, OH 43210, USA\\
$^{4}$Department of Physics, Duke University, Box 90305, Durham, NC 27708, USA\\
$^{5}$McWilliams Center for Cosmology, Department of Physics, Carnegie Mellon University, 5000 Forbes Ave, Pittsburgh, PA 15213, USA\\
$^{6}$NASA Goddard Space Flight Center, Mail Code 665, Greenbelt, MD 20771, USA\\
$^{7}$Department of Astrophysical Sciences, Princeton University, 4 Ivy Lane, Princeton, NJ 08540, USA\\
$^{8}$Center for Computational Astrophysics, Flatiron Institute, 162 5th Ave, New York, NY 10010, USA\\
$^{9}$Argonne National Laboratory, 9700 S Cass Ave, Lemont, IL 60439, USA\\
$^{10}$Department of Physics, University of Texas at Dallas, EC36, 800 W Campbell Rd, Richardson, TX 75080, USA\\
$^{11}$Department of Physics and Astronomy, University of Pennsylvania, 209 South 33rd Street, Philadelphia, PA 19104-6396, USA\\
$^{12}$NSF AI Planning Institute for Data-Driven Discovery in Physics, Carnegie Mellon University, Pittsburgh, PA 15213, USA\\
$^{13}$Institute for Astronomy, University of Edinburgh, Edinburgh EH9 3HJ, UK\\
$^{14}$Physics and Astronomy Department, University of Pittsburgh, 4200 Fifth Ave, Pittsburgh, PA 15260, USA
}

\date{Accepted XXX. Received YYY; in original form ZZZ}

\pubyear{2022}

\begin{document}
\label{firstpage}
\pagerange{\pageref{firstpage}--\pageref{lastpage}}
\maketitle

\begin{abstract}
The upcoming {\slshape Nancy Grace Roman Space Telescope} will carry out a wide-area survey in the near infrared. A key science objective is the measurement of cosmic structure via weak gravitational lensing. {\slshape Roman} data will be undersampled, which introduces new challenges in the measurement of source galaxy shapes; a potential solution is to use linear algebra-based coaddition techniques such as {\sc Imcom} that combine multiple undersampled images to produce a single oversampled output mosaic with a  desired ``target'' point spread function (PSF). We present here an initial application of {\sc Imcom} to 0.64 square degrees of simulated {\slshape Roman} data, based on the {\slshape Roman} branch of the Legacy Survey of Space and Time (LSST) Dark Energy Science Collaboration (DESC) Data Challenge 2 (DC2) simulation. We show that {\sc Imcom} runs successfully on simulated data that includes features such as plate scale distortions, chip gaps, detector defects, and cosmic ray masks. We simultaneously propagate grids of injected sources and simulated noise fields as well as the full simulation. We quantify the residual deviations of the PSF from the target (the ``\changetext{leakage}''), as well as noise properties of the output images; we discuss how the overall tiling pattern as well as Moir\'e patterns appear in the final \changetext{leakage} and noise maps. We include appendices on interpolation algorithms and the interaction of undersampling with image processing operations that may be of broader applicability. The companion paper (``Paper II'') explores the implications for weak lensing analyses.
\end{abstract}


\section{Introduction}

Weak gravitational lensing -- the distortion of the shapes of distant galaxies as their light passes through the  gravitational potential of foreground structures -- has emerged as one of the powerful tools for probing the growth of structure in the Universe (see \citealt{2008ARNPS..58...99H}, \citealt{2013PhR...530...87W}, and \citealt{2018ARA&A..56..393M} for recent reviews). It has now been more than two decades since the early detections of lensing by galaxies \citep{1996ApJ...466..623B} and of the two-point correlations of shear \citep{2000A&A...358...30V, 2000MNRAS.318..625B, 2000Natur.405..143W, 2000astro.ph..3338K}. In that time, both the size of cosmological surveys and the understanding of the instruments and data processing necessary to extract the weak lensing signal has advanced considerably. The current generation of weak lensing surveys -- the Kilo Square Degree survey (KiDS), the Dark Energy Survey (DES), and the Hyper Suprime Cam (HSC) -- have measured the amplitude of cosmic structure $S_8$ to precision better than a few percent \citep{hikage/etal:2019,hamana/etal:2020,asgari/etal:2021,amon/etal:2022,secco/etal:2022}. With these surveys, and the detailed picture of the initial conditions for the growth of structure provided by cosmic microwave background observations \citep{planck/etal:2020}, it is now possible to do detailed comparisons of low redshift structures to the predictions of general relativity in the $\Lambda$CDM model and various alternatives \citep[e.g.][]{2021MNRAS.505.6179L}.

In the 2020s, several major surveys are expected to come online that will lead to large advances in both the statistical constraining power and control of systematic uncertainties in weak lensing measurements. These include the ground-based Vera Rubin Observatory, which will carry out a ``Wide, Fast, Deep'' survey of the optical sky in the $ugrizy$ bands \citep{2012arXiv1211.0310L, 2019ApJ...873..111I}; the {\slshape Euclid} satellite, which features a wide-field visible channel for galaxy shape measurements and a near infrared (NIR) instrument to improve photometric redshifts \citep{2011arXiv1110.3193L}; and the {\slshape Nancy Grace Roman Space Telescope}, whose Wide Field Instrument will survey the sky at high angular resolution in 4 NIR bands spanning 0.9--2.0 $\mu$m (in the Reference Survey design\footnote{The Reference Survey is an example survey used during the design and construction phases to show that the {\slshape Roman} mission meets its requirements. The actual survey conducted will be designed through a community process and could be different.}; \citealt{2015arXiv150303757S, 2019arXiv190205569A}). All of these observatories bring advantages relative to current surveys in terms of systematic error control for galaxy shapes: Rubin will obtain hundreds of images of each field, thus enabling much better internal constraints of instrument systematics, while the space missions provide the high angular resolution and stability possible above the Earth’s atmosphere. Furthermore, the coverage from $u$ band in Rubin through $\sim 2$ $\mu$m for the space missions (with {\slshape Roman} data reaching NIR depths comparable to Rubin sensitivity, and {\slshape Euclid} data being shallower but covering a wider footprint) provides an enormous wavelength baseline for photometric redshifts \citep{2019ApJ...877..117H, 2022ARA&A..60..363N}.

Although space is in many ways an ideal location for a weak lensing experiment, the high angular resolution brings some challenges related to the pixel scale. A diffraction-limited telescope produces a point spread function (PSF) with a characteristic angular width of $\lambda/D$, where $\lambda$ is the wavelength of observation and $D$ is the diameter of the entrance pupil. If the pixel scale $P$ is larger than $\lambda/(2D)$, then the image is {\em undersampled} in the Nyquist sense: there are Fourier modes present in the image with more than $\frac12$ cycle per sample, with the consequence that the images cannot be unambiguously interpolated, and thus the image intensity $I({\bmath r})$ cannot be treated as a continuous field. It also produces biases in the moments of images, including the first moment (centroid, relevant to astrometry) and the second moments (sizes and shapes, relevant to weak lensing), which have been studied in many contexts \citep[e.g.][]{1999PASP..111.1434L, 2000PASP..112.1360A, 2007PASP..119.1295H, 2011PASP..123..470S}. One possible solution to the undersampling problem is to simply use small pixels -- but if one has a fixed pixel count in the focal plane (often limited by available resources or technical considerations), then shrinking the pixels leads to a smaller field of view and a slower survey if we fix the survey depth and hence the required exposure time per pointing.\footnote{If significant, read noise can impose an additional penalty for a well-sampled survey due to the the low sky flux on each pixel.} An alternative, adopted for {\slshape Euclid} and {\slshape Roman}, is to accept undersampling, and use multiple dithered exposures of each field. This increases survey speed, but requires the development of algorithms for each application that take multiple undersampled images as input.

Undersampling can affect several stages of a weak lensing analysis \citep[e.g.][]{2021MNRAS.502.4048K, 2023arXiv230107725F}. One particular step is calibration of the shear estimator -- that is, determining how the measured ellipticity of a galaxy $e_i$ in a catalog responds to an applied shear $\gamma_j$.\footnote{We include an index since $e$ and $\gamma$ are 2-component quantities.} While one might hope for an ellipticity measurement algorithm that has unit shear response $\partial \langle e_i\rangle/\partial\gamma_j=\delta_{ij}$, any stable ellipticity estimator has a response that depends on the galaxy population \citep{2007MNRAS.380..229M, 2011MNRAS.414.1047Z}, and so modern lensing analyses contain a ``shear calibration'' step that determines this response given the ensemble of galaxy morphologies present in that survey (at that depth, resolution, and observed wavelength). Whether the data are well-sampled or not, shear calibration must work with the fact that Fourier modes in the image are only well-measured up through some $k_{\rm max}$, and the shear operation moves modes across the $k=k_{\rm max}$ boundary \citep[e.g.][]{2010MNRAS.406.2793B}, so one cannot take an observed image and infer what the sheared image would look like at the same resolution. In the past decade, several principled shear calibration approaches have been introduced that apply some re-smoothing (or in Fourier space, a cut $k_{\rm cut}<k_{\rm max}$) before measuring galaxy shapes (or moments) and have been successful at mitigating the galaxy population-dependent shear calibration biases in simulations. These include {\sc Metacalibration}, which numerically applies a shear to each object to compute the ensemble response \citep{2017arXiv170202600H, 2017ApJ...841...24S, 2022arXiv220607683Z}; the Bayesian Fourier Domain technique, which builds the probability distribution of Fourier-space moments \citep{2014MNRAS.438.1880B, 2016MNRAS.459.4467B}; and approaches that analytically build shear responses \citep{2022arXiv220810522L}.\footnote{\citet{2012MNRAS.420.1518M} proposed an algorithm that uses higher-resolution images of a small sample of the galaxies in the same wavelength range to determine the statistical properties of the galaxy images at $k>k_{\rm max}$, with the application of using Hubble Space Telescope images to calibrate ground-based shapes.} Further development of these methods is anticipated in support of final analyses of the Stage III ground-based survey data and the upcoming Vera Rubin Observatory. However, all of these techniques rely fundamentally on operations such as cuts in Fourier space that cannot be performed on undersampled data. Indeed, the first attempts to simulate {\sc Metacalibration} for undersampled images resulted in percent-level biases \citep{2021MNRAS.502.4048K, 2023MNRAS.519.4241Y} that exceed requirements for the upcoming surveys. This motivates us to develop image processing techniques to recover full sampling, as a step in the lensing analysis that precedes shape measurement (and even source selection), so that these shear calibration techniques can be applied to {\slshape Roman} data.

The shear response is a property not just of the shape measurement algorithm, but of the sample selection as well, since applying a shear to a galaxy could cause it to cross a selection threshold \citep{2000ApJ...537..555K, 2003MNRAS.343..459H}. The same tools that have been developed to measure the shear response can be extended to incorporate the response from source selection \citep[e.g.][]{2020ApJ...902..138S, 2023arXiv230303947S}. Furthermore, in a tomographic weak lensing analysis, the assignment of a galaxy to a particular tomographic bin is itself a form of selection, and thus one needs to derive the shear response of the photometry in each filter used for constructing the tomographic bins \citep[e.g.][]{2018PhRvD..98d3528T,2021MNRAS.504.4312G, 2021MNRAS.505.4249M}. The next-generation weak lensing surveys have specified ``shape measurement'' filters (J129+H158+F184 in the case of {\slshape Roman}) as well as additional filters that are used for photometric redshifts (Y106 for {\slshape Roman}). This means that although the source galaxies do not need to be {\em resolved} in the photo-$z$-only filters, one still needs to either recover {\em full sampling} in these filters (and use the aforementioned frameworks) or develop an alternative framework for determining the statistical shear response. (This issue was not fully understood at the time the {\slshape Roman} Reference Survey was designed.) Therefore, this paper has investigated recovery of a fully sampled mosaic in {\slshape Roman} Y106 band as well.

\begin{figure*}
    \centering
    \includegraphics[width=6in]{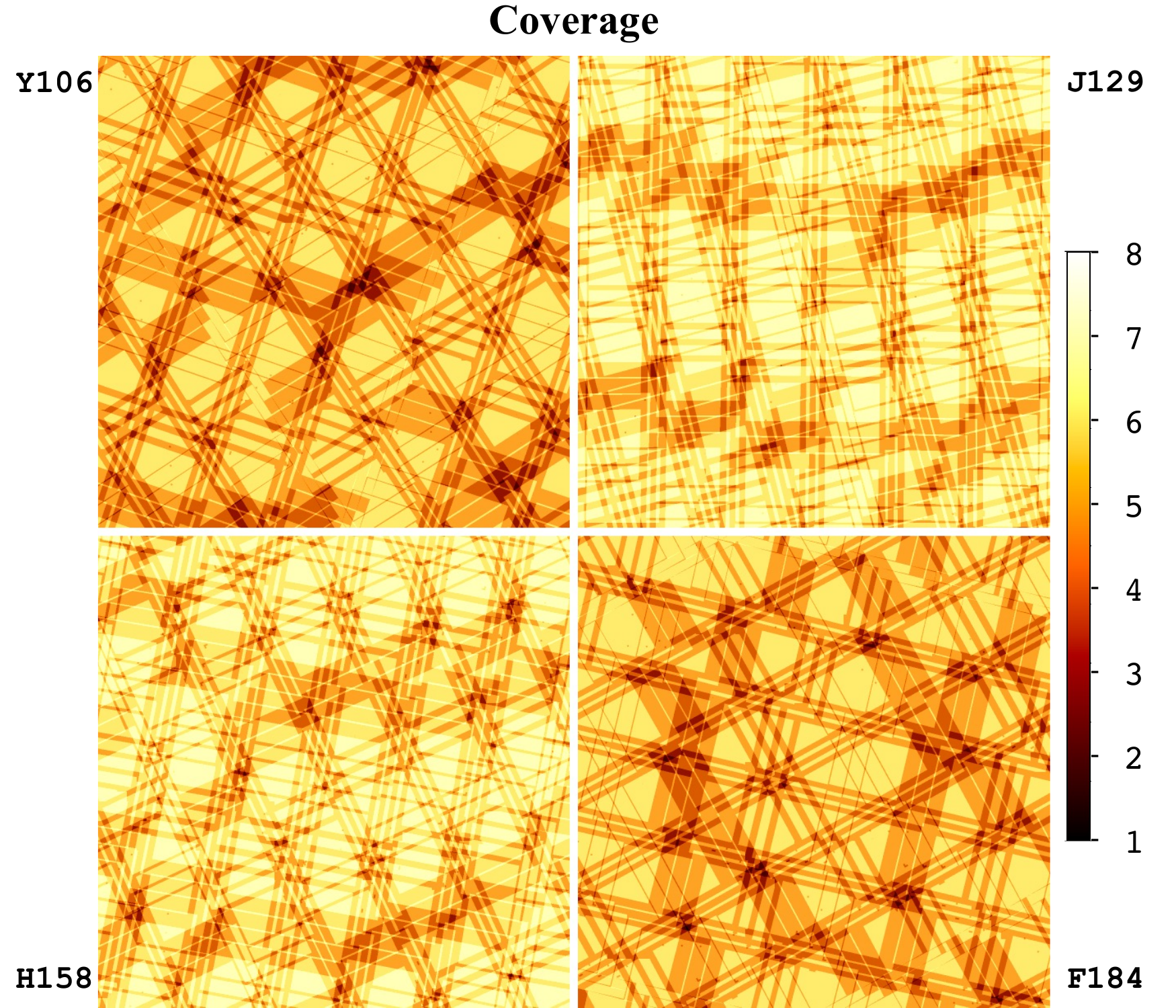}
    \caption{\label{fig:coveragemap}The coverage (number of exposures) in each of the 4 bands in the $48\times 48$ arcmin region considered in this paper. Each sub-panel shows one of the filters. The 18-chip ``pawprint'' feature of the {\slshape Roman} focal plane is easily visible, as is the presence of two roll angles from the two passes in each filter.}
\end{figure*}

The recent joint {\slshape Roman}+Rubin simulations \citep{2022arXiv220906829T}, built on top of the Legacy Survey of Space and Time (LSST) Dark Energy Science Collaboration (DESC) Data Challenge 2 (DC2) simulations \citep{2019ApJS..245...26K, 2021arXiv210104855L, 2021ApJS..253...31L, 2022OJAp....5E...1K}, provide an opportunity to test out algorithms to recover full sampling as part of an integrated simulation and processing pipeline suite, and explore the implications for {\slshape Roman} weak lensing analyses. This is the first in a series of papers that develops and tests an implementation of the {\sc Imcom} algorithm \citep{2011ApJ...741...46R} on a simulation of part of the {\slshape Roman} Reference Survey. In this paper (``Paper I''), we focus on the characteristics of the simulation, relevant mathematical background, image combination machinery, and basic properties of the outputs. The companion paper (``Paper II'') covers statistical analyses of the output images, including the ellipticities of simulated stars, correlation functions, noise power spectra, and noise-induced biases. Both papers use a $48\times 48$ arcmin region from the simulations, large enough to contain $\sim 2$ {\slshape Roman} fields of view and a representative portion of the tiling pattern (see Fig.~\ref{fig:coveragemap}).

This paper is organized as follows. The problem of combining images to create fully sampled output with uniform PSF, and the goals of this simulation effort, are described in more detail in Section~\ref{sec:statement}. The input data, including ancillary information such as masks, are described in Section~\ref{sec:input}. The image coaddition simulation methodology is described in Section~\ref{sec:co-add}. Some of the basic results on output PSF and noise properties are described in Section~\ref{sec:results}, and we conclude in Section~\ref{sec:discussion}. Appendix~\ref{app:interpolation} presents some useful results on optimal interpolation methods for functions with known oversampling rates that we derived for this work, and that may be of more general interest, while a description of the computing resources for the project can be found in Appendix~\ref{app:computing}. A summary of how undersampling interacts with the operations used in modern shear calibration methods is provided in Appendix~\ref{app:sampling}.

\section{Statement of the problem}
\label{sec:statement}

A variety of methods have been explored in the literature for combining multiple images of the sky into a single ``coadded'' image. If the coadded image is to be used as the starting point for a standard weak lensing analysis, one wants it to be both oversampled and have a well-defined PSF in the sense of \citet{2023OJAp....6E...5M}: the output image should be the true astronomical scene convolved with a PSF. We will go further here and ask for an algorithm that produces a specific desired output PSF $\Gamma$ that is uniform and circular --- thus we choose the output PSF and design a coaddition scheme to achieve it, rather than running a coaddition code and accepting the measured PSF at the end. This is advantageous for survey uniformity and mitigation of additive biases, but as we will see this is only possible under certain circumstances.

Thus in the context of this paper, the goal of the coaddition pipeline is to take in several input images of the sky, which are at some native pixel scale $s_{\rm in}$ and in general have their own rotations, distortions, PSFs, and masks; and produce a well-sampled output image at an output pixel scale $s_{\rm out}$, and with a uniform, round output PSF. Implicit in this statement of charge is that we are trying to accomplish at once several tasks that are sometimes distinct steps in an image processing pipeline:
\newcounter{coaddlist}
\begin{list}{\arabic{coaddlist}.\ }{\usecounter{coaddlist}}
    \item\label{it:interp} {\em Interpolation} over masked pixels (e.g., cosmic ray hits, bad columns, or hot, dead, or unstable pixels).
    \item\label{it:resamp} {\em Resampling} onto a common grid.
    \item\label{it:round} {\em Rounding} and {\em homogenization} of the output point spread function (for some weak lensing pipelines; this step could also be performed last or not at all).
    \item\label{it:avg} {\em Averaging} of the intensities from each input image to yield a single output image.
\end{list}
For oversampled data, it may be reasonable to treat these as separate, since operations such as resampling, convolution, and (sometimes) filling in a single missing sample, can be carried out on an oversampled function without introducing biases. However, they may be viewed in a unified framework if all of the operations used are linear. In this case, each step is a matrix operation, leading to an output image that is a linear combination of input images, with the mapping described by an $m\times n$ matrix ${\mathbfss T}$, where $m$ is the number of output pixels and $n$ is the number of input pixels. These statements apply to many of the common algorithms that have been applied to ground-based weak lensing data sets (either for shapes, source selection, or photo-$z$s): for example, linear predictive codes for interpolating bad pixels \citep[\S4.5]{2018PASJ...70S...5B}; Lanczos-3 (\citealt{2002ASPC..281..228B}; \citealt{2018PASJ...70S...5B}, \S3.3) or polynomial \citep[\S4.4]{2014MNRAS.440.1296H} interpolation for resampling; and pixel-domain \citep[\S7]{2002AJ....123..583B} or Fourier-domain \citep[\S4.1]{2014MNRAS.440.1296H} rounding kernels, and PSF Gaussianization (\citealt{2012MNRAS.421.2355H},\S3; \citealt{2015MNRAS.454.3500K}, \S4.2).

None of steps \#\ref{it:interp}--\ref{it:round} are possible individually on undersampled data without introducing biases or making additional assumptions about the astronomical scene. However, the end goal of constructing a well-sampled output image with uniform PSF with some matrix ${\mathbfss T}$ may still be possible, even if ${\mathbfss T}$ cannot be factored into the individual steps. The Fourier-domain algorithm of \citet{1999PASP..111..227L} and the iterative algorithm of \citet{2011PASP..123..497F} are examples that combine some of these steps for some types of input data. The {\sc Imcom} technique of \citet{2011ApJ...741...46R} searches for a matrix ${\mathbfss T}$ that accomplishes all 4 steps with minimum noise and target PSF error (as measured with quadratic metrics).\footnote{We want each output pixel to depend only on input pixels within a few arc seconds of its position. In practice, this is implemented by splitting the image into postage stamps; see Sec.~\ref{ss:imcom-postage} for details.} It is computationally expensive but can work with rolls, geometric distortions, varying input PSFs, and complex masks, and thus is a promising choice for a space-based weak lensing survey with {\slshape Roman}. {\sc Imcom} returns a residual estimate for the output PSF; this allows us to identify cases where the desired output PSF is impossible to build (e.g., due to aliasing with insufficient dithers, or contains Fourier modes not represented in the input image).

We note that the Drizzle algorithm \citep{2002PASP..114..144F} that is commonly used to combine undersampled space-based images is itself a linear operation that can be described using a coaddition matrix ${\mathbfss T}$. In the case of Drizzle, ${\mathbfss T}$ is sparse; the entry $T_{\alpha i}$ is determined by the overlap of output pixel $\alpha$ with a ``shrunken'' version of input pixel $i$. The Drizzle matrix ${\mathbfss T}$ is within the search space for {\sc Imcom}, and therefore by {\sc Imcom}'s target metrics of sum-of-squares error in the PSF and noise, {\sc Imcom} will always perform at least as well as Drizzle (usually much better). But by choosing a particular sparse ${\mathbfss T}$, the Drizzle algorithm has a lower memory footprint and much faster run time, and therefore is likely to remain useful in the weak lensing analysis as a ``quick look'' tool (this is also how it was used in the DC2 simulation; \citealt{2022arXiv220906829T}) and for flagging purposes.

While \citet{2011ApJ...741...46R} demonstrated their method on some test problems with $\sim 6\times 6$ arcsec postage stamps, and the method has also been demonstrated on laboratory data \citep{2013PASP..125.1496S}, the method has not yet been applied to {\slshape Roman} simulations over an area large enough to measure the statistical properties of the coadded images. The availability of the Rubin Data Challenge 2 (DC2) + {\slshape Roman} simulation suite, and updated knowledge of Roman properties including characterization of the flight detectors, makes this an excellent time to embark on such a simulation. The qualitative and quantitative objectives for this simulation are as follows:
\newcounter{objectives}
\begin{list}{\arabic{objectives}.\ }{\usecounter{objectives}}
\item Wrap the algorithm in a driver that can tile the sky and pipe the appropriate (simulated) observations to the linear algebra kernel and re-assemble the output into a community standard format such as FITS.
\item Find an example (not necessarily final) set of parameters for the {\sc Imcom} algorithm that avoid basic problems such as noise amplification or ghosting across postage stamp boundaries (or reduce them to acceptable levels).
\item Characterize how chip gaps and cosmetic defects map into the assembled mosaics.
\item Use injected sources to test the output normalization, astrometry, size, shape, and higher moments of the final coadd PSF after propagation through {\sc Imcom}.
\item Measure the correlation function of the ellipticities at scales overlapping the range likely to be included in the {\slshape Roman} weak lensing analysis.
\item Measure the noise properties of the output image for both uncorrelated and correlated input noise.
\item Compare the noise measured on the output images to the predictions from the Exposure Time Calculator \citep{2013ascl.soft11012H} and analytical descriptions in the literature \citep[e.g.][]{2002PASP..114...98B}.
\item Determine the as-realized computing time requirements for the coadd on a modern computing cluster (in order to inform both resource estimates and priorities for further optimization).
\item Compare the moments (i.e., shape and size) of bright unsaturated stars (suitable for PSF characterization) in the output images with the measurements performed on Drizzled coadd from DC2+{\slshape Roman} simulations. 
\end{list}
Most of these objectives can only be achieved by simulating $>1$ field of view. This paper presents the initial set of coaddition simulations and first look results; follow-on papers will go into more detail on some of the individual objectives.

\begin{table*}
    \centering
    \caption{\label{tab:layers}The layers of input data used in this paper. Data may be in ``all'' of the bands (Y106, J129, H158, and F184), or in the indicated bands only (the decision to add the last two layers was made after processing of Y106 and F184 had started). All layers are generated as $4088\times 4088$ SCA images prior to being fed into the coaddition.}
    \begin{tabular}{ccccc}
\hline\hline
      Index & Name  & Description & Bands & Reference \\ \hline
    0 & SCI & DC2 image simulation (``simple'' model) & All & Sec.~\ref{ss:sim} \\
    1 & truth & DC2 image simulation -- noiseless images & All & Sec.~\ref{ss:sim} \\
    2 & gsstar14 & Injected point source grid (drawn by {\sc GalSim}) & All & Sec.~\ref{ss:inj} \\
    3 & cstar14 & Injected point source grid (drawn by {\tt pyimcom\_croutines}) & All & Sec.~\ref{ss:inj} \\
    4 & whitenoise1 & Uncorrelated input noise & All & Sec.~\ref{ss:noise} \\
    5 & 1fnoise2 & $1/f$ noise (generated in each channel) & All & Sec.~\ref{ss:noise} \\
    6 & err & DC2 image simulation -- noise realization & J129,H158 & Sec.~\ref{ss:sim} \\
    7 & labnoise & Ground test dark frames & J129,H158 & Sec.~\ref{ss:labnoise} \\
\hline\hline
    \end{tabular}
\end{table*}

\section{Input data}
\label{sec:input}

We generate several types of input data: full and noiseless image simulations, noise fields, and grids of injected sources. All of these are common types of inputs to image processing pipelines in weak lensing data analysis. Since the coaddition process is linear, we may use the distributive property to superpose outputs and generate, e.g., an output sky image with additional $1/f$ noise, or an injected object in the real survey \citep[e.g.][]{2016MNRAS.457..786S}. The processing of all layers at the same time allows the ${\mathbfss T}$ matrix to be computed only once, a major advantage since it is both the most computationally demanding step and is so large that only a few examples can be stored to disk rather than the ${\mathbfss T}$ for the whole survey. The layers used in this analysis are shown in Table~\ref{tab:layers}.

\subsection{The {\slshape Roman} + Rubin simulations}
\label{ss:sim}

The principal input data to this study is the {\slshape Roman} arm of the joint {\slshape Roman} + Rubin DC2 simulation, described in \citet{2022arXiv220906829T}. The simulation begins with a cosmological $N$-body simulation \citep{2019ApJS..245...16H} populated with galaxies \citep{2012NewA...17..175B, 2020MNRAS.495.5040H}. This forms the basis for the cosmoDC2 simulated extragalactic sky \citep{2019ApJS..245...26K, 2022OJAp....5E...1K}, which includes a superset of the region simulated in this paper. This is integrated with a local Universe simulation and a telescope + instrument simulation
for the Rubin Observatory \citep{2021arXiv210104855L, 2021ApJS..253...31L}. The {\slshape Roman} arm of the image simulation observes the same sky, but with the current version of the {\slshape Roman} telescope + instrument simulation framework (an update of \citealt{2021MNRAS.501.2044T}).

The simulation includes a portion of the {\slshape Roman} Reference Survey tiling strategy, which observes the sky in four bands: Y106, J129, H158, and F184. Each band has two passes over the sky at different roll angles, with small-step diagonal dithers to cover the chip gaps. The number of dither positions at each roll was based on preliminary estimates of the number of dithers required to mitigate sampling issues \citep{2011ApJ...741...46R, 2015arXiv150303757S}: 4 in J129, and 3 in each of H158 and F184. The Y106 band strategy was not required to achieve full sampling and so used 3 positions. The resulting coverage pattern is shown in Fig.~\ref{fig:coveragemap}. A more in-depth description of the sky tiling can be found in Appendix~A of \citet{2022arXiv220906829T}.

The main input layer used for this simulation is the ``simple'' model sky images from \citet{2022arXiv220906829T}. These include the convolution of sky objects with the PSF and Poisson noise. They do not include the full detector physics model (this version was chosen so that we can test downstream processing steps, such as image combination, before corrections for the detector effects are ready). The two most important parts of the detector physics missing for our purpose are (i) that there is no pixel mask in the ``simple'' simulation (we introduce our own in Sec.~\ref{ss:masks}); and (ii) the charge diffusion is not included. Since charge diffusion smears out the effective PSF before pixelization, it actually improves sampling, so we expect that ignoring it is conservative from the perspective of a code that attempts to reconstruct a fully sampled image. This expectation is explicitly tested for a small subsample of the data in Sec.~\ref{ss:CD}.

The ``simple'' input model also comes with a PSF model (corresponding to a flat \changetext{spectral energy distribution or SED} in $F_\lambda$) and an astrometric solution (the World Coordinate System or WCS in the FITS header). The WCS solution in the simulation is ``perfect'' in the sense that we are working with the same WCS used to draw the image; propagation of systematic errors in the WCS solutions will be considered in a future paper.

We have also included as additional layers the ``truth'' (noiseless) images from the simulation, and --- for J129 and H158 --- the ``err'' HDU (the realization of background Poisson noise used in the simulation).

\subsection{Masks}
\label{ss:masks}

The ``simple'' DC2+{\slshape Roman} simulations produce output for all of the pixels. However, in the real mission there will be masked pixels. These can be divided into two types: pixels that are defective (e.g., disconnected, hot, unstable) and are flagged by the calibration pipeline; and pixels affected by a cosmic ray in that particular exposure. These two classes of masks affect image combination differently, since defective pixels tend to be spatially clustered and are the same pixels when we dither, whereas the cosmic ray impacts randomly knock out small groups of pixels.

For this simulation, we have taken the ``permanent mask'' based on the SCA files constructed from acceptance testing \citep[Appendix B]{2022arXiv220906829T}. Out of the 18 chips initially selected for flight, an average of 3.01\% of the pixels were flagged (BADPIX HDU) by at least one of the steps in the pipeline. The sources of these flags in order are shown in Table~\ref{tab:badpix}. This should be considered only a first approximation to the likely mask used in flight, since the hot pixel populations change as the detector ages or is exposed to radiation \citep[e.g.][]{2019wfc..rept....3S}.

\begin{table}
    \centering
    \caption{\label{tab:badpix}The permanent mask flags based on the BADPIX HDU of the SCA files. The last line shows the union of these flags; since some pixels have multiple flags, the sum of the fractions is greater than ALL.}
    \begin{tabular}{ccc}
    \hline\hline
    Bit & Meaning & Fraction \\
    \hline
    0 & Non-responsive pixel & 0.53\% \\
    1 & Hot pixel (did not use 2 hr dark for dark estimate) & 0.20\% \\
    2 & Very hot pixel (used first read for dark estimate) & 0.11\% \\
    3 & Adjacent to pixel with flagged response & 2.47\% \\
    4 & Low CDS, high total noise pixel & 0.03\% \\
    5 & Failed non-linearity solution & 0.01\% \\
    6 & Failed gain solution & 0.08\% \\
    \hline
    ALL & & 3.01\% \\
    \hline\hline
    \end{tabular}
\end{table}

The cosmic ray mask is based on a random number generator, with the seeds chosen based on the observation ID and SCA so that the same mask is generated for each observation even if it is called in a different block. The density of cosmic ray strikes is taken to be $7.7\times 10^{-4}$ per pixel, which is the product of the $10^{-6}\,$cm$^2$ pixel area, the 140 s exposure time, and the expected rate of 5.5 events\,cm$^{-2}$\,s$^{-1}$ expected for {\slshape Roman} (see \citealt{2016PASP..128c5005K}, and note that at L2 we do not have the trapped electron population). While experience with similar detectors on the {\slshape James Webb Space Telescope} is that usually 1--2 pixels are affected \citep[\S6.6]{2022arXiv220705632R}, {\slshape Roman} has smaller pixels (so potentially a higher leakage into neighbors via charge diffusion) and we are aiming for higher precision. Therefore, for this simulation, we have masked a $3\times 3$ pixel region surrounding each cosmic ray. The less common larger events such as ``snowballs'' \citep[e.g.][Fig. 15]{2023PASP..135b8001R} are not included in the current simulation; we plan to add them in the future once we understand better how they should be implemented. However for the purposes of testing how {\sc Imcom} responds to masked pixels, it is the more frequent events that are likely to have the greatest impact.

\begin{figure}
    \centering
    \includegraphics[width=3.2in]{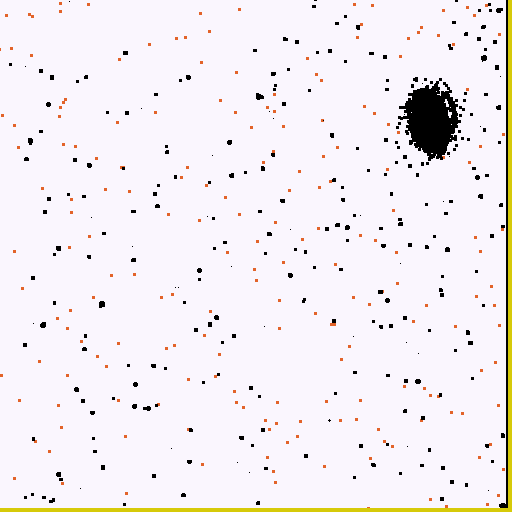}
    \caption{\label{fig:maskfig}An example of a simulated mask. This is the $512\times 512$ lower-right corner of SCA 11 in a H158-band observation (ID 8836). Reference pixels are shown in dark yellow; permanently masked pixels are shown in black; and pixels rejected in this observation only due to cosmic rays are shown in red-orange.}
\end{figure}

An example of a mask is shown in Fig.~\ref{fig:maskfig}.

\subsection{Injected sources}
\label{ss:inj}

We have also created two layers that are grids of injected stars. The stars have unit flux and are injected on a HEALPix\footnote{http://healpix.sourceforge.net} grid \citep{2005ApJ...622..759G} of resolution 14 ({\tt nside}$\,=2^{14}=16384$) in the equatorial coordinate system. The selection of grid points that land on each SCA was performed with {\sc HealPy} routines \citep{2019JOSS....4.1298Z}. Two layers are generated --- one with the external {\sc GalSim} package, and one with the internal interpolation machinery in our pipeline.

We constructed the grid of injected sources with {\sc GalSim} as follows. For each PSF that is characterized by the observation and SCA that overlaps with the output region, we interpolated the PSF made by the DC2+{\slshape Roman} simulation, which is oversampled by the factor of 8, using the interpolant, \emph{lanczos50}. The interpolated PSF image is then convolved with the delta function of unity, and is drawn at each grid point.

The internal simulation was generated from the same PSF, but interpolated using the D5,5,$\tfrac1{12}$ interpolation kernel. In principle one should get the same results from the two methods aside from the choice of interpolation kernel or treatment of edge effects; however, having both serves as an important cross-check on the implementation.

\subsection{Simulated noise fields}
\label{ss:noise}

We construct two types of simulated noise fields: white noise and $1/f$ noise. Since the image combination is a linear process, the proper normalization of these noise fields, or the choice to include both, can be chosen in post-processing. We therefore make the simplest choices to normalize the inputs.

The white noise input is a $4088\times 4088$ Gaussian random field with mean 0 and variance 1.

\begin{figure}
    \centering
    \includegraphics[width=2.75in]{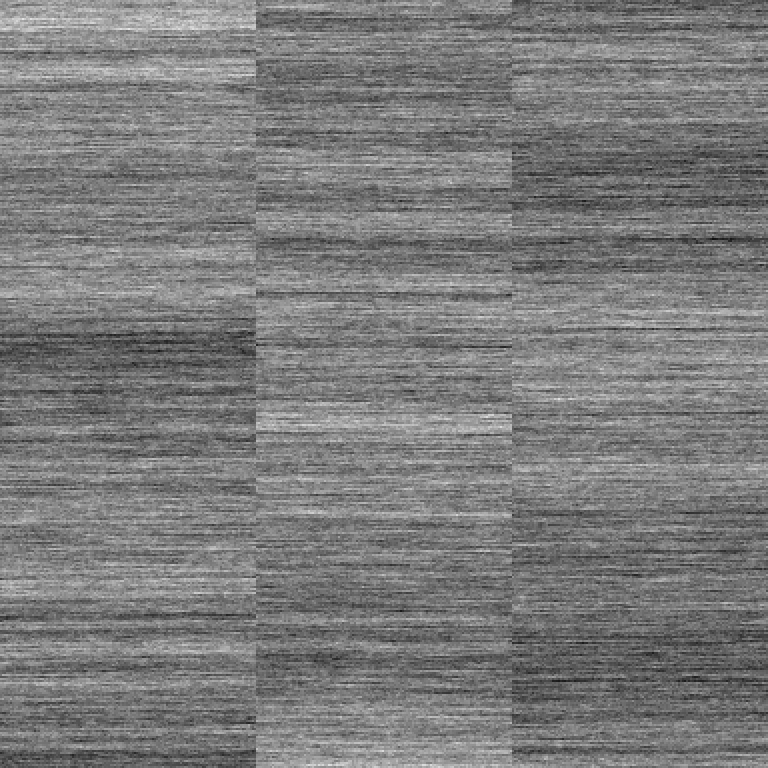}
    \caption{\label{fig:noise1f}A $384\times 384$ cutout of a $1/f$ noise field generated in Sec.~\ref{ss:noise}. The grayscale is a linear stretch from $-16$ to $+16$.}
\end{figure}

The $1/f$ noise input is constructed as follows. For each of the 32 readout channels, we construct a 1-dimensional Fourier domain signal of length $N=2^{20}$: $\{\tilde S_k\}_{k=-N/2}^{N/2-1}$, where the real and imaginary components of $\tilde S_k$ are Gaussians with mean 0 and variance $1/(2|f_k|)$ (except for the constant mode $\tilde S_0=0$), where $f_k = k/N$ is the normalized frequency. This is then discrete Fourier transformed, $S_j = \sum_{k=-N/2}^{N/2-1} \tilde S_k {\rm e}^{2\pi{\rm i}jk/N}$, and we take the real part. This gives a signal whose variance per logarithmic range in frequency is 1, i.e., the variance coming from all modes between $f_{\rm min}$ and $f_{\rm max}$ is $\approx \ln (f_{\rm max}/f_{\rm min})$. We take a length $2^{19}$ portion of this signal (to avoid wrapping effects), reformat it into a $4096\times 128$ array, and broadcast it into the appropriate portion of a $4096\times 4096$ image. The even channels are flipped left-to-right (see, e.g., Fig.~2 of \citealt{2020PASP..132g4504F} for the ordering of the readout pattern). This does not include guide window or row overheads, but it does provide a noise field with the characteristic horizontal banding pattern of $1/f$ noise that we can use to assess the impact on weak lensing with coadded images. An example of a $1/f$ noise field generated in this way is shown in Fig.~\ref{fig:noise1f}.

Both noise inputs use the {\tt numpy.random.default\_rng} random number generator, with a seed chosen based on the observation ID, the SCA, and an integer specified by the user. In order to reduce storage requirements, the noise fields are generated on the fly at the same time the science data is read. The seed construction ensures that when we make a mosaic coadd, and a given SCA image contributes to more than one tile of the mosaic, that the same noise realization is generated each time.

\subsection{Laboratory noise fields}
\label{ss:labnoise}

In addition to simulated noise fields, we construct a layer containing noise from laboratory testing of the SCA detectors. By including detector read noise in our analysis, we are able to test in a new way how the detectors themselves might impact galaxy shape measurement. The lab detector test noise fields are intended for a separate work, and thus a more complete and detailed analysis will be forthcoming in a companion paper to this. Here we will present the basic principles of the laboratory noise fields, and refer the interested reader to this future paper for further details.

The lab noise data, taken at the Detector Characterization Laboratory, are $N_{\rm t}\times 4096\times 4224$ cubes (where $N_{\rm t}$ is the number of time slices), including dark frames and low and high level flat frames for each SCA. Data is split into 32 readout channels, plus one reference output channel and four rows of reference pixels on each side of the grid. For construction of the noise frames, lab data is averaged into six effective time bins (``Multi-Accum'' processing; see e.g., Section 7.7 of \citealt{2022wfci.book...14D}), slope fitted, and reference pixel-corrected, and converted to electrons using the gain determined using {\sc solid-waffle} \citep{2020PASP..132g4504F}. We impose each SCA's lab-tested pixel mask \citep[Appendix B]{2022arXiv220906829T} on each frame so that the mask is the same in all input layers. This ensures that we are still able to process all the layers at the same time and thus only compute the ${\mathbfss T}$ matrix once. For the total focal plane, this additional masking amounted to 0.095\% of the pixels. Lab noise field analysis and results will be presented in detail in the forthcoming paper (Laliotis et al., in prep).

\begin{table*}
    \centering
    \caption{\label{tab:size}Sizes and dimensions used in this work.}
    \begin{tabular}{clcc}
    \hline\hline
        Parameter or variable name & Description & Value & Unit\\ \hline
        {\tt s\_in} & Input (native) pixel scale & 0.11 & arcsec \\
        $\Delta\theta$ ({\tt dtheta}) & Output pixel scale & 0.025 & arcsec \\
        $n_2$ & Postage stamp size in output pixels & 50 & ~ \\
        $k$ ({\tt fade\_kernel}) & Transition width in pixels & 3 & ~ \\
        $n_2\Delta\theta$ & Postage stamp angular size (excluding transition region) & 1.25 & arcsec \\
        $(n_2+2k)\Delta\theta$ & Postage stamp angular size (including transition region) & 1.4 & arcsec \\
        {\tt INPAD} & Acceptance radius for input pixels & 1.25 & arcsec \\
        $n_1$ & Block size in postage stamps (1D) & 48 & ~ \\
        {\tt PAD} & Padding region of block in postage stamps & 2 & ~ \\
        $n_1n_2\Delta\theta$ & Block angular size (excluding extra postage stamps) & 1.0 & arcmin \\
        $(n_1+2{\tt PAD})n_2\Delta\theta$ & Block angular size (including extra postage stamps) & 1.08333 & arcmin \\
        $(n_1+2{\tt PAD})n_2$ & Block image side length in output pixels & 2600 & ~ \\
        {\tt BLOCK} & Mosaic size in blocks (1D) & 48 & ~ \\
        {\tt BLOCK}$\times n_1n_2\Delta\theta$ & Mosaic angular size & 0.80 & degree \\
    \hline\hline
    \end{tabular}
\end{table*}

\section{Image coaddition}
\label{sec:co-add}

The image coaddition process is an updated version of the {\sc Imcom} framework for coaddition of postage stamps \citep{2011ApJ...741...46R}. Major changes include replacing the original Fortran 90 code with a Python interface and C back end; and the new driver script to produce mosaic coadds.

The mosaic process is organized hierarchically, controlled by keywords in the configuration file. The top level is a mosaic, consisting of a single world coordinate system over a region of sky small enough to be treated as ``flat'' to within weak lensing requirements (i.e., the coordinate system shear and plate scale variation should be $<10^{-4}$). The mosaic is divided into a square grid of blocks; the processing of each block in each filter is a single run of the Python script {\tt run\_coadd.py}, and the resulting coadded images and metadata are stored in a single FITS file. The blocks are themselves divided into a square grid of postage stamps, and are extended by some number of postage stamps ({\tt PAD}) so that there is overlap among the blocks. Each postage stamp has a grid of output pixels, and each output pixel is constructed as a linear combination of input pixels. The algorithm allows any input pixels that are un-masked and within a specified radius ({\tt INPAD} arc seconds) of the square stamp. The output pixels in a postage stamp are surrounded by a ring of transition pixels that allow the weights to vary smoothly when going to the next postage stamp.

The sizes of each hierarchical layer used in this paper are shown in Table~\ref{tab:size}. These are of course subject to refinement as we prepare for the eventual {\slshape Roman} science analysis.

\begin{figure*}
\includegraphics[width=6.75in]{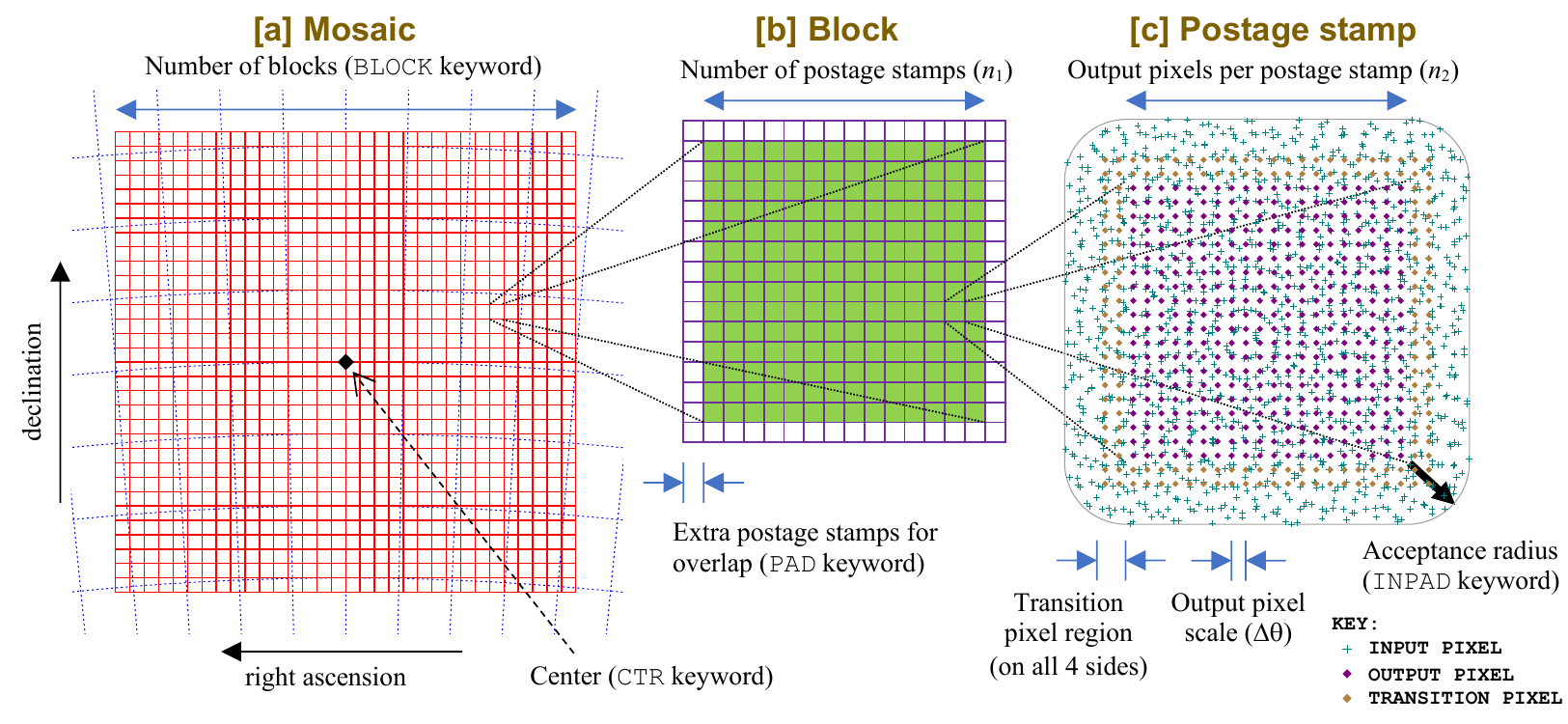}
\caption{\label{fig:tilefig}The hierarchical structure of mosaic coadds in this paper. The mosaic (panel [a]) is defined by a center, a map projection, and a number of blocks ({\tt BLOCK}$\times${\tt BLOCK}). Each block (panel [b]) is itself composed of postage stamps; we make an $n_1\times n_1$ array, with padding of {\tt PAD} postage stampps around the rim so that the blocks overlap. The postage stamps (panel [c]) are composed of an $n_2\times n_2$ grid of output pixels, with a transition region around the edge that is merged at the block processing level before writing to a FITS file. The postage stamp is built from all un-masked input pixels in all input images within a given acceptance radius of the stamp.}
\end{figure*}

We briefly review the postage stamp coaddition problem in Sec.~\ref{ss:imcom-postage}, and indicate where there have been algorithm updates. We describe the block and mosaic construction in Sec.~\ref{ss:imcom-mosaic}. The choice of output PSFs --- essentially determining the resolution of the coadded images --- is discussed in Sec.~\ref{ss:imcom-output}.

\begin{figure*}
\centering
\includegraphics[width=6.5in]{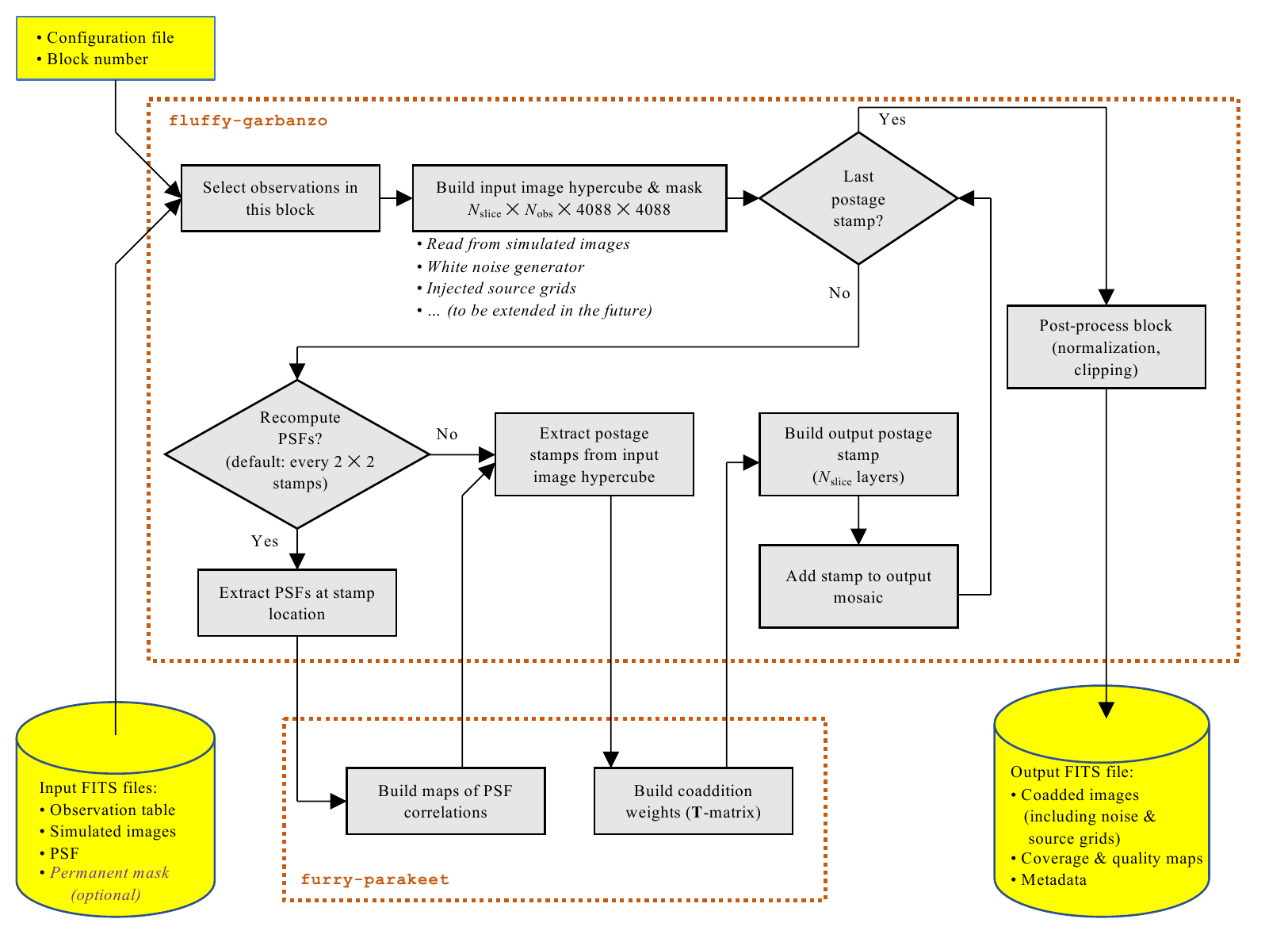}
\caption{\label{fig:workflow}The workflow for coaddition of a block in this paper. There are two repositories: the postage stamp coaddition ({\sc furry-parakeet}) and the mosaic driver with interfaces to the simulations ({\sc fluffy-garbanzo}).}
\end{figure*}

A diagram of the block coaddition script is shown in Fig.~\ref{fig:workflow}.

\subsection{Postage stamp coaddition}
\label{ss:imcom-postage}

The {\sc Imcom} formalism is developed in \citet{2011ApJ...741...46R}\footnote{See also \citet{2010PASP..122..248B}, who develop a related formalism for combining the pixels in 2D fiber spectra into a 1D spectrum.}, and here we summarize only the key points that are needed here. {\sc Imcom} produces a coadded postage stamp starting from a set of input postage stamps: these images can be flattened and concatenated into a length $n$ vector $\{I_i\}_{i=0}^{n-1}$.\footnote{Since the new implementation is in Python+C, in this paper we follow the Python+C convention of indices starting at 0.} Masked pixels can simply be removed from this vector with a corresponding decrease in $n$. These are to be combined into an output image with $m$ pixels: $H_\alpha = \sum_{i=0}^{n-1} T_{\alpha i}I_i$, where ${\mathbfss T}$ is an $m\times n$ coaddition matrix. We denote the positions of each pixel center by $\{{\bmath r}_i\}_{i=0}^{n-1}$ (input images) or $\{{\bmath R}_\alpha\}_{\alpha=0}^{m-1}$ (output image). If the input images have effective PSF $G_i$, then the output pixel $\alpha$ has an effective PSF \changetext{at offset ${\bmath s}$} given by\footnote{There is a subtlety in defining the ``PSF'' of a pixel when the PSF does not act exactly as a convolution, i.e., Eq.~(5) of \citet{2023OJAp....6E...5M} is not satisfied. In general a pixel at position ${\bmath R}_\alpha$ responds to a point source at some position ${\bmath r}$ according to a PSF $G({\bmath R}_\alpha-{\bmath r})$. The term ``point spread function'' literally refers to this function at fixed ${\bmath r}$ and varying ${\bmath R}_\alpha$. Equation~(\ref{eq:PSFout}) fixes the output pixel ${\bmath R}_\alpha$ and varies ${\bmath r}$, which is a more natural choice when pixels are discrete and positions on the sky are continuous. One might think of this object as a ``reverse'' of the usual PSF.}
\begin{equation}
{\rm PSF}_{{\rm out},\alpha}({\bmath s}) = \sum_{i=0}^{n-1} T_{\alpha i} G_i({\bmath R}_\alpha - {\bmath r}_i + {\bmath s}).
\label{eq:PSFout}
\end{equation}
It is a common difficulty with image combination algorithms that PSF$_{{\rm out},\alpha}$ varies from pixel to pixel in the coadded image.
{\sc Imcom} takes as input a ``target PSF'' $\Gamma$, and attempts to find coaddition weights $T_{\alpha i}$ that make PSF$_{{\rm out},\alpha}$ uniform and close to $\Gamma$. Quantitatively, we wish to minimize the leakage function\footnote{We \changetext{intended to take $\tilde\Upsilon({\bmath u})=1$. As noted in Section~4.4, the production runs were run with an alternative weighting in the objective function, $\tilde\Upsilon({\bmath u})=[1 + A_\Upsilon {\rm e}^{-2\pi^2\sigma^2_\Upsilon(u^2+v^2)}]^2$, with $A_\Upsilon=1$ and $\sigma_\Upsilon=\frac32\times 0.11\,$arcsec.}},
\begin{equation}
\frac{U_\alpha}C = \frac{\lVert {\rm PSF}_{{\rm out},\alpha} - \Gamma \rVert^2}{\lVert \Gamma\rVert^2},
~~~\lVert f \rVert^2 \equiv \int \tilde\Upsilon({\bmath u}) |\tilde f({\bmath u})|^2 \,d^2{\bmath u},
\label{eq:UC}
\end{equation}
\changetext{where the square norm is written in Fourier space with weighting $\tilde\Upsilon({\bmath u})$.
The numerator $U_\alpha$ is the square norm of the difference between the as-realized output PSF and the target, while the denominator $C$ is the square norm of the target (thus forming a dimensionless ratio). We will refer to $U_\alpha/C$ itself as the ``leakage'' (smaller is better), and to the quantity $-10\log_{10} (U_\alpha/C)$ as the ``fidelity'' (larger is better: one can think of this as a suppression of PSF error in decibels).}
If the $n\times n$ noise covariance matrix of the input is ${\mathbfss N}$, then the $m\times m$ output covariance is ${\boldsymbol\Sigma} = {\mathbfss T}{\mathbfss N}{\mathbfss T}^{\rm T}$. {\sc Imcom} also tries to minimize the output covariance $\Sigma_{\alpha\alpha}$ for input white noise (${\mathbfss N} = {\mathbb I}_{n\times n}$).\footnote{It is possible that there could be a benefit to tuning the algorithm to optimize for non-white noise. We plan to assess the utility of this after studying the noise properties with the flight electronics.} 

Having two objective functions, leakage $U_\alpha/C$ and noise $\Sigma_{\alpha\alpha}$, implies a trade-off: usually one can be improved at the expense of the other. We thus optimize row $\alpha$ of the ${\mathbfss T}$-matrix with respect to the objective function $U_\alpha+\kappa_\alpha\Sigma_{\alpha\alpha}$, where $\kappa_\alpha$ is a Lagrange multiplier that controls the trade-off (decreasing $\kappa_\alpha$ decreases leakage and increases noise; increasing $\kappa_\alpha$ increases leakage and decreases noise). The method of solving for $T_{\alpha i}$ given $\kappa_\alpha$ is described in \citet{2011ApJ...741...46R}; the algorithm performs a bisection search in $\log_{10}\kappa_\alpha$, with the user specifying the criterion for whether to increase or decrease $\kappa_\alpha$.

Our postage stamp coaddition pipeline implements almost the same algorithm as in the original {\sc Imcom}, but with a Python interface and using NumPy routines for the linear algebra and Fast Fourier Transform (FFT) steps; the {\tt for} loop over $\kappa_\alpha$ and the interpolation are coded in C and wrapped using the NumPy C Application Programming Interface. Aside from changes in language, the substantive differences in the relative to the original {\sc Imcom} are:
\begin{itemize}
\item The construction of ${\mathbfss T}$ in \citet{2011ApJ...741...46R} requires the construction of the correlations $[G_i\circ G_j]({\bmath s})$ for every pair of input PSFs. These are now computed and saved in the {\tt PSF\_Overlap} class, because these correlations only need to be re-computed on the scale of the PSF variation, not for every single postage stamp. This saves time when constructing a large-area mosaic.
\item The interpolation of the correlations $[G_i\circ G_j]({\bmath s})$ in the original {\sc Imcom} was performed using bivariate polynomial interpolation. In our case, since the PSFs are band-limited, we know that their correlation is band-limited and given a grid size we know the oversampling factor (i.e., sample spacing relative to the Nyquist spacing). We have derived optimal interpolation kernels for this case in Appendix~\ref{app:interpolation}; we used the ``D5,5,$\tfrac1{12}$’’ kernel here.
\item Rather than simply specifying a target leakage $U_\alpha/C$ or a target noise $\Sigma_{\alpha\alpha}$, we specify a maximum noise $(\Sigma_{\alpha\alpha})_{\rm max}$ as our first priority and a maximum leakage $(U_\alpha/C)_{\rm max}$ as our second priority. That is, if both the leakage and noise targets can be met, the Lagrange multiplier search looks for the minimum noise that meets the leakage requirement; but if they cannot both be met, it treats the noise target as a hard constraint and finds the minimum leakage subject to that constraint. Of course in the latter case we would consider masking that pixel in a weak lensing analysis, but it avoids spectacular noise amplifications that could confuse some downstream analysis packages.
\end{itemize}
The postage stamp coaddition tools are in a separate GitHub repository (furry-parakeet) from the block driver.

\subsection{Blocks and mosaics}
\label{ss:imcom-mosaic}

The coaddition of a block (Fig.~\ref{fig:workflow}) begins with the selection of observation ID/SCA pairs that overlap with the block. Once we have these observations, a 4D numpy array is constructed to include all of the input data: {\tt in\_data[$i$,$j$,$y$,$x$]} contains the pixel in layer $i$, observation ID/SCA $j$, and pixel $(x,y)$. Each layer consists of simulated images, noise fields, or grids of injected sources (Table~\ref{tab:layers}). The input data are stored as 32-bit floating point numbers.

The algorithm then executes a loop over the postage stamps in that block. To save computing time, the PSFs are extracted and a {\tt PSF\_Overlap} object created every 4th postage stamp (it is treated as uniform in a block of $2\times 2$ postage stamps, based on the PSF for the position at the common corner of the stamps). Then a suite of input postage stamps is created by mapping the centers of the output postage stamp back to each input image; these input stamps are deliberately oversized, since they can then be reduced by combining the input image mask with the logical test for whether a pixel is within the acceptance region (right panel of Fig.~\ref{fig:tilefig}). Then we build the ${\mathbfss T}$-matrix (\S\ref{ss:imcom-postage}), and multiply to get an output cube (postage stamp for each layer).

A complication arises when the postage stamps are tiled to make a block: if the postage stamps are simply placed next to each other, there are noise discontinuities in the output image since one suddenly jumps to a different set of input pixels from which one can construct the output pixels. For some applications, these discontinuities may present no difficulties. However, some common applications such as peak finders that run at the resolution of the output image may be confused by these effects \citep[e.g.][]{1996A&AS..117..393B}. Therefore, we use the transition pixels to smoothly transition from one postage stamp to the next. The overlap of the output postage stamps (including transition pixels) is $2k$ pixels; we define a sequence $\{a_m\}_{m=1}^{2k}$. If we have a ``left'' postage stamp $I_{\rm left}(x,y)$ (whose output region, excluding the transition pixels, is $x<x_{\rm cut}$) and a ``right'' postage stamp $I_{\rm right}(x,y)$ (whose output region, again excluding the transition pixels, is $x\ge x_{\rm cut}$), then in the transition region $x_{\rm cut}-k\le x<x_{\rm cut}+k$, we may make a merged image,
\begin{equation}
I_{\rm merged}(x,y) = a_{x_{\rm cut}-x+k}I_{\rm left}(x,y)+ a_{x-x_{\rm cut}+k+1}I_{\rm right}(x,y).
\label{eq:I-merge}
\end{equation}
This approach does not affect the output PSF as long as the transition sequence satisfies $a_m + a_{2k+1-m} = 1$. We use the truncated sine function \citep{2022arXiv220810522L},
\begin{equation}
a_m = \frac{m}{2k+1} - \frac1{2\pi}\sin \frac{2\pi m}{2k+1},
\end{equation}
which avoids the discontinuities of the first derivative of the noise that would occur without the second term. The merging technique in Eq.~(\ref{eq:I-merge}) is trivially extendable to the $y$ direction as well.

The mosaics are built using a single map projection with an array of blocks. In principle, one could select from any of the projections in the FITS WCS standard \citep{2002A&A...395.1077C}. We have used the stereographic projection here since it introduces no shear distortion, has zero plate scale gradient at the projection center, and the {\slshape Roman} footprint can be efficiently tiled with square regions\footnote{The Mercator projection satisfies the first two criteria, and would be appropriate to a great circle strip, e.g., \citet{2014MNRAS.440.1296H}; but since strips converge on a sphere, the unique area must be tapered.}, although we may revisit this choice in the future. At an angle $\theta$ from the projection center, the stereographic projection introduces a lowest-order plate scale error of $\frac14\theta^2$, or $2.4\times 10^{-5}$ at the corner of a 0.8 degree square ($\theta=0.8/\sqrt2\times\pi/180$). Thus for this test region we only need one mosaic.

We refer to a block within a mosaic by its array position: $(0,0)$ is in the lower-left (southeast) corner, and $({\tt BLOCK}-1,0)$ is in the lower-right (southwest corner).

\subsection{Output PSFs}
\label{ss:imcom-output}

\begin{table}
    \centering
    \caption{\label{tab:outpsf}Properties of the input and output PSFs. The PSF effective area is defined in the square-norm sense $\Omega_{\rm psf} = 1/\int [G({\bf s})]^2\,d^2{\bf s}$ and expressed in input pixels. The sampling factors are in the convention of $\lambda/Ds_{\rm in}$; this is the same convention used in \citet{2011ApJ...741...46R}, but is a factor of 2 larger than the convention of \citet{2021MNRAS.502.4048K}. \changetext{The output smearing is listed as the full width at half maximum of the Gaussian in units of native pixels.}}
    \begin{tabular}{ccccc}
    \hline\hline
    Band & \multicolumn2c{Input PSF} & \multicolumn2c{Output PSF} \\
         \cmidrule(lr){2-3}\cmidrule(lr){4-5}
    & Effective  & Sampling  & Output & Full width at \\
     & area & factor & smearing & half maximum \\
    ~ & $\Omega_{\rm psf}/s_{\rm in}^2$ & $Q=\lambda/Ds_{\rm in}$ & $\sqrt{8\ln 2}\,\sigma/s_{\rm in}$ & $\theta_{\rm FWHM}$ [arcsec] \\
         \hline
    Y106 & \textcolor{white}07.06 & 0.834 & 2.25 & 0.279 \\
    J129 & \textcolor{white}08.60 & 1.021 & 1.75 & 0.230 \\
    H158 & 10.96 & 1.250 & 1.50 & 0.210 \\
    F184 & 15.28 & 1.456 & 1.25 & 0.200 \\
    \hline\hline
    \end{tabular}
\end{table}

\begin{figure}
    \centering
    \includegraphics[width=3.2in]{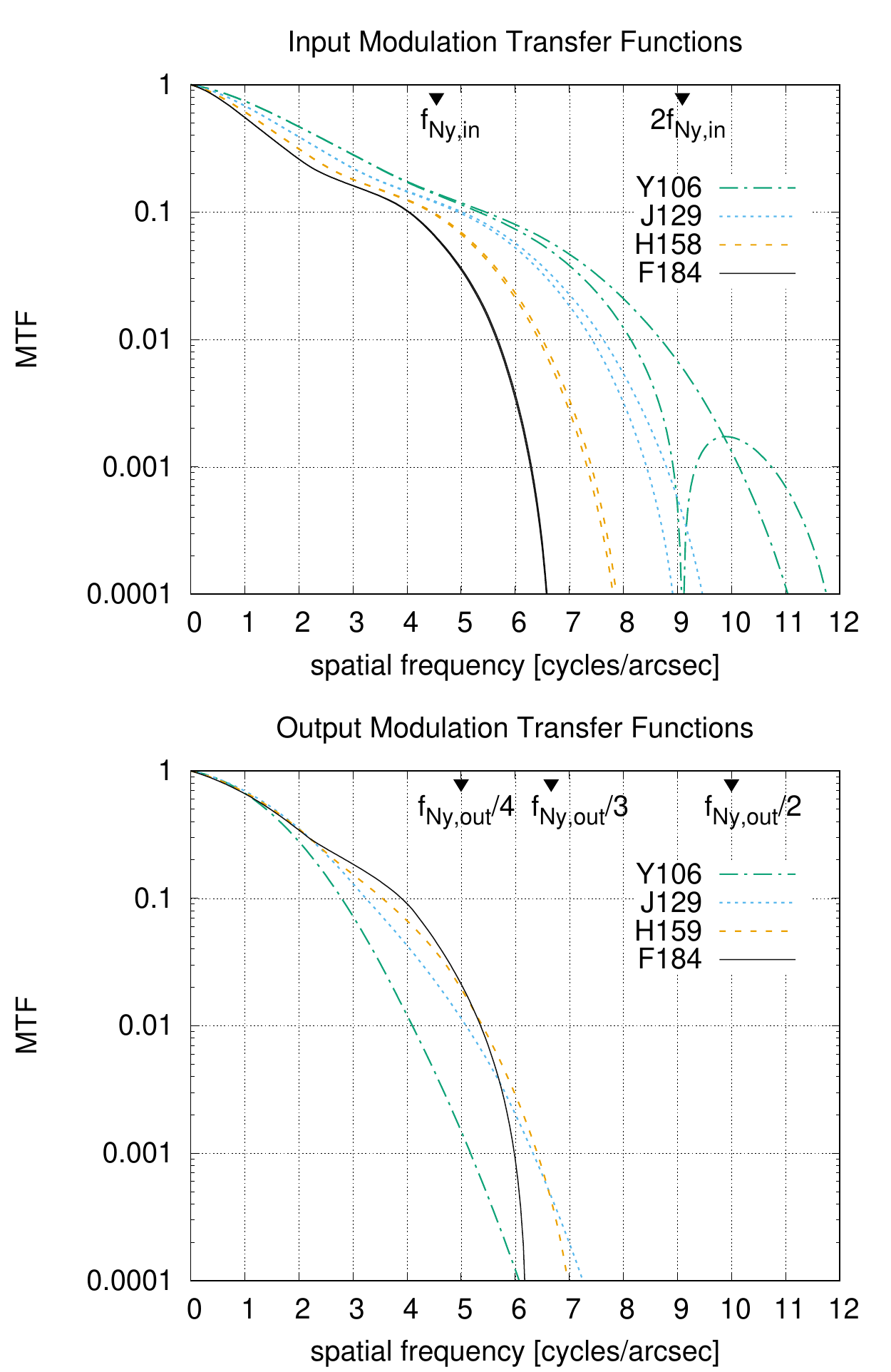}
    \caption{\label{fig:tf}The input and output modulation transfer functions (MTFs: absolute value of the Fourier transform of the PSF). The upper panel shows the MTFs as computed in the Exposure Time Calculator \citep{2013ascl.soft11012H} based on the requirement aberrations. Two curves are shown for each filter, either for wave vector aligned with the pixel grid or at a 45$^\circ$ angle; for F184, the difference is not visible on the plot. The Nyquist frequency at the native (input) pixel scale is marked. The lower panel shows the output PSFs, with fractions of the Nyquist frequency at the output pixel scale $s_{\rm out}$ marked. Note that all of the input images are undersampled (the MTF is nonzero for spatial frequencies extending past $f_{\rm Ny,in}$) but the output images are well sampled, with the MTF dropping to negligible levels ($<10^{-4}$) by $\sim 0.36f_{\rm Ny,out}$ in all filters.
    }
\end{figure}

The {\sc Imcom} formalism is different from many other coaddition algorithms in that the target output PSF $\Gamma$ is specified by the user and this is used to determine the input-to-output mapping (${\mathbfss T}$-matrix), rather than the other way around. It is therefore advantageous to choose an output PSF that is circularly symmetric, so that the image coaddition and ``rounding kernel'' are accomplished in a single step.

A straightforward choice is the convolution of an obstructed Airy disc \citep[e.g.][]{1986ApOpt..25.2404R} with a Gaussian (whose width determines the resolution of the coadded image):
\begin{equation}
\Gamma({\bmath s}) = \int_{{\mathbb R}^2}
\frac{[J_1(\pi s'/\xi) - \varepsilon J_1(\pi \varepsilon s'/\xi)]^2}{\pi (1-\varepsilon^2) s'{^2}}
\frac{{\rm e}^{-({\bmath s}-{\bmath s}')^2/(2\sigma^2)}}{2\pi\sigma^2}
{\rm d}^2{\bmath s}',
\end{equation}
where $\xi=\lambda/D$ is the diffraction scale at the central wavelength of that filter, $J_1$ is the Bessel function of the first kind, and $\varepsilon$ is the linear obstruction fraction. This profile has an exact band limit at the diffraction scale, just like all of the input PSFs, and it has the usual $\propto 1/s^3$ diffraction wings with the same amplitude as for the input PSFs. Being circularly symmetric, it does not contain the diffraction spikes.

The choice of $\sigma$ involves a trade-off. If $\sigma$ is larger, then the output PSF is larger, which makes it more difficult to measure the shapes of small galaxies and increases blending. If $\sigma$ is smaller, then the algorithm will attempt to reconstruct the higher spatial frequency Fourier modes in the image, which usually suffer more aliasing. This means that the output PSF \changetext{leakage} $U_\alpha/C$ is larger (worse). This effect is more significant with small numbers of dithers where not all combinations of input Fourier modes can be de-aliased. The implication is that for smaller $\sigma$ (hence smaller target PSF), we may need to mask a larger portion of the final output image to control shear measurement systematics. (A possible way around this, which we have not explored in this paper, is to build several mosaics at different output resolutions, with the lowest resolution being available everywhere and the highest resolution available in select regions with more dithers.)

\begin{figure*}
    \centering
    \includegraphics[width=7in]{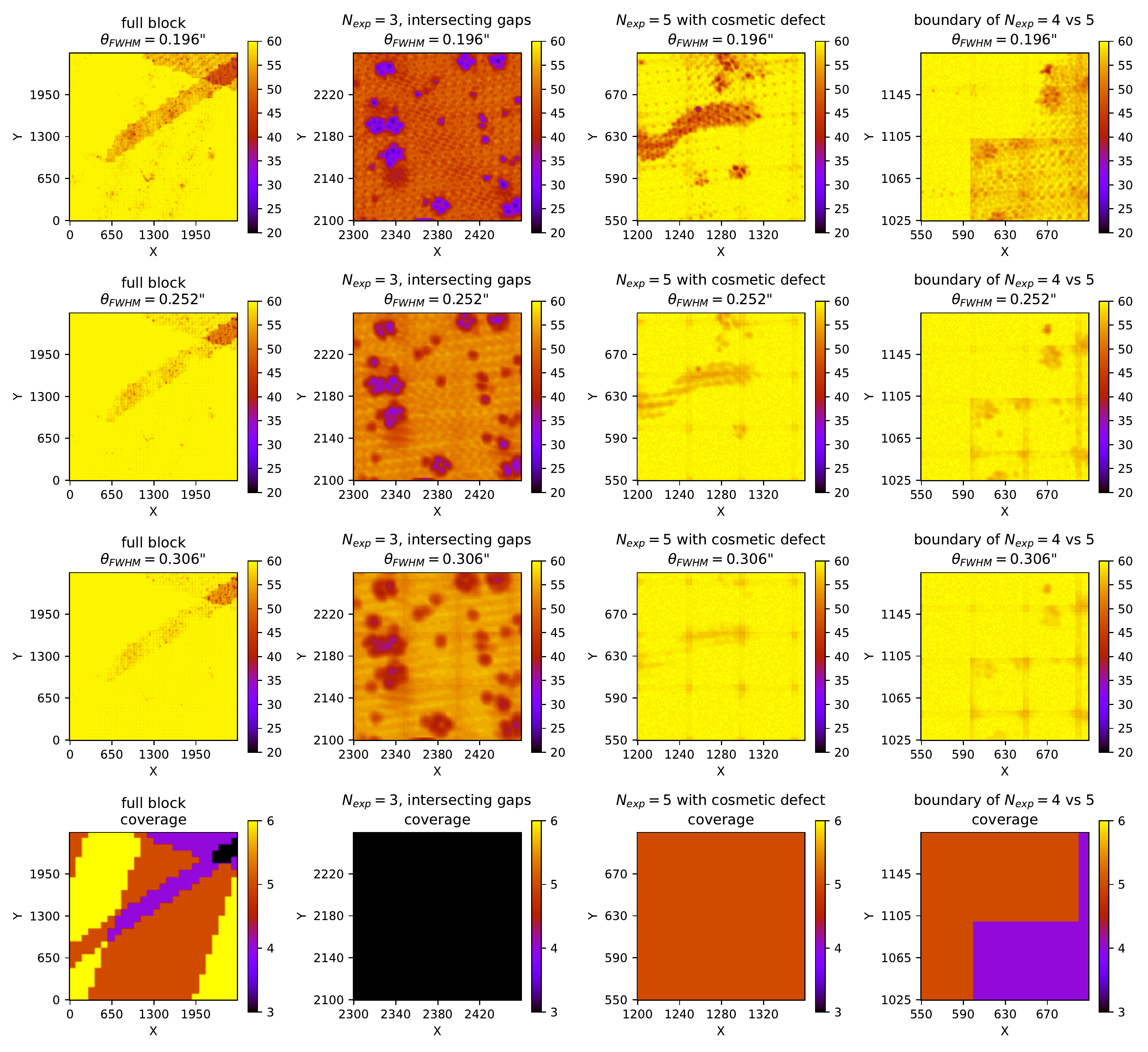}
    \caption{\label{fig:yresfig}Examples of how output fidelity depends on the output PSF and on the number of exposures covering a given region. The first three rows show the fidelity maps, $-10\log_{10} (U_\alpha/C)$, for a block of the {\slshape Roman} Y106 band centered at right ascension 53.5142$^\circ$ and declination $-40.3898^\circ$. On this scale, 60 corresponds to the specified leakage $U_\alpha/C = 10^{-6}$; lower values correspond to worse matching of the output PSF to the target. We show three resolutions of the target PSF ($\theta_{\rm FWHM}$ indicated), with sharpest resolution at the top. The bottom row shows the coverage map (number of exposures). The first column shows the full $1.08\times 1.08$ arcmin block; the other three columns show zoom-ins of particular regions (the tics are in units of output pixels). Note the particularly poor reconstruction when $N_{\rm exp}=3$ (second column), corresponding to an intersection of chip gaps. Even when $N_{\rm exp}=5$ (third column), if the target PSF is too small (top), one sees imprinting of the 0.11 arcsec (4.4 output pixel) input pixel grid where there is difficulty interpolating from the small number of samples to certain sub-pixel positions. We also see losses where a cosmetic defect reduces the effective number of exposures to 4. Similar behavior can be seen in the right column, which overlaps with a 4-exposure region.}
\end{figure*}

An example of the resolution versus output fidelity trade-off in the {\slshape Roman} Y106 band is shown in Fig.~\ref{fig:yresfig}. The smoothing scales chosen for each of the {\slshape Roman} filters in this simulation are shown in Table~\ref{tab:outpsf}, and the modulation transfer functions are shown in Fig.~\ref{fig:tf}.

It may seem strange to additionally smooth the image to mitigate sampling effects; after all, the high angular resolution is one of the reasons to build a space mission in the first place. However, it is actually quite similar to the suggestion in \citet{2021MNRAS.502.4048K} to use a larger weighting scale for galaxy ellipticities\footnote{As an example of how these ideas are related, let us consider the quadrupole moment of an image $I$ at radius $\sigma$, $Q_\sigma[I] = \int I({\bmath r}) (x+{\rm i}y)^2 {\rm e}^{-r^2/2\sigma^2}\,{\rm d}^2{\bmath r}$. Let us think about what happens once full sampling is recovered so that we can really think of this operation as an integral. If a new image $I_{\rm sm}$ is constructed as the convolution of $I$ with a Gaussian of scale $\sigma_0$, then $Q_\sigma[I_{\rm sm}] = (\sigma/\sigma_{\rm tot})^6 Q_{\sigma_{\rm tot}}[I]$ where $\sigma_{\rm tot}^2 = \sigma^2+\sigma_0^2$. That is, the second moment of the smeared image is (aside from a scaling factor) the second moment of the original image at a larger scale.} and the convolution suggested by \citet{2022AJ....164..214S}. And we note that the output PSFs in Table~\ref{tab:outpsf} are still much smaller than those obtained in ground-based surveys.

\subsection{Known issues}

There are several issues that arose during the production runs for this project. These were not considered significant enough to warrant re-running the simulations, nor will they affect any of our primary results. However, we do plan to address them in a future version of the simulation:
\begin{list}{$\bullet$}{}
\item There was an indexing error that led to the wrong cosmic ray map being read in the Y106 and F184 simulations (this was fixed prior to the J129 and H158 runs). When a postage stamp was being created using the $i$th input image overlapping that postage stamp, the cosmic ray mask was read from the $i$th input image for the overall block. Since the cosmic ray maps are realizations of the same random process, this almost always has no effect. But it does mean that in a transition region between two postage stamps at the edge of a chip, there could be cases where the two postage stamps have inconsistent cosmic ray masks.
\item The lookup algorithm that chooses input images to be used for a block used a search radius that did not account for plate scale variations. This means that there are a few instances where the corners of a chip overlap the block but that input image was not used. In the few cases where this happened, the sense of the effect is that the chip gaps in the coaddition simulation are larger than what we will have with the real data. \changetext{In addition, sometimes the PSF computation points (at the center of every $2\times 2$ group of postage stamps) went off the edge of the SCA and hence that SCA was rejected from all 4 of the surrounding postage stamps; this also has the effect of increasing the effective chip gaps.}
\item The focal plane layout in the image simulations has some chip spacings that are different from the as-built {\slshape Roman} focal plane. The simulation and the WCS used to coadd the images are however internally consistent.
\item \changetext{The output WCS is shifted 2 postage stamps (2.5 arcsec) from the intended output region due to an error in writing the WCS. This has no effect on the validity of the results since all downstream steps consistently used this WCS; we simply coadded a slightly different region than intended.}
\item \changetext{The weighting of Fourier modes in the leakage function was intended to be uniform, $\tilde\Upsilon({\bmath u})=1$ in the language of \citet{2011ApJ...741...46R}. An experimental feature to place more weight on the longer-wavelength modes, setting $\tilde\Upsilon({\bmath u}) = [1 + A_\Upsilon {\rm e}^{-2\pi^2\sigma_\Upsilon^2(u^2+v^2)}]^2$, was accidentally left on for the production runs with $A_\Upsilon=1$ and $\sigma_\Upsilon = \frac32\times 0.11$\,arcsec. This placed more weight in the optimization than intended on the lowest spatial frequencies (relative to the highest spatial frequencies) by a factor of $\approx 4$; since the resulting ${\mathbfss T}$ is still a valid coaddition matrix, the results on {\sc Imcom} performance and applicability to weak lensing shape measurement remain valid.}
\end{list}

\begin{figure*}
\includegraphics[width=5.4in]{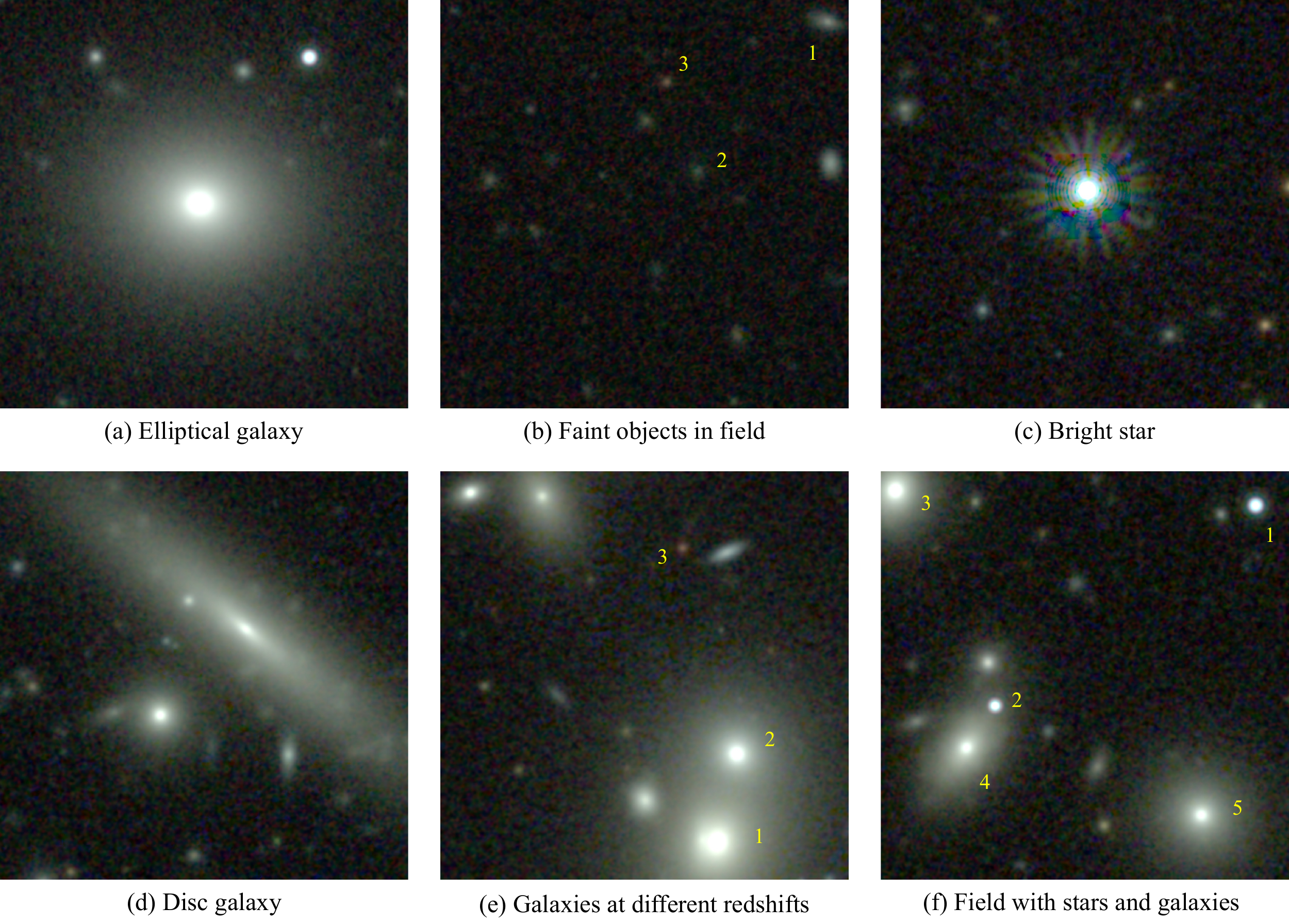}
\caption{\label{fig:interestingobjects}An assortment of objects in the simulated coadded images. Each panel is a Y106 (blue) + J129 (green) + F184 (red) composite, with a field of 18 arcsec on a side, with a scale stretched from $-8$ to $+1200$ $e/s_{\rm in}^2/$exposure. (a) An elliptical galaxy at $z=0.18$. (b) A deep field; the objects labeled are galaxies at $z=0.81$ (object 1), $z=0.79$ (2), and $z=0.93$ (3). (c) A bright star (visible magnitude $m_{550\,\rm nm}=16.4$); note the square pattern imprint of the postage stamp boundaries and the diffraction spikes. (d) A disc galaxy at $z=0.22$. (e) An assortment of galaxies; objects 1 and 2 are at $z=0.34$, and object 3 at $z=2.81$ appears red since the Balmer + 4000\AA\ feature is redshifted into the H158 band. (f) An assortment of stars (1 and 2) and galaxies (3--5 are in a group at $z=0.28$).}
\end{figure*}

\begin{figure*}
    \centering
    \includegraphics[width=6in]{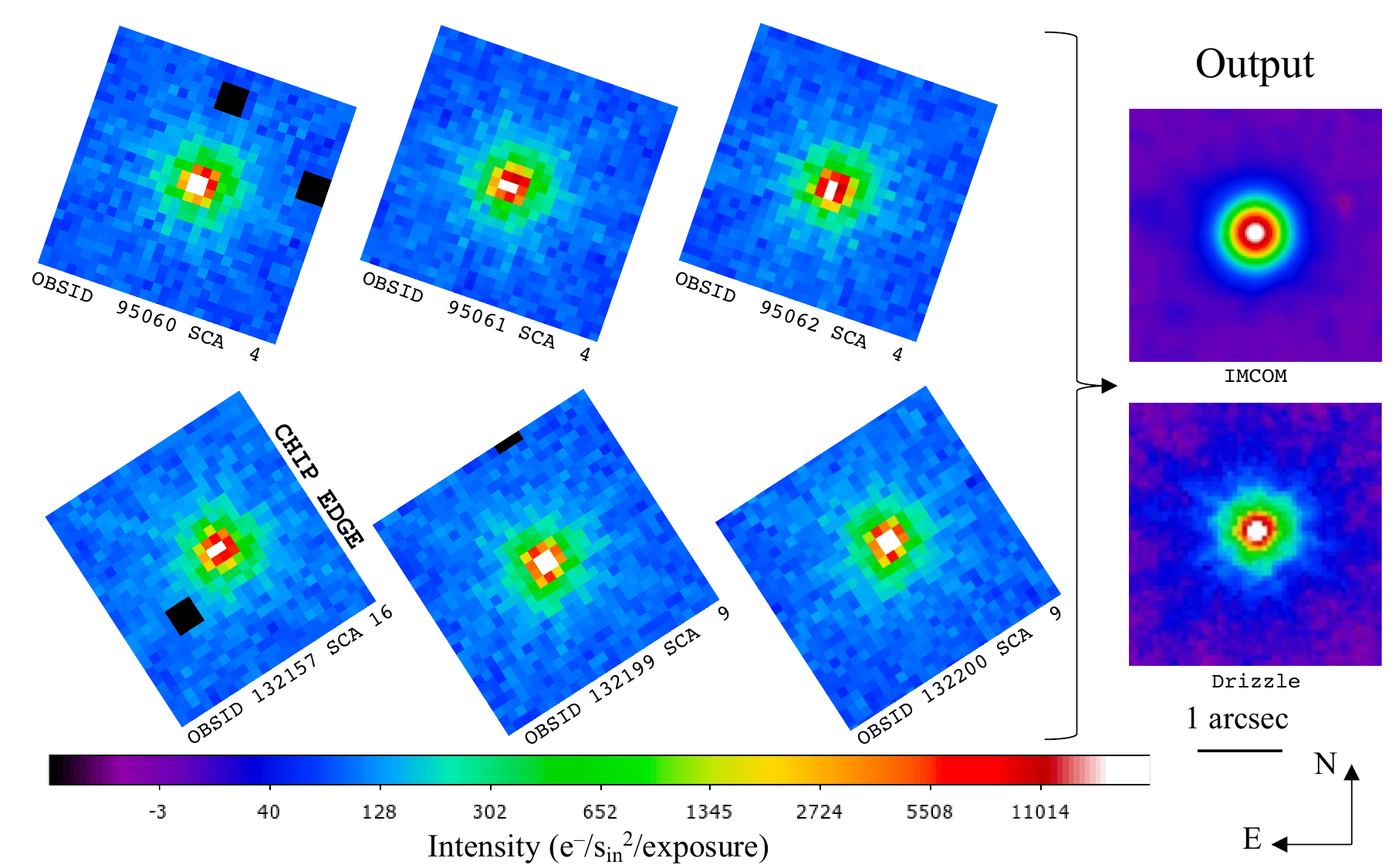}
    \caption{\label{fig:stackstar}An example of a simulated coadded star in the Y106 band. The left part of the figure shows the 6 images of the star, at the as-observed roll angles in the two passes, in the ``simple'' DC2 simulation. The right panel shows the coadded map. The {\sc Imcom} coadd (this project) is shown at top; the drizzled coadd is shown below. A logarithmic color stretch is used. The sky level (61 e/p) is present in the inputs but has been removed in the outputs. The black squares in the inputs are masked pixels (for the {\sc Imcom} run).}
\end{figure*}

\section{Results}
\label{sec:results}

We present some general results, examples, and statistics of the output PSF \changetext{leakage} $U_\alpha/C$ and features in the noise here. Statistical analyses of the results in the context of weak lensing are presented in the companion Paper II.

Some examples of regions in the coadded images are shown in Fig.~\ref{fig:interestingobjects}.

\subsection{Stars}

An example of a star propagating through the coadd pipeline is shown in Fig.~\ref{fig:stackstar}. We show both the {\sc Imcom} coaddition as well as the same star coadded in with the commonly used ``Drizzle'' algorithm \citep{2002PASP..114..144F} as described in Appendix C of \citep{2022arXiv220906829T}. The Drizzle output has pixel scale 0.0575 arcsec and {\tt pixfrac}=0.7. Note that the {\sc Imcom} algorithm attempts to produce a round output PSF --- this means, in particular, that the diffraction spikes are suppressed to the extent possible by adjusting the ${\mathbfss T}$-matrix. Some imperfections at the level of a few$\times 10^{-4}$ of the central surface brightness remain visible. Drizzle is a more local operation, and thus the asymmetry of the original PSFs as well as the diffraction spikes are more prominent. 

\subsection{Output PSF fidelity}

We describe the output PSF fidelity in terms of the quantity
$-10 \log_{10}(U_\alpha/C)$. This is a measure of how well the {\sc Imcom} algorithm thinks it has done on matching the PSF of the output image to the target PSF. The requested fidelity on this scale is 60, which corresponds to a 0.1\% error (in a root-sum-square sense) of all of the moments of the PSF.

\begin{figure*}
\centering
\includegraphics[width=6in]{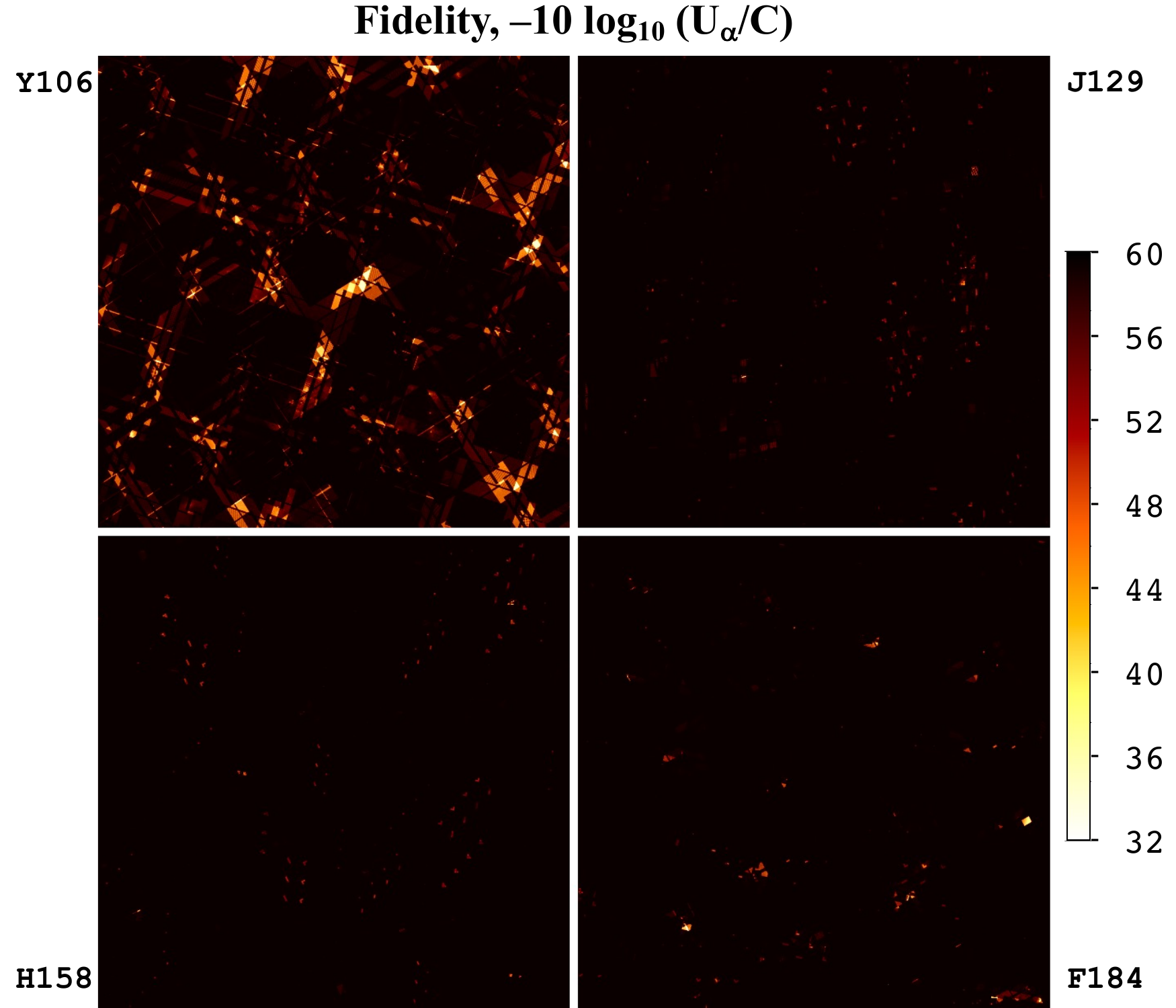}
\caption{\label{fig:fidelitymap}The fidelity map over the $48\times 48$ arcmin region simulated, in each of the 4 bands. The lowest values are seen in Y106, since it has the most undersampled PSF and the number of dither positions is lower than in J129. Note that the same features in the coverage map (Fig.~\ref{fig:coveragemap}) also appear here: the output PSF is not as well matched to the target in regions of intersecting chip gaps.}
\end{figure*}

\begin{figure}
    \centering
    \includegraphics[width=3.2in]{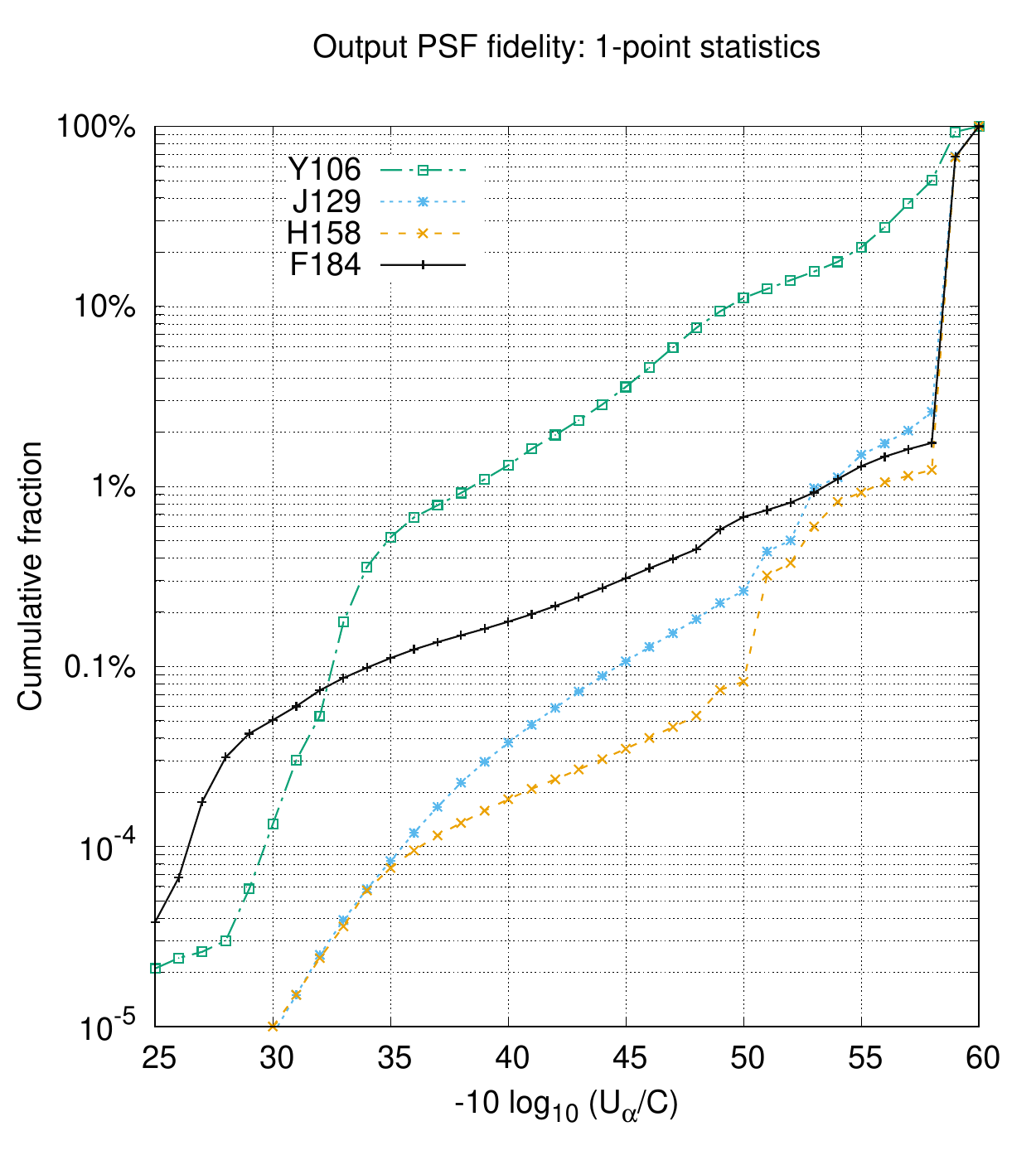}
    \caption{\label{fig:fcum}The cumulative distribution of the output PSF fidelity $-\log_{10}(U_\alpha/C)$ for each of the Roman filters. The vertical axis shows the fraction of the output $0.8\times 0.8$ degree mosaic with PSF fidelity worse than the indicated value.
    }
\end{figure}

The fidelity maps over the full simulation region for this paper are shown in Fig.~\ref{fig:fidelitymap}. The chip gaps are easily visible as regions of lower fidelity: in these cases, there were insufficient dither positions to break the degeneracies among the various Fourier modes in the image.
Equation~(\ref{eq:KQ}) gives a mode-counting argument for the minimum number of dithers required to disambiguate all Fourier modes in the astronomical scene (note that this is a necessary but not sufficient condition): this argument leads us to a number of dithers of 5 (Y106), 4 (J129), 3 (H158), and 2 (F184). However, these numbers could prove insufficient due to accidental degeneracies in the dithering pattern (e.g., two dithers at the same roll have an integer pixel offset) or masked pixels; and the required number may also be lower if the high spatial frequency modes have low enough transfer function that they can be ignored even if they are formally present in the image. The {\sc Imcom} simulations in this paper provide a more robust determination of the needed number of dithers in each filter. The worst performance is in Y106 band, since the Reference observing strategy was not originally designed for shape measurement using Y106\footnote{Recall this pre-dated the realization that assigning a galaxy to a photometric redshift bin is itself a part of source sample selection and therefore a part of shear measurement.} and so we accepted fewer dither positions. The trouble comes mainly from regions with 4 or fewer dithers in Y106; given this result, and the importance of the Y106 filter in photometric redshifts \citep[e.g.][]{2019ApJ...877..117H}, in a future version of the tiling strategy we might consider adding a dither position in Y106 even if we have to use shorter exposure times or accept one less dither in F184. The statistics of the PSF fidelity maps are shown in Figure~\ref{fig:fcum}.

\subsection{Output noise and Moir\'e patterns}

\begin{figure}
\includegraphics[width=3.2in]{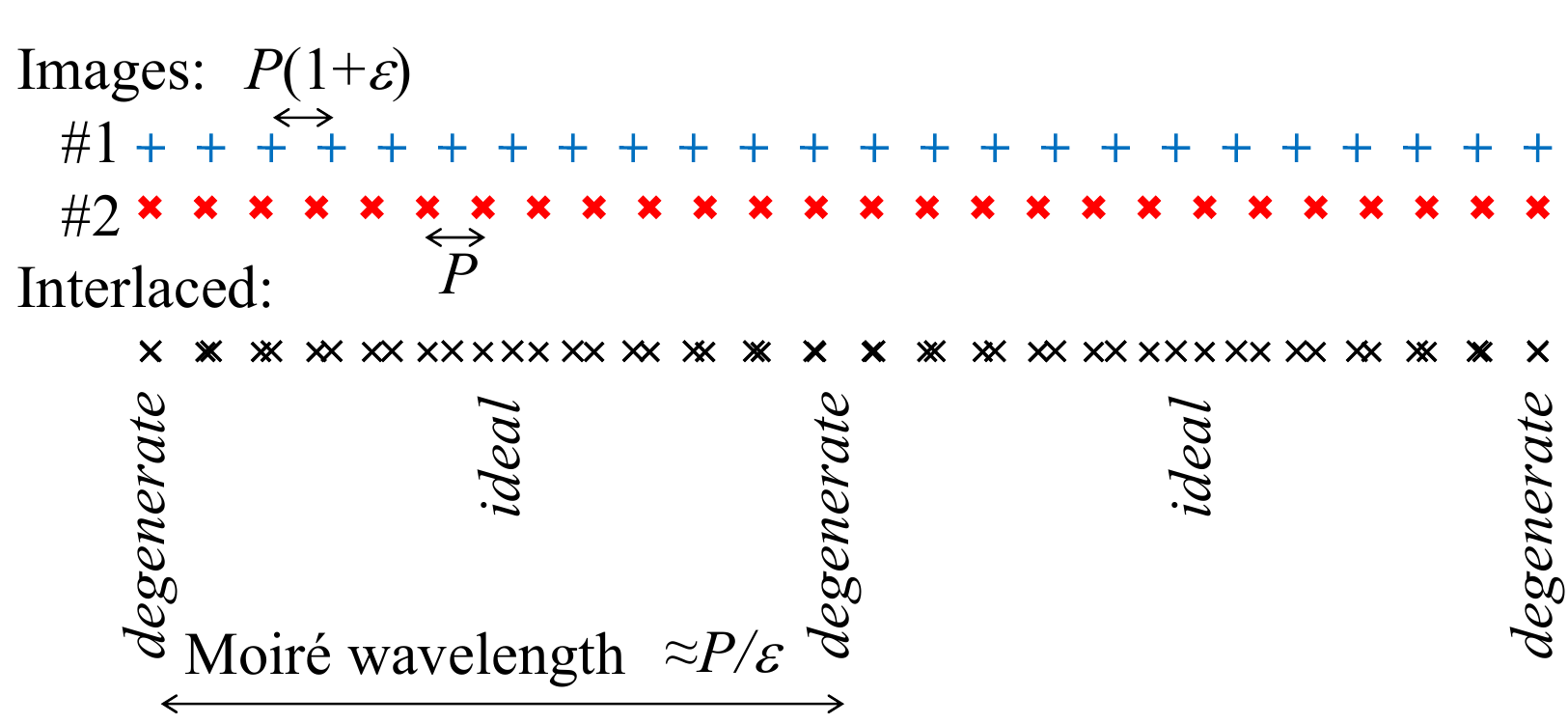}
\caption{\label{fig:moire_cartoon}An illustration of a Moir\'e pattern, with 2 input images in 1 dimension. The top row (blue + signs) represents the pixel positions in image \#1, and the middle row (red $\times$ signs) represents the pixel positions in image \#2. The interlaced pixel positions (from both \#1 and \#2) are shown in the bottom row, and one sees alternating regions of ``degenerate'' combination (in the linear algebra sense: the two input images give information at the same locations) and ``ideal'' combination (half pixel offsets, so effectively achieving twice the sampling rate). The wavelength of the pattern is the harmonic difference of the two input wavelengths, $1/[1/P - 1/P(1+\varepsilon)] \approx P/\varepsilon$, where $\varepsilon\ll 1$ is the fractional difference in plate scale.}
\end{figure}

\begin{figure*}
\includegraphics[width=6.75in]{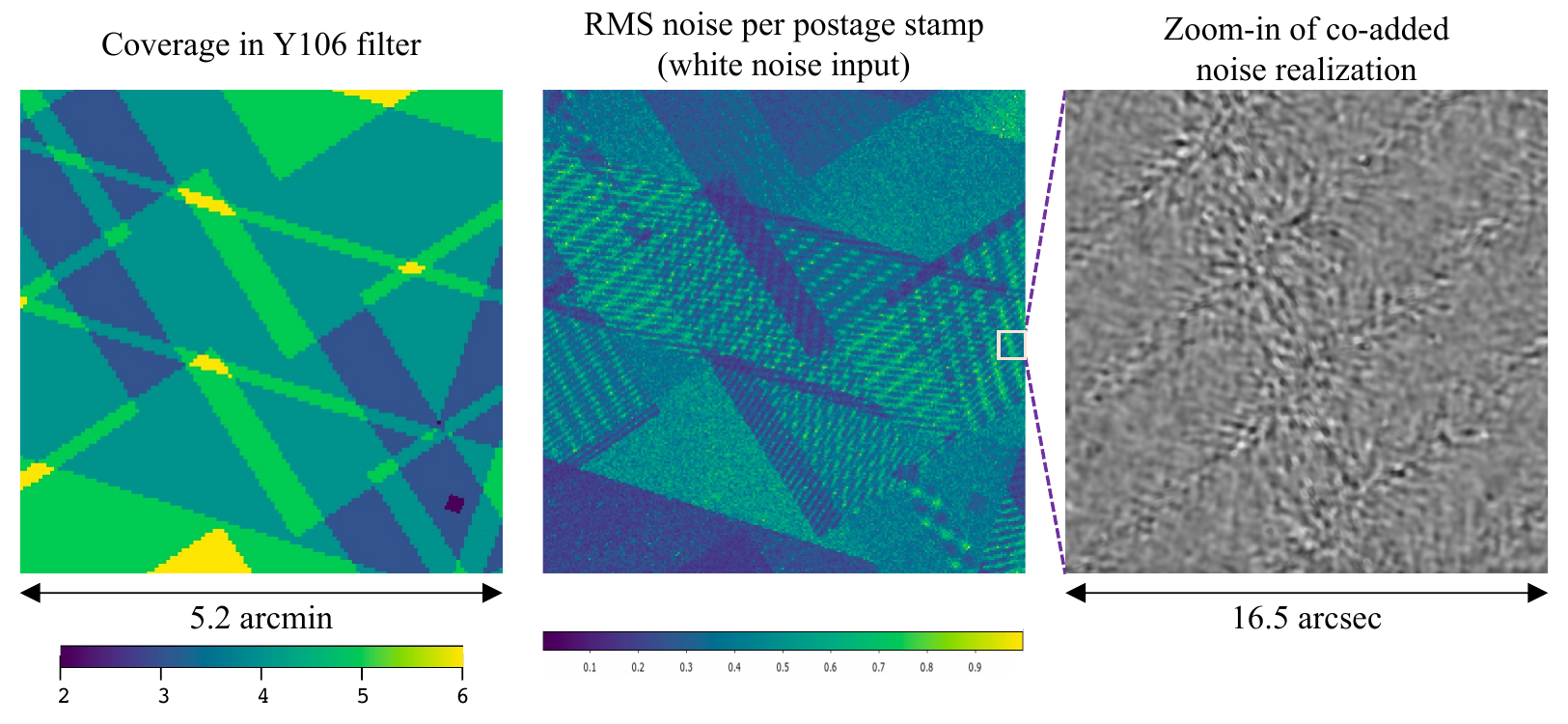}
\caption{\label{fig:moire}An example of Moir\'e patterns in the simulated coadd in the Y106 filter. The left panel is a coverage map in a region with intersecting chip gaps in the two passes. The middle panel shows the RMS of the ``whitenoise1'' noise field, as measured in $1.25\times 1.25$ arcsec postage stamps. The wave-like features are Moir\'e patterns; note that these patterns usually start and end at a chip gap, and that a range of wave vectors are present. The right panel shows a zoom-in of a $16.5\times 16.5$ arcsec ($660\times 660$ output pixel) region of the coadded ``whitenoise1'' noise field. Increased variance in the form of ripples with specific wave vectors can be seen in the degeneracy bands.}
\end{figure*}

Another aspect of working with multiple undersampled input images is the existence of Moir\'e patterns. If one combines two input images with no roll, but with a fractional difference $\varepsilon$ in plate scale \changetext{(so the plate scales are $P$ and $P(1+\varepsilon)$)}, then the images interlace to produce alternating regions where the samples land on top of each other, and where the samples are ideally interlaced ($\tfrac12$ pixel offset), with a wavelength of $P/\varepsilon$. This behavior is shown in Fig.~\ref{fig:moire_cartoon} for the simple case of 2 input images in 1 dimension. One can see that in some parts of the Moir\'e pattern, the image is effectively sampled at the native pixel scale, whereas in other regions the sampling rate has been improved by a factor of 2; and an obvious concern for a survey is the resulting heterogeneity of the survey data. In particular, in the ``degenerate'' regions, reconstructing a fully sampled image at output pixel positions in between the samples may prove impossible, or at least result in a large amplification of the noise in the input images \citep[e.g.][]{1999PASP..111..227L}. For Roman, the plate scale is 0.11 arcsec, and the fractional variation in plate scale over a $\sim \frac14$ SCA dither in the $y$-direction is $\sim 1.2\%$, leading to a predicted Moir\'e wavelength of $\sim 9$ arcsec; this should be thought of as representative since the distortion gradient itself is spatially variable. An example of such a feature in the Roman image combination simulations is shown in Fig.~\ref{fig:moire}. Such features are more common in regions with multiple intersecting chip gaps. While the Moir\'e patterns are intrinsic to the survey pattern --- they represent spatial variations in what degeneracies are present in the data --- the specific way they appear in higher-level products such as coadded images or ellipticity biases depends on the algorithm.

\begin{figure}
    \centering
    \includegraphics[width=3.25in]{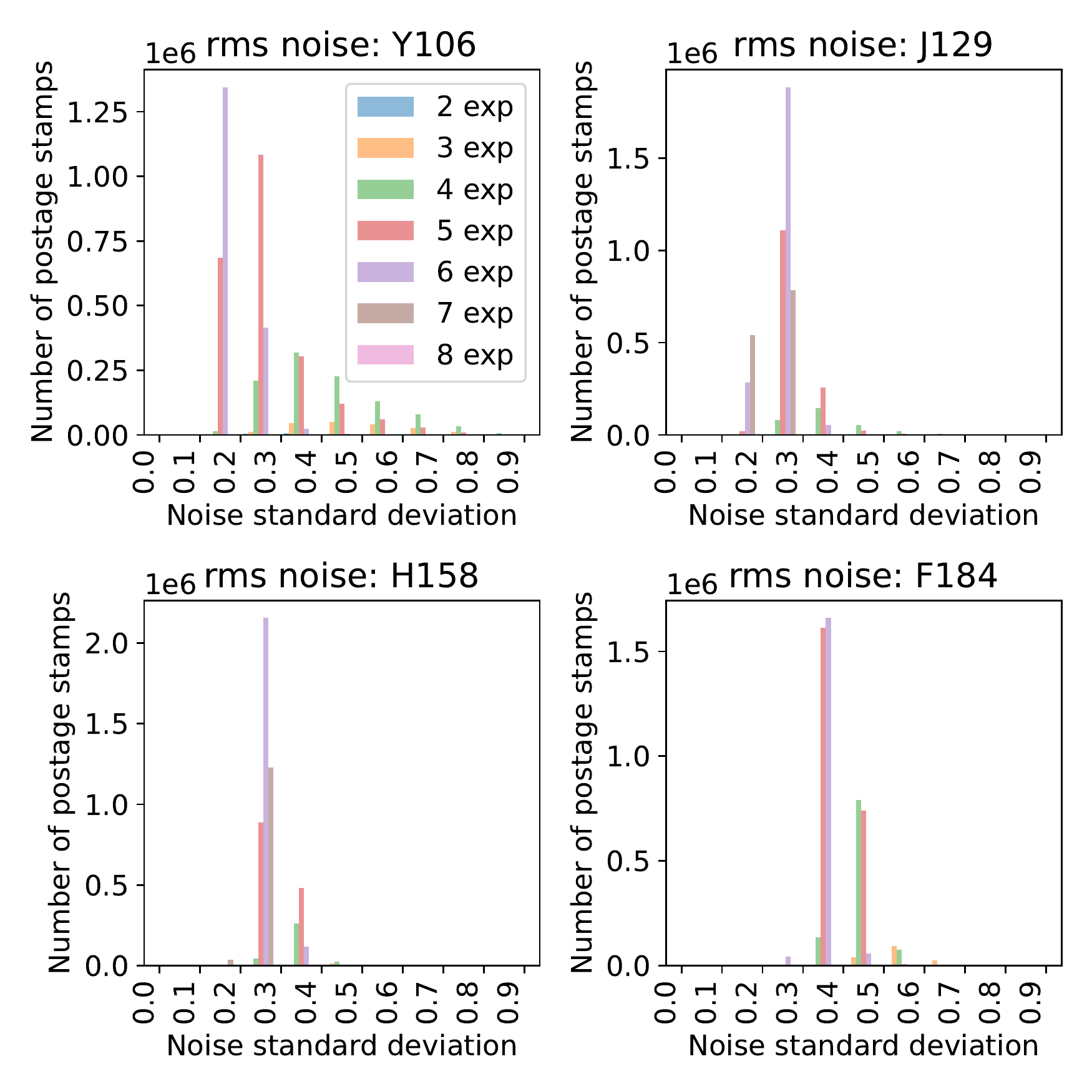}
    \caption{\label{fig:noisehist}The root-mean-square noise in each output postage stamp in the ``whitenoise1'' layer (this is normalized to unit variance in the input noise maps). Each panel shows a histogram for one of the four bands; the histograms are broken down by number of input exposures. Note the tail to larger noise values in Y106 with small numbers of exposures; this is driven by regions with strong Moir\'e patterns.}
\end{figure}

The full situation with the Moir\'e patterns is more complicated with $N>2$ dithered images since then $N(N-1)/2$ Moir\'e wavelengths are present and the pattern is not strictly periodic. In some regions, the dither positions will be closer to the degenerate case in Fig.~\ref{fig:moire_cartoon}; the fraction of the area where of all the dither positions land within, say, $\frac14$ pixels of each other can be expressed analytically\footnote{This can be thought of as a probability for random dithers to cluster in this way. The probability is $N/4^{N-1}$: there are $N$ choices for which dither is on the ``left'' side of the cluster, and then a probability $\frac14$ for each later dither to be between 0 and $\frac14$ pixel to its right.}, and is $\frac3{16}=18.75\%$ for $N=3$; $\frac1{16}=6.25\%$ for $N=4$; and $\frac5{256}\approx 1.95\%$ for $N=5$. If one goes up to 2 dimensions but with no rolls, then this argument applies in both $x$ and $y$ directions: for example, for $N=3$, one would expect $1-(1-\frac3{16})^2 \approx 34\%$ of the area to have samples clustered within $\frac14P$ in $x$, $y$, or both (this drops to 12\% for $N=4$ and 4\% for $N=5$). For the case with arbitrary rolls, we are unaware of any practical analytical results on these degeneracies, so numerical simulation remains the best approach to assessing the sampling impact of survey strategy choices for Roman. The histograms of the output noise properties from this simulation are shown in Fig.~\ref{fig:noisehist}.

The banded features in Fig.~\ref{fig:moire} are visually striking and are a potential concern because coadded noise with an anisotropic power spectrum could introduce an additive noise bias (e.g., the centroid bias, \citealt{2000ApJ...537..555K, 2002AJ....123..583B}, although this is just one of a hierarchy of biases related to noise; \citealt{2012MNRAS.425.1951R, 2012MNRAS.424.2757M}). While the separation of the bands themselves is at the Moir\'e scale of order 10 arcsec, the band structures are coherent over scales of $>1$ arcmin (see the middle panel of Fig.~\ref{fig:moire}) and one may wonder whether they introduce power at scales of interest for the weak lensing analysis. A quantitative investigation of the additive noise biases is carried out in the companion Paper II.

\begin{figure*}
\centering
\includegraphics[width=6in]{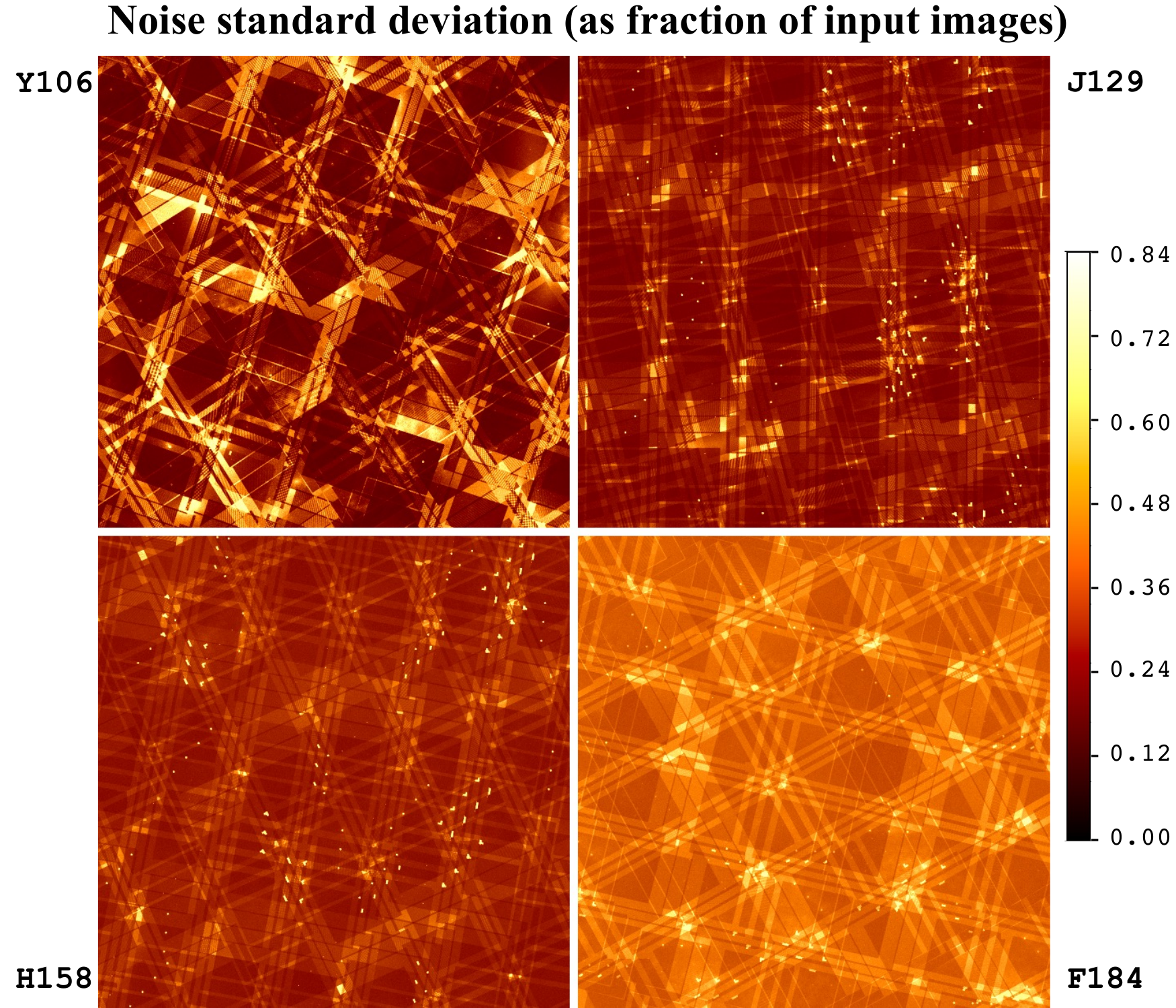}
\caption{\label{fig:noisemap}The noise map over the $48\times 48$ arcmin region simulated, in each of the 4 bands, measured as the RMS of the ``whitenoise1'' noise field, as measured in $1.25\times 1.25$ arcsec postage stamps. (The Y106 panel is a zoom-out of the center panel in Fig.~\ref{fig:moire}.) Once again, we can see the features in the coverage map (Fig.~\ref{fig:coveragemap}). The regions of large cosmetic defects in the SCAs can be seen in the output noise maps as strings of higher-noise splotches corresponding to the dithering sequence.}
\end{figure*}

A fully zoomed-out output noise map is shown in Fig.~\ref{fig:noisemap}.

\subsection{Impact of charge diffusion}
\label{ss:CD}

The ``simple'' image simulations did not include a charge diffusion, and so we have not included charge diffusion in the main results of this paper. We intuitively expect that this is a ``stress test'' of the application of {\sc Imcom}, since charge diffusion occurs before pixelization \citep{2020JATIS...6d6001M} and thus a PSF including charge diffusion is better sampled than one without. However, it is important to check this expectation. We tested this by re-running one block in the most difficult filter (Y106) with several values of the charge diffusion, ranging from 0.0 to 0.4 pixels RMS.
As an intuitive guide to the importance of charge diffusion, the least-squares fit Gaussian to an Airy disc has width $\sigma_{\rm Airy} = 0.41\lambda/D$, so by root-sum-square addition we would expect that inclusion of 0.24 pix RMS charge diffusion effectively smears a diffraction-limited Y106 PSF to the resolution of J129, and 0.38 pix RMS smears it to the resolution of H158; however, since the Airy disc and features induced by aberrations are not Gaussian, these should be viewed as rough estimates and the full output of the {\sc Imcom} should be used for planning purposes. We chose block location $(44,34)$ because it includes some intersecting chip gaps (the right panel of Fig.~\ref{fig:moire} is in this block).

\begin{figure}
    \centering
    \includegraphics[width=3.25in]{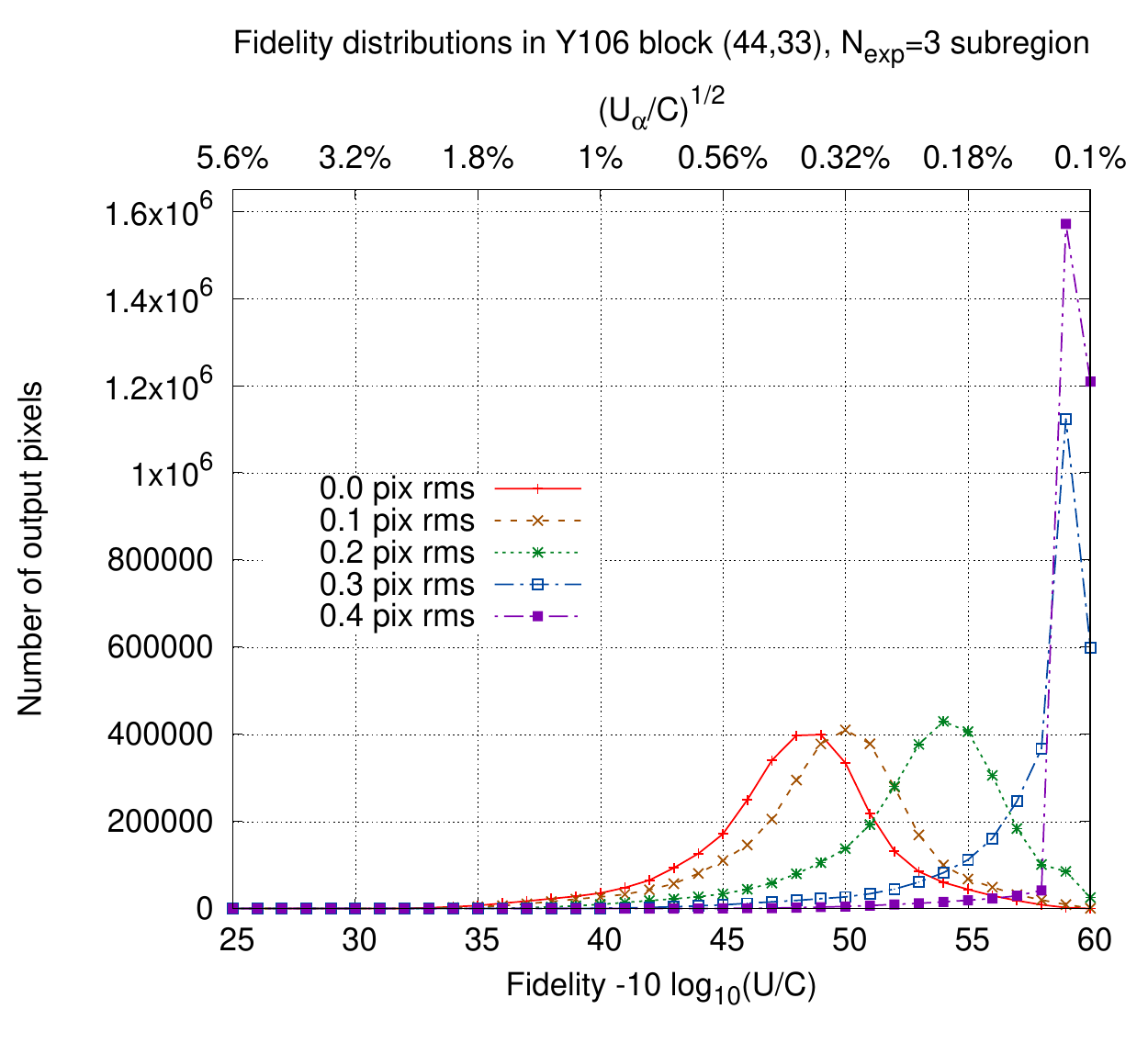}
    \caption{\label{fig:alt-CD}The histogram of the output image fidelity in the Y106 filter in the 3-exposure region of block $(44,34)$. Results are shown for 5 values of the charge diffusion length: 0 (no charge diffusion), 0.1, 0.2, 0.3, and 0.4 pixels rms per axis. Note the improvement in PSF fidelity as the charge diffusion is increased since the input images are better sampled.}
\end{figure}

Block $(44,34)$ contains a total of 1180 postage stamps (0.51 arcmin$^2$) with a coverage of only 3 exposures. This generally consists of 2 exposures at one roll angle and 1 exposure at the other. For these postage stamps, we show the histogram of output PSF fidelity in Fig.~\ref{fig:alt-CD}. The fidelity improves with increasing charge diffusion. The measured charge diffusion for the {\slshape Roman} detectors is 0.27--0.35 pixels \citep{2022PASP..134a4001G}. We caution that those measurements were performed at an illumination wavelength of 500 nm, and it is possible that the charge diffusion length could be shorter at the longer wavelengths.

\section{Conclusion}
\label{sec:discussion}

We have presented an implementation of the {\sc Imcom} image combination algorithm \citep{2011ApJ...741...46R} and its initial application to the {\slshape Roman} simulations of \citet{2022arXiv220906829T}. The algorithm attempts to create a coadded image with a specified target PSF (generally taken to be circular and constant). We find a set of parameters and choice of target PSF where the algorithm is successful on the simulated {\slshape Roman} wide-area imaging survey data with the reference survey dithering pattern and the sampling in the four reference survey filters (Y106, J129, H158, and F184), including allowing for chip gaps and cosmetic defects. The quality of coadded PSF is reported in an output ``fidelity'' map. The output noise maps show Moir\'e patterns characteristic of combining undersampled images with plate scale variations, especially in regions with small numbers of exposures. We have found that this effect is reduced if charge diffusion is included, which motivates more detailed characterization of the charge diffusion in {\slshape Roman} detectors before the final tiling strategy is selected. While the algorithm is computationally expensive, a mitigating factor is that simulated noise realizations and grids of injected sources can be processed at the same time at little additional cost.

Weak lensing applications of {\sc Imcom} coadds require not just ``one-point'' statistics of the PSF and noise properties, but the correlation functions of the PSF residuals and the noise power spectrum and its spatial variation. These properties are investigated in the companion Paper II.

\section*{Acknowledgements}

During the preparation of this work, C.H. and E.M. were supported by NASA contract 15-WFIRST15-0008, Simons Foundation  grant 256298, and David \& Lucile Packard Foundation grant 2021-72096. M.Y. and M.T. were also supported by NASA award 15-WFIRST15-0008 as part of the Roman Cosmology with the High-Latitude Survey Science Investigation Team and by NASA under JPL Contract Task 70-711320, ``Maximizing Science Exploitation of Simulated Cosmological Survey Data Across Surveys.'' Work at Argonne National Laboratory was supported under the U.S. Department of Energy contract DE-AC02-06CH11357. This research used resources of the Argonne Leadership Computing Facility, which is supported by DOE/SC under
contract DE-AC02-06CH11357. R.M. was supported by a grant from the Simons Foundation (Simons Investigator in Astrophysics, Award ID 620789).

We thank Paul Martini, Peter Taylor, Chun-Hao To, and David Weinberg for feedback on the methodology and analysis choices for this project.

The detector data files used in the {\slshape Roman} image simulations are based on data acquired in the Detector Characterization Laboratory (DCL) at the NASA Goddard Space Flight Center. We thank the personnel at the DCL for making the data available for this project. Computations for this project used the Pitzer cluster at the Ohio Supercomputer Center \citep{Pitzer2018} and the Duke Compute Cluster.

This project made use of the numpy \citep{2020Natur.585..357H} and astropy  \citep{2013A&A...558A..33A,2018AJ....156..123A, 2022ApJ...935..167A} packages.
Some of the figures were made using ds9 \citep{2003ASPC..295..489J} and {\sc Matplotlib} \citep{Hunter:2007}.
Some of the results in this paper have been derived using the healpy and HEALPix package \citep{2005ApJ...622..759G, 2019JOSS....4.1298Z}.

\section*{Data Availability}

The codes for this project, along with sample configuration files and setup instructions, are available in the two GitHub repositories:
\begin{itemize}
\item https://github.com/hirata10/furry-parakeet.git (postage stamp coaddition)
\item https://github.com/hirata10/fluffy-garbanzo.git (mosaic driver)
\end{itemize}
The version of the code used for this project is in tag 0.1.

\bibliographystyle{mnras}
\bibliography{mainbib}

\appendix

\section{Interpolation kernels}
\label{app:interpolation}

The computation of the PSF overlap matrices for the image combination algorithm requires interpolation of a band-limited function from a 2D grid, ${\mathbb Z}^2$. A standard linear algorithm\footnote{In the linear algebra sense: interpolation of a function commutes with scalar multiplication and is distributive over addition.} for interpolating a point based on $(2K)^2$ neighbors, separable in $x$ and $y$, is
\begin{equation}
\hat f(\alpha + \xi_\alpha, \gamma+\xi_\gamma) = \sum_{\mu=1-K}^K w_\mu(\xi_\alpha) \sum_{\nu=1-K}^K w_\nu(\xi_\gamma)f(\alpha+\mu,\gamma+\nu),
\label{eq:t-interp}
\end{equation}
where $\alpha$ and $\gamma$ are integers, $0\le\xi_\alpha,\xi_\gamma\le 1$ are the fractional parts of the points to which we interpolate, \changetext{and the functions $w_\mu$ are interpolation weights}. We use $\hat f$ to denote the interpolated function and $f$ to denote the original (known only on the grid). The functions $w_\mu(\xi)$ are the interpolation weights and differ by interpolation scheme. This appendix is dedicated to the choice of weights. We summarize the major options in Table~\ref{tab:interp_methods}.

\subsection{Common choices}

Choices common in the image processing literature include polynomial interpolation of order $2K-1$:
\begin{equation}
w_\mu(\xi|{\rm P}\,K) = 
\frac{(-1)^{K-\mu}}{(K-\mu)!(K-1+\mu)!(\xi-\mu)} \prod_{\sigma=1-K}^K (\xi-\sigma)
\label{eq:poly-K}
\end{equation}
(this is a consequence of the polynomial interpolation formula of \citealt{1779RSPT...69...59W}; see, e.g., Eq.~A7 of \citealt{2004PhRvD..70j3501H} for this form)
and Lanczos-$K$ interpolation:
\begin{equation}
w_\mu(\xi|{\rm L}\,K) = K\frac{
\sin [\pi(\xi - \mu)]\sin [\frac\pi K(\xi - \mu)]}{[\pi(\xi - \mu)]^2}.
\label{eq:lanczos-K}
\end{equation}
The Lanczos kernel does not preserve a constant background, i.e., it has $\sum_\mu w_\mu(\xi)\neq 1$, so one may also construct the background-conserving version
\begin{equation}
w_\mu(\xi|{\rm L}'\,K) = \frac{w_\mu(\xi|{\rm L}\,K)}{
\sum_{\sigma=1-K}^K w_\sigma(\xi|{\rm L}\,K)}.
\label{eq:lanczosP-K}
\end{equation}
All of these interpolation schemes are exact in the limit that $K\rightarrow\infty$ and the image is Nyquist-sampled (i.e., has Fourier wavevectors ${\mathbfit u}$ with $-\frac12<u_x,u_y<\frac12$). However for moderate values of $K$, they suffer from multiplicative errors as well as introduction of aliased Fourier modes \citep{2014PASP..126..287B}. The fundamental issue is that these kernels are optimized for specific purposes: the polynomial kernel is best for functions with extremely low spatial frequencies, and the Lanczos kernel is a well-tested choice when moderate accuracy is required for functions that contain frequencies within a factor of $\sim 2$ of the Nyquist limit \citep{Turkowski90}. Neither of these is especially well suited to the needs of weak lensing or other precision photometry programs, where one works with functions that are a few times Nyquist sampled (e.g., $4\times$ is common) but where 4 or more significant digits of accuracy in the interpolated function is required.

\subsection{Schemes optimized using a least-square error metric}

\begin{table*}
    \caption{Interpolation methods considered in this appendix. The allowed parameters are $K$ (kernel size: based on $2K$ nearest pixels), $L$ (number of abscissae, for the discretely optimized kernels), and $R$ (band limit for which the kernel is optimized).}
    \label{tab:interp_methods}
    \begin{tabular}{l|cccccc}
    \hline
    Code & Name & Parameters & Background & Equation \\ 
     & & & conserving? &  \\ \hline
    ~~~P & polynomial & $K$ & Yes & Eq.~(\ref{eq:poly-K}) \\
    ~~~L & Lanczos & $K$ & No & Eq.~(\ref{eq:lanczos-K}) \\
    ~~~L$'$ & rescaled Lanczos & $K$ & Yes & Eq.~(\ref{eq:lanczosP-K}) \\
    ~~~S & square window LSE & $K$, $R$ & No & Eqs.~(\ref{eq:wt-S}, \ref{eq:sys-S}, \ref{eq:C-S}) \\
    ~~~S$'$ & rescaled square window LSE & $K$, $R$ & Yes &  Eqs.~(\ref{eq:wt-S}, \ref{eq:sys-S}, \ref{eq:C-S}, \ref{eq:wprime}) \\
    ~~~T & triangle window LSE & $K$, $R$ & No & Eqs.~(\ref{eq:wt-S}, \ref{eq:sys-S}, \ref{eq:C-T}) \\
    ~~~T$'$ & rescaled triangle window LSE & $K$, $R$ & Yes & Eqs.~(\ref{eq:wt-S}, \ref{eq:sys-S}, \ref{eq:C-T}, \ref{eq:wprime}) \\
    ~~~D & discrete window LSE & $K$, $L$, $R$ & No & Eqs.~(\ref{eq:WD}, \ref{eq:WH}) \\
    \hline
    \end{tabular}
\end{table*}

To define an optimal interpolation scheme, we return to the 1-dimensional interpolation problem (since we have enforced separability) and define the various errors that may occur when a function is interpolated.\footnote{The error metrics defined here as $m(u)$ and $\bar\varepsilon^2(u)$ can be expressed in the notation of \citet{2014PASP..126..287B} as $-E_0(u)$ and $\sum_{j\neq 0}|\tilde K_u(j+u)|^2$.} We suppose that we are interpolating a complex exponential function \changetext{of spatial frequency $u$}, $f_u(x) = {\rm e}^{2\pi{\rm i}ux}$ (any function can be constructed as a superposition of these). Then there is an interpolated function $\hat f_u(x)$. The original spatial frequency is present with some multiplicative error $m(u)$:
\begin{equation}
1 + m(u) = \int_0^1 \hat f_u(x)\, {\rm e}^{-2\pi{\rm i}ux}\,{\rm d}x.
\label{eq:mdef}
\end{equation}
There is also a leakage into the other modes (or ``ghosting'' in the terminology of \citealt{2014PASP..126..287B}),
\begin{equation}
\bar\varepsilon^2(u) = \int_0^1 \left| \hat f_u(x) - [1+m(u)] {\rm e}^{2\pi{\rm i}ux} \right|^2\,{\rm d}x.
\label{eq:edef}
\end{equation}
The total mean squared error in the reconstruction is the sum of these:
\begin{equation}
\varepsilon^2(u) \equiv
|m(u)|^2 + \bar\varepsilon^2(u) = \int_0^1 \left| \hat f_u(x) - {\rm e}^{2\pi{\rm i}ux} \right|^2\,{\rm d}x.
\label{eq:edef2}
\end{equation}
That is, $\varepsilon(u)$ is the root-sum-square of the multiplicative bias and all of the ghost amplitudes.
We are now ready to define some conditions for an optimized kernel. We have tested several choices; these are summarized in Table~\ref{tab:interp_methods}.

If we are working with a band-limited function with spatial frequencies $|u|<R$, then we might choose to minimize
\begin{equation}
\Omega = \frac1{2R}\int_{-R}^R
\rho(u) \varepsilon^2(u)\,{\rm d}u,
\label{eq:Omega}
\end{equation}
where $\rho$ is a non-negative even window function.
Setting the functional derivative of $\Omega$ to zero gives:
\begin{equation}
0 = \frac{\delta\Omega}{\delta w^\ast_\mu(\xi)}
= \int_{-R}^R
{\rm e}^{-2\pi{\rm i}u\mu}
\left[ \sum_{\nu=1-K}^K w_\nu(\xi)\,{\rm e}^{-2\pi{\rm i}u\nu} - {\rm e}^{2\pi{\rm i}u\xi} \right]
\,{\rm d}u.
\end{equation}
This is a linear equation for $w_\nu(\xi)$. It has a solution of the form:
\begin{equation}
\left( \begin{array}c 
w_{1-K}(\xi) \\ w_{2-K}(\xi) \\ \vdots \\ w_K(\xi)
\end{array} \right)
={\mathbfss S}^{-1}{\mathbfit b},
~~~~{\mathbfit b}=
\left( \begin{array}c 
C(\xi+K-1) \\ C(\xi+K-2) \\ \vdots \\ C(\xi-K)
\end{array} \right),
\label{eq:wt-S}
\end{equation}
where the system matrix is
\begin{equation}
{\mathbfss S} = \left( \begin{array}{cccc}
C(0) & C(1) & \hdots & C(2K-1) \\
C(-1) & C(0) & \hdots & C(2K-2) \\
\vdots & \vdots & \ddots & \vdots \\
C(-2K+1) & C(-2K+2) & \hdots & C(0)
\end{array} \right)
\label{eq:sys-S}
\end{equation}
and
\begin{equation}
C(x) = \frac 1R\int_{-R}^R \rho(u)\, {\rm e}^{2\pi{\rm i}ux}\,{\rm d}x.
\label{eq:C-general}
\end{equation}
Since $C$ is even, ${\mathbfss S}$ is symmetric.

The most obvious choice is to weight all Fourier modes in the band limit equally: $\rho(u)=1$.
This leads to what we call the ``square window least-squares error (LSE)'' interpolation method (``square'' because the weighting of different input Fourier modes $u$ is constant for $|u|<R$ and 0 otherwise).
\begin{equation}
C_{\rm square~window}(x) = \frac 1R\int_{-R}^R  {\rm e}^{2\pi{\rm i}ux}\,{\rm d}x = 2 {\,\rm sinc\,}(2Rx),
\label{eq:C-S}
\end{equation}
where we use the numpy sinc convention, ${\rm sinc\,}z=\sin(\pi z)/(\pi z)$.
Since the matrix ${\mathbfss S}^{-1}$ does not depend on $\xi$, but only on the parameters $K$ and $R$, it can be computed once for an interpolation scheme and saved.

In some cases, we might want some control over the errors for input Fourier modes out to the band limit $R$, but we want the tightest control for small $u$. This is the case, for example, of a diffraction limited image whose power spectrum only truly drops to zero at $u=R$ but is small at $u\gtrsim R/2$. We might then modify the objective function to be
$\rho(u) = 1 - |u|/R$.
This leads to
\begin{equation}
C_{\rm triangle~wave}(x) = \frac1R\int_{-R}^R \left( 1 - \frac{|u|}R\right)
{\rm e}^{2\pi{\rm i}ux}\,{\rm d}x
= {\rm sinc^2\,}(Rx).
\label{eq:C-T}
\end{equation}
This leads to the ``triangle window LSE.''

\subsubsection{Background conservation}

The aforementioned interpolation algorithms do not conserve a constant background, i.e., $\sum_{\mu=1-K}^K w_\mu(\xi)$ is not exactly equal to unity. This is not surprising since we optimized the weights so that the integral of the LSE over a range of frequencies is minimized. For interpolation of images with a large background, one might want the constant mode to be represented perfectly. Thus we are led to considering the lowest cost ($\Omega$) $2K$-point interpolation scheme that has zero error for the constant mode, i.e., the $u=0$ mode.

This problem can be solved by adding a formally infinite $\delta$-function to $\rho(u)$, i.e., we construct a new objective function:
\begin{equation}
\rho'(u) = \rho(u) + R\lambda\delta(u),
~~~
{\mathbfss S}' = {\mathbfss S}+\lambda{\mathbfit e}{\mathbfit e}^{\rm T},
~~~
{\mathbfit b}' = {\mathbfit b} + \lambda{\mathbfit e},
\end{equation}
where ${\mathbfit e}$ is a length $2K$ column vector of all ones, and taking the limit of $\lambda\rightarrow\infty$. This leads to a new weight that can be obtained from the Sherman-Morrison formula, in the form of \citet{bartlett51}:
\begin{eqnarray}
{\mathbfit w}' &=&
({\mathbfss S} + \lambda{\mathbfit e}{\mathbfit e}^{\rm T})^{-1}({\mathbfit b} + \lambda{\mathbfit e})
\nonumber \\
&=&
\left( {\mathbfss S}^{-1}
- \frac{\lambda {\mathbfss S}^{-1}{\mathbfit e}{\mathbfit e}^{\rm T}{\mathbfss S}^{-1}}{1+\lambda {\mathbfit e}^{\rm T}{\mathbfss S}^{-1}{\mathbfit e}}
\right)({\mathbfit b} + \lambda{\mathbfit e})
\nonumber \\
&=&
{\mathbfss S}^{-1}{\mathbfit b} + \frac{1-{\mathbfit e}^{\rm T}{\mathbfss S}^{-1}{\mathbfit b}}{\lambda^{-1} + {\mathbfit e}^{\rm T}{\mathbfss S}^{-1}{\mathbfit e}}{\mathbfss S}^{-1}{\mathbfit e}
\nonumber \\
&\rightarrow &
{\mathbfss S}^{-1}{\mathbfit b} + \frac{1-{\mathbfit e}^{\rm T}{\mathbfss S}^{-1}{\mathbfit b}}{{\mathbfit e}^{\rm T}{\mathbfss S}^{-1}{\mathbfit e}}{\mathbfss S}^{-1}{\mathbfit e}.
\end{eqnarray}
Now at this stage, we see that ${\mathbfss S}^{-1}{\mathbfit b}$ is simply ${\mathbfit w}$, the weight vector that is obtained without enforcing background conservation. Then we have a new weight vector given by
\begin{equation}
w'_\mu = w_\mu +\left( 1 - \sum_{\nu=1-K}^K w_\nu \right)\eta_\mu,
\label{eq:wprime}
\end{equation}
where
\begin{equation}
\eta_\mu = \frac{\sum_{\nu = 1-K}^K [{\mathbfss S}^{-1}]_{\mu\nu}}{
\sum_{\sigma=1-K}^K
\sum_{\nu = 1-K}^K [{\mathbfss S}^{-1}]_{\sigma\nu}
}.
\label{eq:eta}
\end{equation}
The vector ${\bmath\eta}$ depends only on the choice of interpolation scheme and can be pre-tabulated. Note that by construction, the entries in ${\bmath\eta}$ sum to unity, guaranteeing that $\sum_{\mu=1-K}^K w'_\mu=1$. The use of any vector ${\bmath\eta}$ that sums to unity would fix the background conservation issue (for example, the rescaling in \citealt{2014PASP..126..287B} corresponds to $w_\mu/\sum_\sigma w_\sigma$), but Eq.~(\ref{eq:eta}) does so at the least cost to the weighted mean square error over the specified range in $u$.

Interpolation kernels that are ``improved'' using this approach are denoted with a $'$ (e.g., S$'$ or T$'$).

\subsubsection{Rounding error issues}

A disadvantage of the LSE methods for large $K$ and small $R$ is that the ${\mathbfss S}$ matrix becomes nearly singular, resulting in amplification of roundoff error when one takes the inverse; eventually, these errors can exceed the formal analytical error of the interpolation method. Tools such as direct system solution without the inverse (e.g., {\tt numpy.linalg.solve}) reduce the problem but do not eliminate it. The cases we have shown in the figures do not suffer from these problems when floating point computations are performed using IEEE 754 standard 64-bit arithmetic. But for the most demanding applications (where we aim for $\varepsilon \lesssim 10^{-8}$), we aim for alternatives to direct computation of ${\mathbfss S}$.

\subsection{Discretized error metric}

We recall that an integral is often evaluated by taking a linear combination of values of the functions at specified abscissae. Applying this idea to Eq.~(\ref{eq:Omega}), we may define an error metric that is not an integral over Fourier modes, but rather a discrete sum:
\begin{equation}
\rho(u) = R\sum_{j=1}^L \alpha_j[\delta(u-Rz_j) + \delta(u+Rz_j)],
\end{equation}
where $\{z_j\}_{j=1}^L$ are the abscissae (rescaled so that $0<z_j<1$); $\{\alpha_j\}_{j=1}^L$ are the weights, assumed to be positive; and we have enforced symmetry of positive and negative frequencies. We require $L\ge K$ so that the matrix ${\mathbfss S}$ has full rank. Then
\begin{equation}
C(x) = 2 \sum_{j=1}^L \alpha_j \cos(2\pi Rz_jx)
\equiv 2 \sum_{j=1}^L \alpha_j \cos (\zeta_jx),
\end{equation}
where we defined $\zeta_j = 2\pi Rz_j$.
The advantage of this is that the matrix ${\mathbfss S}$ can be factorized as ${\mathbfss S} = {\mathbfss M}^{\rm T}{\mathbfss M}$, where ${\mathbfss M}$ is the $2L\times 2K$ matrix
\begin{equation}
{\mathbfss M} = \left(\begin{array}{cccc}
\beta_1\cos (\zeta_1s_1) & \beta_1\cos (\zeta_1s_2) & \hdots & \beta_1\cos (\zeta_1s_{2K}) \\
\beta_2\cos (\zeta_2s_1) & \beta_2\cos (\zeta_2s_2) & \hdots & \beta_2\cos (\zeta_2s_{2K}) \\
\vdots & \vdots & \ddots & \vdots \\
\beta_L\cos (\zeta_Ls_1) & \beta_L\cos (\zeta_Ls_2) & \hdots & \beta_L\cos (\zeta_Ls_{2K}) \\
\beta_1\sin (\zeta_1s_1) & \beta_1\sin (\zeta_1s_2) & \hdots & \beta_1\sin (\zeta_1s_{2K}) \\
\beta_2\sin (\zeta_2s_1) & \beta_2\sin (\zeta_2s_2) & \hdots & \beta_2\sin (\zeta_2s_{2K}) \\
\vdots & \vdots & \ddots & \vdots \\
\beta_L\sin (\zeta_Ls_1) & \beta_L\sin (\zeta_Ls_2) & \hdots & \beta_L\sin (\zeta_Ls_{2K}) \\
\end{array}\right),
\end{equation}
and we have defined $\beta_j \equiv \sqrt{2\alpha_j}$ and  $s_j \equiv -K-\frac12+j$. The length $2K$ vector ${\mathbfit b}$ can be expressed as ${\mathbfit b} = {\mathbfss M}^{\rm T}{\mathbfit c}$ where
\begin{equation}
{\mathbfit c} = \left(\begin{array}{c}
\beta_1\cos [\zeta_1(\xi-\frac12)] \\
\beta_2\cos [\zeta_2(\xi-\frac12)] \\
\vdots \\
\beta_L\cos [\zeta_L(\xi-\frac12)] \\
\beta_1\sin [\zeta_1(\xi-\frac12)] \\
\beta_2\sin [\zeta_2(\xi-\frac12)] \\
\vdots \\
\beta_L\sin [\zeta_L(\xi-\frac12)] \\
\end{array}\right).
\end{equation}
(These results can be verified by direct multiplication, as well as the identity $\cos(x-y)=\cos x\cos y+\sin x\sin y$.) Next we perform the singular value decomposition of ${\mathbfss M}$, so that ${\mathbfss M}=\mathbfss {UDV}$, where ${\mathbfss U}$ is a $2L\times 2L$ orthogonal matrix; ${\mathbfss D}$ is a $2L\times 2K$ matrix with all zeros except for the diagonal $\{D_{jj}\}_{j=1}^{2K}$; and ${\mathbfss V}$ is a $2K\times 2K$ orthogonal matrix. In this case, Eq.~(\ref{eq:wt-S}) simplifies to
\begin{eqnarray}
{\mathbfss S}^{-1}{\mathbfit b}
&=& ({\mathbfss M}^{\rm T}{\mathbfss M})^{-1}{\mathbfss M}^{\rm T}{\mathbfit c}
\nonumber \\
&=& ({\mathbfss V}^{\rm T}{\mathbfss D}^{\rm T}{\mathbfss U}^{\rm T}{\mathbfss U}{\mathbfss D}{\mathbfss V})^{-1}{\mathbfss V}^{\rm T}{\mathbfss D}^{\rm T}{\mathbfss U}^{\rm T}{\mathbfit c}
\nonumber \\
&=& {\mathbfss V}^{\rm T}({\mathbfss D}^{\rm T}{\mathbfss D})^{-1}{\mathbfss D}^{\rm T}{\mathbfss U}^{\rm T}{\mathbfit c}.
\end{eqnarray}
This can be expressed as
\begin{equation}
w_{j-K}(\xi) = \sum_{l=1}^L \left\{H_{jl} \cos\left[\zeta_l(\xi-\tfrac12)\right]
+ H_{j,l+L} \sin\left[\zeta_l(\xi-\tfrac12)\right] \right\}.
\label{eq:WD}
\end{equation}
where the $2K\times 2L$ matrix ${\mathbfss H}$ can be written as
\begin{equation}
H_{jl}= \beta_l
\sum_{k=1}^{2K}  \frac{V_{kj}U_{lk}}{D_{kk}},
\label{eq:WH}
\end{equation}
where in the prefactor we define $\beta_{l+L}=\beta_l$.
Note that ${\mathbfss H}$ depends on $K$, $L$, and the weights $z_j$ and $\alpha_j$, but not on $\xi$; thus once a scheme is chosen, it can be pre-computed. The expression contains a division by a singular value $D_{kk}$, but since ${\mathbfss S}={\mathbfss M}^{\rm T}{\mathbfss M}$, the condition ratio of ${\mathbfss M}$ can be better than that of ${\mathbfss S}$ (if $L=K$, it it the square root). This means that roundoff errors are correspondingly less pernicious.

The choice remains of the weights. The simplest choice, which we adopt here, is to use $z_j$ and $\alpha_j$ corresponding to $2L$-point Gauss-Legendre quadrature; we expect it to be the best discrete approximation to the ``square wave LSE'' scheme. Furthermore, while the method works for $L>K$, we have not found significant gains that offset the increased computational cost (we must compute $2L$ trigonometric functions for each 1D interpolation). Thus we have retained $L=K$.

The discrete window interpolators with $L=K$ are in principle exact at the spatial frequencies $u_j$, since then ${\mathbfss M}$ is invertible so ${\mathbfss S}^{-1}{\mathbfit b}={\mathbfss M}^{-1}{\mathbfit c}$ and one can show that an input function $f(x)$ that is a cosine or sine wave leads to a vector ${\mathbfit f}^{\rm T} = (f(1-K),\,f(2-K)\,...f(K))$ that is a constant times a row of ${\mathbfss M}$. Then the  interpolated value, $\hat f(\xi)={\mathbfit f}^{\rm T}{\mathbfss M}^{-1}{\mathbfit c} = ({\mathbfss M}^{{\rm T}-1}{\mathbfit f}){\mathbfit c}$, collapses. However, in finite-precision arithmetic, this property may not hold.

\subsection{Examples}

We now present some specific choices of kernel that are optimized for various choices of oversampling factor and precision. Recall again that in what follows, ``sinc'' follows the numpy convention, ${\rm sinc\,}z = \sin(\pi z)/(\pi z)$. Also the oversampling relative to Nyquist is $1/(2R)$: thus $R=\frac14$ is $2\times$ Nyquist, $R=\frac18$ is $4\times$ Nyquist, etc.

\subsubsection{6-point interpolation}

A well-studied choice in weak lensing image processing is $4\times$ Nyquist sampling (e.g., by zero-padding an FFT) and 6-point ($K=3$) interpolation. \citet{2014PASP..126..287B} found that multiplicative errors and ghosting could be reduced to the $<0.1\%$ level with quintic polynomial interpolation (P3). The square window LSE method gives the interpolation weights for the ``S3,$\tfrac18$'' scheme:
\begin{equation}
w_{\mu}(\xi|{\rm S}3,\tfrac18) = \sum_{\sigma=-2}^3 q^{\rm (S)}_{\mu\sigma} {\rm \,sinc\,}\frac{\xi - \sigma}4,
\label{eq:S3_1__8}
\end{equation}
where the $q^{\rm(S)}_{\mu\sigma}$ coefficients are given in Table~\ref{tab:S3}. Similarly, for the triangle window LSE method or ``T3,$\tfrac18$'' scheme, we get:
\begin{equation}
w_{\mu}(\xi|{\rm T}3,\tfrac18) = \sum_{\sigma=-2}^3 q^{\rm (T)}_{\mu\sigma} {\rm \,sinc^2\,}\frac{\xi - \sigma}8.
\label{eq:T3_1__8}
\end{equation}
The ${\bmath\eta}$ vectors are also shown in the table, so one may use Eq.~(\ref{eq:wprime}) to construct the background-conserving kernels $w_{\mu}(\xi|{\rm S}'3,\tfrac18)$ and $w_{\mu}(\xi|{\rm T}'3,\tfrac18)$.

\begin{table*}
    \caption{Interpolation weight coefficients $q^{\rm (S)}_{\mu\sigma}$ for the S3,$\tfrac18$ and T3,$\tfrac18$ schemes (Eqs.~\ref{eq:S3_1__8} and \ref{eq:T3_1__8}). The $\eta$ coefficients are also given so that one may build the S$'$3,$\tfrac18$ and T$'$3,$\tfrac18$ schemes. The data in these tables are included as {\tt Tables\_A2A3.txt} in the {\tt furry-parakeet} GitHub repository.}
    \label{tab:S3}
    \begin{tabular}{rrrrrr}
\hline
\multicolumn6c{$q_{\mu\sigma}^{\rm(S)}$ coefficients for S3,$\tfrac18$ scheme} \\
\multicolumn1c{$-2$} &
\multicolumn1c{$-1$} &
\multicolumn1c{$0$} &
\multicolumn1c{$1$} &
\multicolumn1c{$2$} &
\multicolumn1c{$3$} \\
\hline
\scriptsize{  1.01495532029840815E+4 } &
\scriptsize{ $-$4.33911434207969724E+4 } &
\scriptsize{  7.98533070813937666E+4 } &
\scriptsize{ $-$7.86589121627444692E+4 } &
\scriptsize{  4.14384977985485020E+4 } &
\scriptsize{ $-$9.36878701410286885E+3 } \\
\scriptsize{ $-$4.33911434207969724E+4 } &
\scriptsize{  1.87006319343270035E+5 } &
\scriptsize{ $-$3.46527427216041717E+5 } &
\scriptsize{  3.43526149682672811E+5 } &
\scriptsize{ $-$1.82105354947356158E+5 } &
\scriptsize{  4.14384977985485020E+4 } \\
\scriptsize{  7.98533070813937666E+4 } &
\scriptsize{ $-$3.46527427216041717E+5 } &
\scriptsize{  6.46080867636206094E+5 } &
\scriptsize{ $-$6.44241725441453862E+5 } &
\scriptsize{  3.43526149682672811E+5 } &
\scriptsize{ $-$7.86589121627444692E+4 } \\
\scriptsize{ $-$7.86589121627444692E+4 } &
\scriptsize{  3.43526149682672811E+5 } &
\scriptsize{ $-$6.44241725441453862E+5 } &
\scriptsize{  6.46080867636206094E+5 } &
\scriptsize{ $-$3.46527427216041717E+5 } &
\scriptsize{  7.98533070813937666E+4 } \\
\scriptsize{  4.14384977985485020E+4 } &
\scriptsize{ $-$1.82105354947356158E+5 } &
\scriptsize{  3.43526149682672811E+5 } &
\scriptsize{ $-$3.46527427216041717E+5 } &
\scriptsize{  1.87006319343270035E+5 } &
\scriptsize{ $-$4.33911434207969724E+4 } \\
\scriptsize{ $-$9.36878701410286885E+3 } &
\scriptsize{  4.14384977985485020E+4 } &
\scriptsize{ $-$7.86589121627444692E+4 } &
\scriptsize{  7.98533070813937666E+4 } &
\scriptsize{ $-$4.33911434207969724E+4 } &
\scriptsize{  1.01495532029840815E+4 } \\
\hline
\multicolumn6c{$\eta_{\mu}^{\rm(T)}$ coefficients for S$'$3,$\tfrac18$ scheme} \\
\multicolumn1c{$-2$} &
\multicolumn1c{$-1$} &
\multicolumn1c{$0$} &
\multicolumn1c{$1$} &
\multicolumn1c{$2$} &
\multicolumn1c{$3$} \\
\hline
\scriptsize{6.198154413828288689E+0} &
\scriptsize{-1.457870288360315847E+1} &
\scriptsize{8.880548469769111719E+0} &
\scriptsize{8.880548469773117404E+0} &
\scriptsize{-1.457870288361116984E+1} &
\scriptsize{6.198154413827788645E+0} \\
\hline
\multicolumn6c{$q_{\mu\sigma}^{\rm(T)}$ coefficients for T3,$\tfrac18$ scheme} \\
\multicolumn1c{$-2$} &
\multicolumn1c{$-1$} &
\multicolumn1c{$0$} &
\multicolumn1c{$1$} &
\multicolumn1c{$2$} &
\multicolumn1c{$3$} \\
\hline
\scriptsize{  2.69093477729340666E+4 } &
\scriptsize{ $-$1.18760717924963697E+5 } &
\scriptsize{  2.21738698701371002E+5 } &
\scriptsize{ $-$2.18473923214483541E+5 } &
\scriptsize{  1.13535864318182255E+5 } &
\scriptsize{ $-$2.49229688603067771E+4 } \\
\scriptsize{ $-$1.18760717924963697E+5 } &
\scriptsize{  5.27960278438832145E+5 } &
\scriptsize{ $-$9.92219132336361217E+5 } &
\scriptsize{  9.83596672992079286E+5 } &
\scriptsize{ $-$5.14178380302917794E+5 } &
\scriptsize{  1.13535864318182255E+5 } \\
\scriptsize{  2.21738698701371002E+5 } &
\scriptsize{ $-$9.92219132336361217E+5 } &
\scriptsize{  1.87610361960852286E+6 } &
\scriptsize{ $-$1.87070559473955585E+6 } &
\scriptsize{  9.83596672992079286E+5 } &
\scriptsize{ $-$2.18473923214483541E+5 } \\
\scriptsize{ $-$2.18473923214483541E+5 } &
\scriptsize{  9.83596672992079286E+5 } &
\scriptsize{ $-$1.87070559473955585E+6 } &
\scriptsize{  1.87610361960852286E+6 } &
\scriptsize{ $-$9.92219132336361217E+5 } &
\scriptsize{  2.21738698701371002E+5 } \\
\scriptsize{  1.13535864318182255E+5 } &
\scriptsize{ $-$5.14178380302917794E+5 } &
\scriptsize{  9.83596672992079286E+5 } &
\scriptsize{ $-$9.92219132336361217E+5 } &
\scriptsize{  5.27960278438832145E+5 } &
\scriptsize{ $-$1.18760717924963697E+5 } \\
\scriptsize{ $-$2.49229688603067771E+4 } &
\scriptsize{  1.13535864318182255E+5 } &
\scriptsize{ $-$2.18473923214483541E+5 } &
\scriptsize{  2.21738698701371002E+5 } &
\scriptsize{ $-$1.18760717924963697E+5 } &
\scriptsize{  2.69093477729340666E+4 } \\
\hline
\multicolumn6c{$\eta_{\mu}^{\rm(T)}$ coefficients for T$'$3,$\tfrac18$ scheme} \\
\multicolumn1c{$-2$} &
\multicolumn1c{$-1$} &
\multicolumn1c{$0$} &
\multicolumn1c{$1$} &
\multicolumn1c{$2$} &
\multicolumn1c{$3$} \\
\hline
\scriptsize{1.071761416473430018E+1} &
\scriptsize{-2.665663946077842539E+1} &
\scriptsize{1.643902529601151130E+1} &
\scriptsize{1.643902529607080965E+1} &
\scriptsize{-2.665663946079621383E+1} &
\scriptsize{1.071761416473430018E+1} \\
\hline
    \end{tabular}
\end{table*}
\begin{table*}
    \caption{Coefficient formulae (${\mathbfss H}$ matrix entries and $\zeta_j$) for the 10-point D5,5,$\tfrac1{12}$ interpolation scheme. Weights are shown for $w_\mu(\xi)$ over the valid range $\mu = -4, -3, ... 5$.\label{tab:D5512}}
    \begin{tabular}{crrrrr}
    \hline
    $l$ & \multicolumn1c1 & \multicolumn1c2 & \multicolumn1c3 & \multicolumn1c4 & \multicolumn1c5 \\
    $\zeta_l$ & 
    \scriptsize{7.795042160878816462E$-$2} &
    \scriptsize{2.269252977160159945E$-$1} &
    \scriptsize{3.557380180911379752E$-$1} &
    \scriptsize{4.529461196132943956E$-$1} &
    \scriptsize{5.099362658787808256E$-$1} \\
    \hline
    \multicolumn6c{Coefficients of $\cos \zeta_l(\xi-\tfrac12)$} \\
    $\mu=-4$ &
    \scriptsize{ 1.912402678501005084E+3} &
    \scriptsize{ $-$4.927004100469148398E+3} &
    \scriptsize{ 5.835905613163729868E+3} &
    \scriptsize{ $-$4.322722449499965478E+3} &
    \scriptsize{ 1.501418877063505988E+3} \\
     $\mu=-3$ &
    \scriptsize{$-$1.217699386087176026E+4} &
    \scriptsize{3.159481253500127059E+4} &
    \scriptsize{$-$3.785475949046354799E+4} &
    \scriptsize{2.836995380606032268E+4} &
    \scriptsize{$-$9.933019743019915040E+3} \\
    $\mu=-2$ &
    \scriptsize{3.246330126827143977E+4} &
    \scriptsize{$-$8.462064453664862958E+4} &
    \scriptsize{1.021891619223469461E+5} &
    \scriptsize{$-$7.724212924975770875E+4} &
    \scriptsize{2.721034680139671400E+4} \\
    $\mu=-1$ &
    \scriptsize{$-$4.342904595880888519E+4} &
    \scriptsize{1.135281945624359505E+5} &
    \scriptsize{$-$1.377799817082234076E+5} &
    \scriptsize{1.047254797516023391E+5} &
    \scriptsize{$-$3.704478343160567601E+4} \\
    $\mu=0$ &
    \scriptsize{2.123096597676863894E+4} &
    \scriptsize{$-$5.557555073910982173E+4} &
    \scriptsize{6.760976692036786699E+4} &
    \scriptsize{$-$5.153062474798672338E+4} &
    \scriptsize{1.826604930348573544E+4} \\
    $\mu=1$ &
    \scriptsize{2.123096597676863894E+4} &
    \scriptsize{$-$5.557555073910982173E+4} &
    \scriptsize{6.760976692036786699E+4} &
    \scriptsize{$-$5.153062474798672338E+4} &
    \scriptsize{1.826604930348573544E+4} \\
    $\mu=2$ &
    \scriptsize{$-$4.342904595880888519E+4} &
    \scriptsize{1.135281945624359505E+5} &
    \scriptsize{$-$1.377799817082234076E+5} &
    \scriptsize{1.047254797516023391E+5} &
    \scriptsize{$-$3.704478343160567601E+4} \\
    $\mu=3$ &
    \scriptsize{3.246330126827143977E+4} &
    \scriptsize{$-$8.462064453664862958E+4} &
    \scriptsize{1.021891619223469461E+5} &
    \scriptsize{$-$7.724212924975770875E+4} &
    \scriptsize{2.721034680139671400E+4} \\
    $\mu=4$ &
    \scriptsize{$-$1.217699386087176026E+4} &
    \scriptsize{3.159481253500127059E+4} &
    \scriptsize{$-$3.785475949046354799E+4} &
    \scriptsize{2.836995380606032268E+4} &
    \scriptsize{$-$9.933019743019915040E+3} \\
    $\mu=5$ &
    \scriptsize{1.912402678501005084E+3} &
    \scriptsize{$-$4.927004100469148398E+3} &
    \scriptsize{5.835905613163729868E+3} &
    \scriptsize{$-$4.322722449499965478E+3} &
    \scriptsize{1.501418877063505988E+3} \\
    \hline
    \multicolumn6c{Coefficients of $\sin \zeta_l(\xi-\tfrac12)$} \\
    $\mu=-4$ &
    \scriptsize{$-$4.904230619110763655E+4} &
    \scriptsize{4.323751412374444772E+4} &
    \scriptsize{$-$3.246339075532347488E+4} &
    \scriptsize{1.875968952461114532E+4} &
    \scriptsize{$-$5.760491925503920356E+3} \\
    $\mu=-3$ &
    \scriptsize{4.103555938606222626E+5} &
    \scriptsize{$-$3.637390867586739478E+5} &
    \scriptsize{2.755014430530755781E+5} &
    \scriptsize{$-$1.606389018624043674E+5} &
    \scriptsize{4.963098826150417153E+4} \\
    $\mu=-2$ &
    \scriptsize{$-$1.555126539182490204E+6} &
    \scriptsize{1.383601738371265586E+6} &
    \scriptsize{$-$1.054523731121562887E+6} &
    \scriptsize{6.189727207369643729E+5} &
    \scriptsize{$-$1.921388961754927877E+5} \\
    $\mu=-1$ &
    \scriptsize{3.501335761714349966E+6} &
    \scriptsize{$-$3.122480589327111840E+6} &
    \scriptsize{2.389400028862954117E+6} &
    \scriptsize{$-$1.408673167242457159E+6} &
    \scriptsize{4.386664738897074712E+5} \\
    $\mu=0$ &
    \scriptsize{$-$5.159499149133198895E+6} &
    \scriptsize{4.606470743687568232E+6} &
    \scriptsize{$-$3.531921534316427074E+6} &
    \scriptsize{2.086791191489480436E+6} &
    \scriptsize{$-$6.508774765937846387E+5} \\
    $\mu=1$ &
    \scriptsize{5.159499149133198895E+6} &
    \scriptsize{$-$4.606470743687568232E+6} &
    \scriptsize{3.531921534316427074E+6} &
    \scriptsize{$-$2.086791191489480436E+6} &
    \scriptsize{6.508774765937846387E+5} \\
    $\mu=2$ &
    \scriptsize{$-$3.501335761714349966E+6} &
    \scriptsize{3.122480589327111840E+6} &
    \scriptsize{$-$2.389400028862954117E+6} &
    \scriptsize{1.408673167242457159E+6} &
    \scriptsize{$-$4.386664738897074712E+5} \\
    $\mu=3$ &
    \scriptsize{1.555126539182490204E+6} &
    \scriptsize{$-$1.383601738371265586E+6} &
    \scriptsize{1.054523731121562887E+6} &
    \scriptsize{$-$6.189727207369643729E+5} &
    \scriptsize{1.921388961754927877E+5} \\
    $\mu=4$ &
    \scriptsize{$-$4.103555938606222626E+5} &
    \scriptsize{3.637390867586739478E+5} &
    \scriptsize{$-$2.755014430530755781E+5} &
    \scriptsize{1.606389018624043674E+5} &
    \scriptsize{$-$4.963098826150417153E+4} \\
    $\mu=5$ &
    \scriptsize{4.904230619110763655E+4} &
    \scriptsize{$-$4.323751412374444772E+4} &
    \scriptsize{3.246339075532347488E+4} &
    \scriptsize{$-$1.875968952461114532E+4} &
    \scriptsize{5.760491925503920356E+3} \\
    \hline
    \end{tabular}
\end{table*}

\begin{figure}
    \includegraphics[width=3.25in]{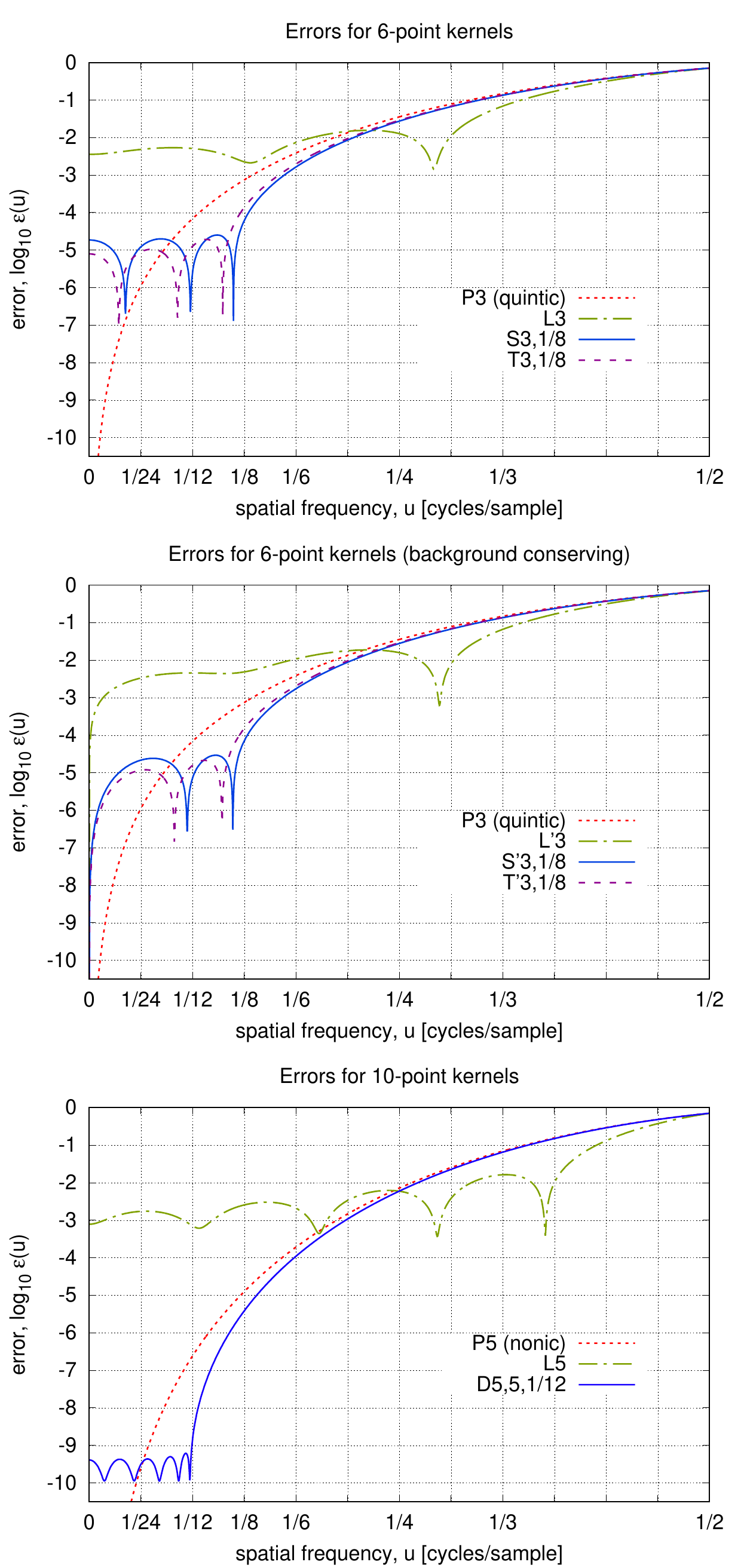}
    \caption{Comparison of the root-mean-square error as a function of the input spatial frequency for the 6-point ($K=3$) interpolation schemes (upper panel); 6-point background-conserving interpolation schemes (middle panel, note P3 is the same); and 10-point ($K=5$) schemes (lower panel). Note that $u=\tfrac12$ corresponds to critical sampling.\label{fig:errK3}}
\end{figure}

A comparison of the errors for the various 6-point interpolation schemes is shown in the upper panel of Fig.~\ref{fig:errK3}. The quintic interpolation scheme performs best at very low spatial frequencies ($u<\tfrac1{16}$). The Lanczos-3 kernel has large errors, several tenths of a percent, across all of the low spatial frequencies, but with this sacrifice it gains improved behavior at intermediate spatial frequencies ($u\sim 0.3$). The S3,$\tfrac18$ kernel is best at showing small ($<10^{-4}$) errors over the widest range. In particular, all the modes in a $4\times$ Nyquist sampled image ($u<\tfrac18$) are reconstructed with errors $<10^{-4}$, whereas the quintic has an error of $\varepsilon(\tfrac18)= 7.4\times 10^{-4}$. The T3,$\tfrac18$ kernel behaves similarly, although as expected the lowest spatial frequencies behave better (error $<10^{-5}$) at the expense of slightly worse performance right at the band limit (reaching error of $1.4\times 10^{-4}$ at $u=\tfrac18$).

\subsubsection{10-point interpolation}

For very high precision applications --- for example, if interpolation is to be used as a step in constructing the ${\mathbfss A}$ matrix \changetext{in \citet{2011ApJ...741...46R}}, where small errors can be magnified in subsequent linear algebra steps --- even the S3,$\tfrac18$ and T3,$\tfrac18$ schemes may not be sufficient. We have therefore implemented the 10-point D5,5,$\tfrac1{12}$ scheme, which is to be applied to images that are at least $6\times$ Nyquist sampled.

The coefficients $w_\mu(\xi)$ are a sum of 10 trigonometric terms, as given by Eq.~(\ref{eq:WD}). The coefficients are given in Table~\ref{tab:D5512}. The performance is shown in the lower panel of Fig.~\ref{fig:errK3}; note that at $u<\tfrac1{12}$, the error is very small, $\varepsilon(u)<1.5\times 10^{-9}$. The nonic polynomial interpolator (P5), by contrast, ``only'' achieves $\varepsilon(\tfrac1{12}) \approx 2.4\times 10^{-7}$; but at the very smallest values of $u$ ($<\tfrac1{24}$) it has the best performance. The Lanczos interpolator (L5), as expected, has significant errors ($>10^{-3}$) at low frequencies, but has good performance ($<1.63\times 10^{-2})$ all the way out to $u=0.38$ (i.e., modes that are only $1.3\times$ Nyquist sampled).

The D5,5,$\tfrac1{12}$ interpolator should be ``exact'' at the five frequencies $u_j$ that are $\frac1{12}$ times the Gauss-Legendre abscissae, i.e., the roots of the $P_{10}$ Legendre polynomial. One can see these as ``dips'' in the bottom panel of Fig.~\ref{fig:errK3}. However, in IEEE 754 64-bit (double precision) arithmetic, one never reaches exactly zero; the $\varepsilon(u)$ curves have minima at $\sim 1.1\times 10^{-10}$ (inspection of the data file shows these are actual minima, not simply the closest point to $u_j$ that was plotted). This is unsurprising since the amplitudes in Table~\ref{tab:D5512} reach $\sim 10^6$ and given the limitations of double precision representations of numbers (53 bits of precision: $2^{-53} \sim 10^{-16}$). Further improvement would probably require the construction of alternate basis sets that are more orthogonal than cosines and sines; this presents no fundamental challenge but we suspect on many architectures the computation time would be increased. Since errors of ``a few $\times 10^{-10}$'' are well below our tolerance, and are likely good enough for the vast majority of image processing needs in astronomy, we stop here and recommend $6\times$ Nyquist sampling and the D5,5,$\tfrac1{12}$ interpolation scheme for the most demanding applications.

\section{Computing resources}
\label{app:computing}

Most of the computing for the image combination runs was carried out at the Pitzer cluster \citep{Pitzer2018} at the Ohio Supercomputer Center (OSC), and the Duke Compute Cluster (DCC). A small number of test runs were carried out on the Amazon Web Services (AWS) Elastic Compute Cloud (EC2), or on a MacBook Pro laptop (``Mac''). The statistics for the computing costs are in Table~\ref{tab:computing}. Note that several experiments were carried out near the beginning of the runs, including running the code on multiple cores using numpy multithreading; this turned out not to be efficient for the current code structure since only some operations benefited from the multithreading, and we dropped it for the remainder of this project. \changetext{For most of the runs, memory ($\approx 10$ GB) rather than cores was the limiting factor; therefore the rate at which the runs could be completed (and, depending on platform, billing rates) were set by the memory usage rather than core-hours.}

The central processing units (CPUs) used were the Intel Xeon Gold 6148 (2.4 GHz) or Platinum 8268 (2.9 GHz) on OSC; the Intel Xeon E5-2680 v3 (2.5 GHz) on DCC; the Intel Xeon Platinum 8124M (3.0 GHz) on AWS; and the Apple M1 Pro18,3 on the Mac laptop.

A change partway through the J129-band runs was to introduce memory mapping (via the {\tt numpy.memmap} function) for the input data cube to reduce the memory usage with many input layers. The input data cube is used least frequently, and on some nodes memory rather than core count limited the rate of processing. We found that using {\tt numpy.memmap} for the input data typically reduced the memory usage by $\sim 30$\%.

Note that the number of operations can vary by band: J129 has the largest number of dither positions in the Reference Survey, which means that the matrices used to construct ${\mathbfss T}$ are larger. Thus one expects J129 to require more core-hours than the other bands, even for the same code and computing architecture.

\begin{table}
    \centering
    \caption{\label{tab:computing}Computing resources for the image coaddition runs in this paper.}
    \begin{tabular}{cccccc}
    \hline\hline
      Band   & \!Number of\! & Platform & Cores & Memory  & Compute time  \\
      & blocks & & & \!mapping?\! & \![1000 core hr]\! \\
      \hline
     \multirow3*{Y106} & \textcolor{white}{00}28 & AWS & 8 & No & \multirow3*{\textcolor{white}095.7} \\
     & \textcolor{white}0129 & OSC & 2 & No &  \\
     & 2147 & OSC & 1 & No & \\ 
     \hline
     \multirow2*{J129} & 1230 & OSC & 1 & No &  \multirow2*{109.5} \\
      & 1074 & OSC & 1 & Yes & \\
    \hline
\multirow3*{H158} 
 & \textcolor{white}{000}4 & Mac & 8 & Yes & \multirow3*{140.6} \\
 & \textcolor{white}0740 & OSC & 1 & No & \\
 & 1297 & OSC & 1 & Yes & \\
\hline
F184 & 2304 & DCC & 1 & No & 134.8 \\
    \hline
Total & 9216 & & & & 480.6\\
    \hline\hline
    \end{tabular}
\end{table}

\section{The effect of pixel-level operations on undersampled images}
\label{app:sampling}

In this appendix, we review the mathematical properties of undersampled data, and how these properties affect the operations commonly used to calibrate shear estimators. \changetext{Sampling is parameterized by $Q$, which is the number of pixels corresponding to one cycle at the maximum spatial frequency; $Q<2$ corresponds to undersampling by the usual Nyquist crtierion, and smaller $Q$ represents more severe undersampling.} We show that the usual implementation of Metacalibration \citep{2017arXiv170202600H, 2017ApJ...841...24S} suffers from aliasing when acting on undersampled images. If the undersampling is not too severe (``weak undersampling'' or $1<Q<2$), then Metacalibration can be implemented by re-convolving the image with a PSF that excludes the problematic Fourier modes. However, the {\slshape Roman} Y106 band and the near infrared bands on {\slshape Euclid} have ``strong undersampling'' ($Q<1$), so the mathematical framework of standard Metacalibration does not apply to these cases. The alternative Deep Metacalibration algorithm \citep{2022arXiv220607683Z} in principle could address these problems when $Q<1$, but many Fourier modes criss-crossing the $(u,v)$-plane must be excluded from the re-convolution PSF. The analytic shear response method of \citet{2022arXiv220607683Z} faces similar issues: in the weakly undersampled case, convolution in pre-processing can eliminate the aliased Fourier modes, but in the strong undersampling case the formalism as currently implemented cannot be applied.

We leave open the possibility that some new mathematical framework might be developed in the future to deal with these issues in the general undersampled case. But given these results, we are motivated to focus our attention on image processing algorithms that can recover a fully sampled image.

To avoid clutter, we work in units of the input pixel scale ($s_{\rm in}=1$).

\subsection{Fourier transforms and operators}

We perform continuous Fourier transforms in the ``waves per pixel'' convention:
\begin{eqnarray}
f(x,y) &=& \int_{{\mathbb R}^2} \tilde f(u,v)\,{\rm e}^{2\pi{\rm i}(ux+vy)}\,{\rm d}u\,{\rm d}v
~~~{\rm and}
\nonumber \\
\tilde f(u,v) &=& \int_{{\mathbb R}^2} f(x,y)\,{\rm e}^{-2\pi{\rm i}(ux+vy)}\,{\rm d}x\,{\rm d}y.
\end{eqnarray}
Convolution in one domain (real or Fourier) is equivalent to multiplication in the other; we denote the convolution of two functions by $\ast$ and multiplication by $\cdot$.

Shear measurement and shear calibration algorithms often use the finite translation operator ${\mathcal T}_{\boldsymbol s}$ with displacement ${\boldsymbol s}$,
\begin{equation}
{\mathcal T}_{{\boldsymbol s}}f(x,y) = f(x+s_x, y+s_y),
\end{equation}
and the distortion operator ${\mathcal Y}_{\kappa,{\boldsymbol\gamma},\varphi}$ with magnification $\kappa$, shear ${\boldsymbol\gamma}$, and rotation $\varphi$,
\begin{equation}
{\mathcal Y}_{\kappa,{\boldsymbol\gamma},\varphi} f(x,y) =
f({\mathbfss M}_{\kappa,{\boldsymbol\gamma},\varphi}(x,y)),
\end{equation}
where
\begin{equation}
{\mathbfss M}_{\kappa,{\boldsymbol\gamma},\varphi} = \left( \begin{array}{cc} \!1-\kappa-\gamma_1\! & -\gamma_2 \\ -\gamma_2 & \!1-\kappa+\gamma_1\! \end{array} \right)
\left( \begin{array}{cc} \cos\varphi & -\sin\varphi \\ \sin\varphi & \cos\varphi \end{array} \right)
.
\end{equation}
(This is the description of shear in terms of the polar decomposition of the $2\times 2$ matrix ${\mathbfss M}_{\kappa,{\boldsymbol\gamma},\varphi}$. The polar decomposition is convenient because it \changetext{allows us to} rotate a galaxy image by angle $\varphi$, then shear and magnify it, and then insert it into an image. It is thus convenient for problems where we want to know what a galaxy looks like at a random orientation $\varphi$, or if we want to insert both a galaxy and its $90^\circ$-rotated image to cancel shape noise in a simulation.)
These operate on Fourier transforms in accordance with
\begin{equation}
\widetilde{{\mathcal T}_{\boldsymbol s}f}(u,v) = {\rm e}^{2\pi{\rm i}(us_x + vs_y)}\tilde f(u,v)
\end{equation}
and
\begin{equation}
\widetilde{{\mathcal Y}_{\kappa,{\boldsymbol\gamma},\varphi}f}(u,v) = \frac1{\det{\mathbfss M}_{\kappa,{\boldsymbol\gamma},\varphi}}\tilde f({\mathbfss M}_{\kappa,{\boldsymbol\gamma},\varphi}^{-1\,\rm T}(u,v)
).
\end{equation}
When we discuss the magnitudes of the wave vectors, it is helpful to keep in mind that in the weak lensing regime $|\kappa|+|{\boldsymbol\gamma}|<1$, the maximum singular value of ${\mathbfss M}_{\kappa,{\boldsymbol\gamma},\varphi}^{-1\,\rm T}$ is 
\begin{equation}
\Lambda = \frac1{1-\kappa-|{\boldsymbol\gamma}|};
\label{eq:Lambda}
\end{equation}
that is, for $(u,v)$ in a circle of radius $r$, the farthest that ${\mathbfss M}_{\kappa,{\boldsymbol\gamma}}^{-1\,\rm T}(u,v)$ can be from the origin is $\Lambda r$.

A discretely sampled function $f_{x,y}$, measured only at integer pixel positions $(x,y)\in{\mathbb Z}^2$, instead has a Fourier transform:
\begin{eqnarray}
f_{x,y} &=& \int_{\mathcal B} \check f(u,v)\,{\rm e}^{2\pi{\rm i}(ux+vy)}\,{\rm d}u\,{\rm d}v
~~~{\rm and}
\nonumber \\
\check f(u,v) &=& \sum_{(x,y)\in{\mathbb Z}^2} f_{x,y} \,{\rm e}^{-2\pi{\rm i}(ux+vy)},
\end{eqnarray}
where ${\mathcal B} = \{ (u,v): -\frac12\le u<\frac12, -\frac12\le v<\frac12 \}$ is the first Brillouin zone or the unit square centered on the origin (shaded square in Fig.~\ref{fig:uvfig}).
The two types of Fourier transforms are related to each other via the $\Sha$ \changetext{(``Shah'' or Dirac comb)} function:
\begin{equation}
\Sha(x) = \sum_{(m,n)\in{\mathbb Z}^2} \delta(x-m) \delta(y-n),
\end{equation}
which is its own Fourier transform ($\widetilde\Sha=\Sha$).
If we discretely sample a function $f$ at the pixel positions, $f_{x,y} = f(x,y)$ for integer $x,y$, then
\begin{eqnarray}
\check f(u,v) &=& [\widetilde{f\cdot\Sha}](u,v)
= [\tilde f\ast\Sha](u,v)
\nonumber \\
&=& \sum_{(\Delta u, \Delta v)\in{\mathbb Z}^2} \tilde f(u+\Delta u,v+\Delta v).
\end{eqnarray}
One sees that the Fourier transform of the discretely sampled data contains a term associated with the original field ($\Delta u = \Delta v = 0$), as well as other terms offset by an integer numbers of cycles per pixel.

We define the fractional part operator
\begin{equation}
{\mathbb F}u = u - {\rm floor}\left( u+\frac12 \right)
\end{equation}
that maps $u$ to the range $-\frac12$ to $+\frac12$. Then $({\mathbb F}u,{\mathbb F}v)$ is the single mode in the first Brillouin zone that aliases to $(u,v)$.

\subsection{Astronomical scenes and available information}

Let us define an image which is a convolution of intrinsic projected scene $S$ and the effective point spread function $G$ (which includes the pixel tophat), which is then sampled at integer positions:
\begin{equation}
  f_{x,y} =  f(x, y) = [ S \ast G](x, y).
\end{equation}
The Fourier transform of the discretely sampled data is then related to the Fourier transform of the sky scene by:
\begin{equation}
\label{eqn:f_aliased}
      \check f(u, v) = \sum_{(\Delta u, \Delta v)\in{\mathbb Z}^2} \tilde S(u+\Delta u, v+\Delta v) \tilde G(u+\Delta u, v+\Delta v). 
\end{equation}
While in principle the sum in Eq.~(\ref{eqn:f_aliased}) is infinite, in practice $G$ has a band limit or maximum spatial frequency present: $\tilde G(u,v)=0$ if $\sqrt{u^2+v^2}\ge 1/Q$, where $Q$ is a sampling parameter (equal to the number of pixels across a single cycle at the maximum spatial frequency). For a space-based observatory, the optics usually operate at the diffraction limit: thus $Q = \lambda/(Ds_{\rm in})$, where $\lambda$ is the wavelength of light, $D$ is the diameter of the entrance pupil, and $s_{\rm in}$ is the pixel scale. For a ground-based observatory, atmospheric smearing usually suppresses the highest spatial frequencies. While turbulent contributions to $\tilde G$ usually decline smoothly rather than going to zero at a ``hard'' cutoff in spatial frequency, there is still some radius in the $(u,v)$-plane beyond which $\tilde G(u,v)$ is negligible.

\begin{figure}
\includegraphics[width=3.25in]{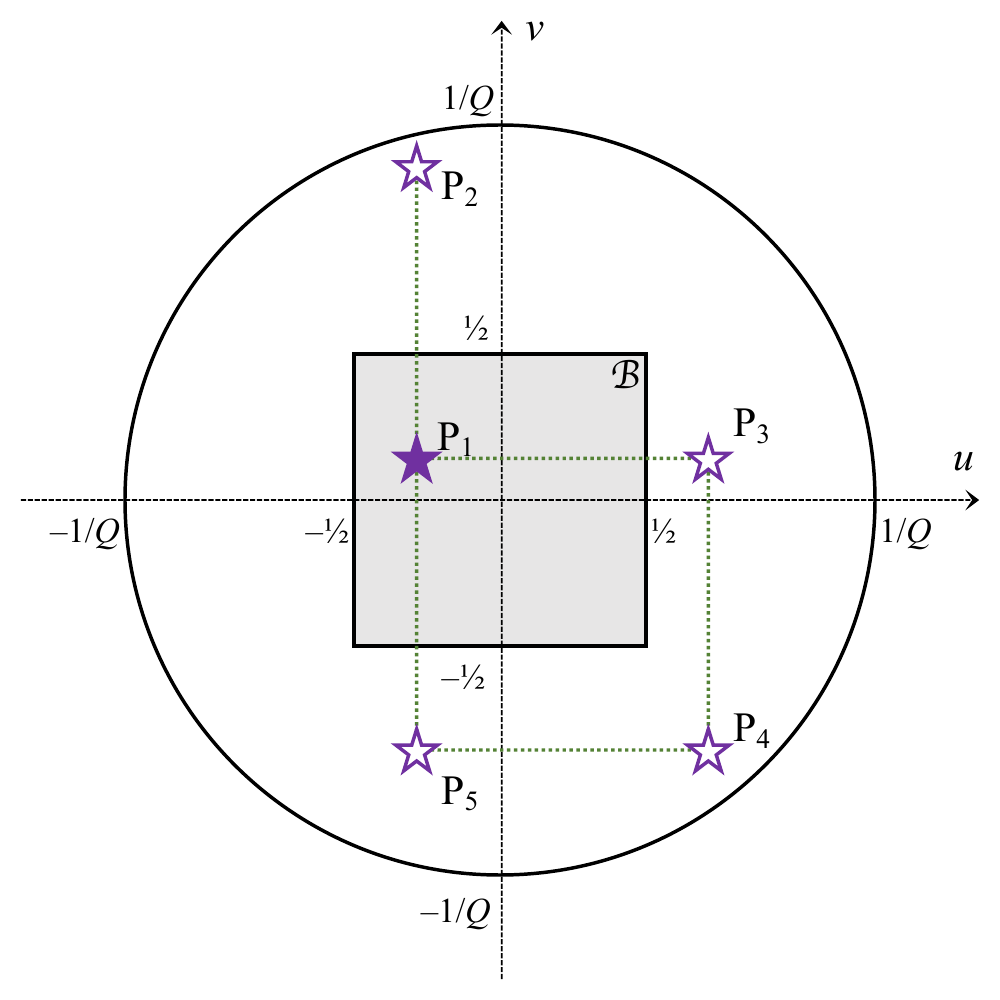}
\caption{\label{fig:uvfig}Computation of the number of modes ${\mathcal N}(u,v)$ aliasing to a given mode P$_1=(u,v)$ (solid star). The first Brillouin zone ${\mathcal B}$ is shown as the shaded box. The circle of radius $1/Q$ shows the range of Fourier modes where $\tilde G$ is non-zero. We use the stars to indicate the modes $(u+\Delta u,v+\Delta v)$ that alias to the original mode, i.e., that have integer $\Delta u$ and $\Delta v$. In this case, the modes with $(\Delta u,\Delta v) = (0,0)$, $(0,1)$, $(1,0)$, $(1,-1)$, and $(0,-1)$ contribute to the sum, indicated by the labels P$_1$ through P$_5$ respectively. So in this case, the number of modes is ${\mathcal N}({\rm P}_1)=5$.}
\end{figure}

\begin{figure*}
    \centering
    \includegraphics[width=6.5in]{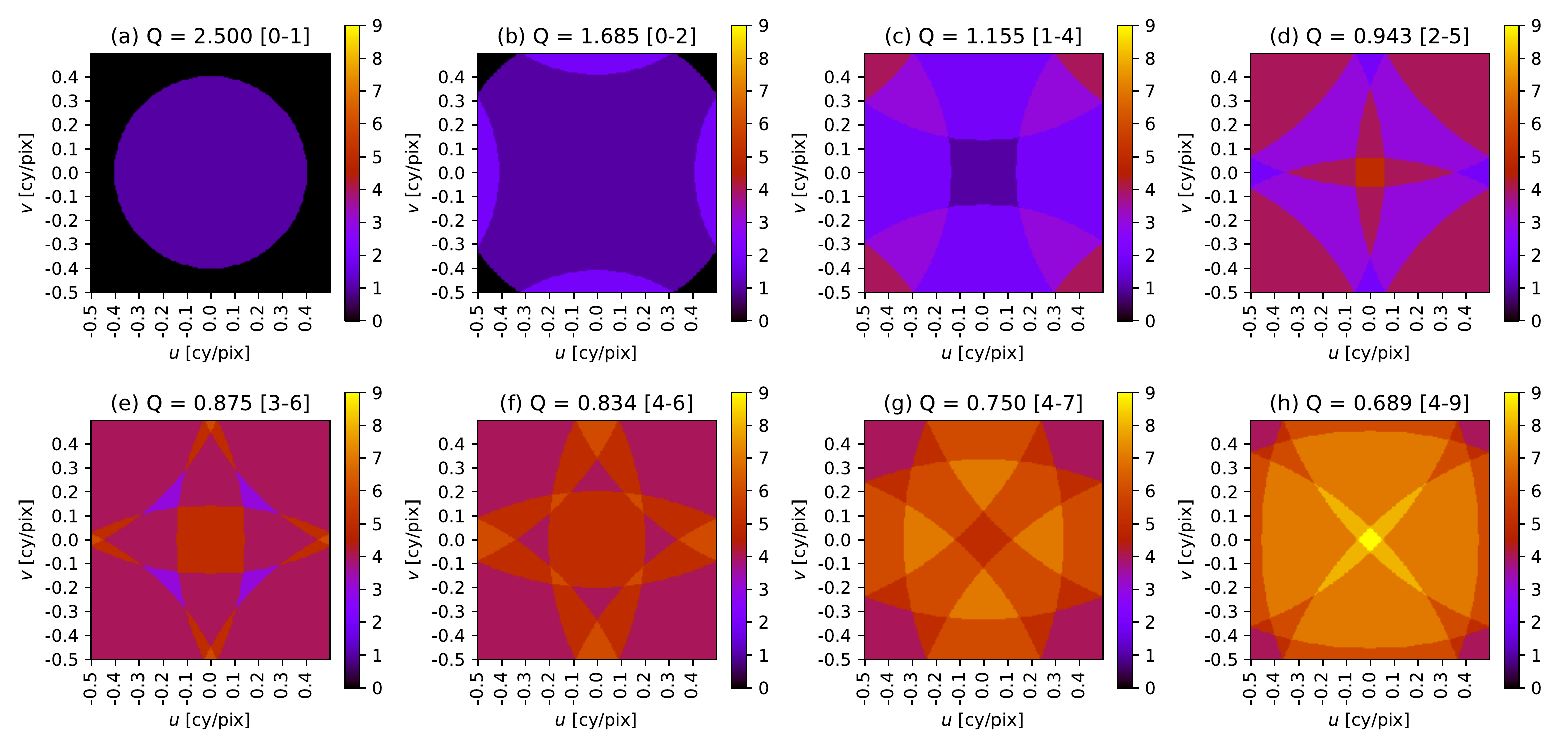}
    \caption{\label{fig:alias}The number ${\mathcal N}(u,v)$ of Fourier modes contributing to each part of the $(u,v)$-plane for discretely sampled data. A sequence of cases is shown depending on the sampling parameter $Q$, ranging from oversampled (a) to increasingly undersampled cases (b,c,d,...). The numbers in square brackets in the caption are the minimum and maximum number of Fourier modes that contribute.}
\end{figure*}

The sampling parameter $Q$ restricts the number ${\mathcal N}(u,v)$ of non-zero terms in Eq.~(\ref{eqn:f_aliased}), with the number of terms growing as $Q$ decreases (more undersampled). To determine ${\mathcal N}(u,v)$, we draw a square lattice with unit spacing centered at $(u,v)$, and count the number of lattice points in a circle of radius $1/Q$ centered at the origin (Fig.~\ref{fig:uvfig}). The results are visualized in Figure~\ref{fig:alias}. Each panel shows the region ${\mathcal B}$ in $(u,v)$-space covered by a discrete Fourier transform centered on (0,0) and this goes from $-\frac12$ to $\frac12$ cycles per pixel on both axes. The color scale shows the number of terms that contribute in Eq.~(\ref{eqn:f_aliased}), i.e., the number of Fourier modes in the original image that alias to $(u, v)$ in that region. The top-left panel shows the case where the image is Nyquist-sampled and each Fourier mode in the image is unique (the maximum allowed frequency can be seen as the radius of the circle $\sqrt{u^2 + v^2}$). It can be seen that more Fourier modes are aliased as the sampling factor $Q$ decreases (i.e., increasing the radius $1/Q$ of the circle). Intuitively, as the circle begins to overflow outside the region, neighboring circles centered at $(u, v) \in \mathbb{Z}^2$ begin to overlap and the frequency in the overlapped region is aliased. 

If ${\mathcal N}(u,v)=1$, then only $(\Delta u,\Delta v)=(0,0)$ contributes, which means that Eq.~(\ref{eqn:f_aliased}) becomes $\check f(u, v) = \tilde G(u, v) \tilde S(u, v)$, and $\tilde S(u,v)$ can be reconstructed by division in Fourier space. However, when ${\mathcal N}(u,v)\ge 2$, there are more than one $\tilde S(u+\Delta u, v+\Delta v)$ to solve but only one constraint $\check f(u, v)$, so $\tilde S(u,v)$ is not recoverable. Thus the region that can be reconstructed unambiguously is the region ${\mathcal N}^{-1}(1) = \{ (u,v): {\mathcal N}(u,v)=1\}$ where exactly one mode contributes.
If $Q>2$ (case a in Fig.~\ref{fig:alias}), then we say that the image is {\em oversampled} and ${\mathcal N}^{-1}(1)$ is the disc ${\mathcal D}_{1/Q} = \{ (u,v): \sqrt{u^2+v^2}<1/Q\}$ of radius $1/Q$. If $1<Q<2$ (cases b and c), then the image is {\em weakly undersampled}: while some modes are aliased, others are not and ${\mathcal N}^{-1}(1)$ is non-empty; the largest disc contained in it is ${\mathcal D}_{1-1/Q}$. If $Q<1$ (cases d and beyond), then the image is {\em strongly undersampled} and ${\mathcal N}^{-1}(1)$ is the null set: there are no Fourier modes $\tilde S(u,v)$ that can be recovered.

\subsection{Recovering information with translational dithers}
\label{ss:dither}

\citet{1999PASP..111..227L} discusses the linear solution to de-alias these modes for the case where multiple dithered images are taken with no rolls: if there are $N_{\rm exp}$ images with dither offsets ${\boldsymbol s}_1 ... {\boldsymbol s}_{N_{\rm exp}}$, then the $j$th image can be thought of as measuring the displaced scene ${\mathcal T}_{{\boldsymbol s}_j}S$. In this case, the discrete Fourier transform of the $j$th image is
\begin{eqnarray}
\check f_j(u,v) &=& {\rm e}^{2\pi{\rm i}(s_{jx}u+s_{jy}v)}
\sum_{(\Delta u, \Delta v)\in{\mathbb Z}^2}
{\rm e}^{2\pi{\rm i}(s_{jx}\Delta u+s_{jy}\Delta v)}
\nonumber \\ && \times
\tilde G(u+\Delta u, v+\Delta v) \tilde S(u+\Delta u, v+\Delta v)
\label{eq:network}
\end{eqnarray}
(compare to Eq.~\ref{eqn:f_aliased}). This is a system of $N_{\rm exp}$ equations with ${\mathcal N}(u,v)$ unknowns, and it can be solved if the $N_{\rm exp}\times {\mathcal N}(u,v)$ matrix ${\mathbfss L}$ with coefficients
\begin{equation}
L_{jm} = {\rm e}^{2\pi{\rm i}(s_{jx}\Delta u_m+s_{jy}\Delta v_m)},
\end{equation}
where $j$ represents one of the $N_{\rm exp}$ images and $m$ represents one of the $(\Delta u,\Delta v)$ terms in the sum, has rank ${\mathcal N}(u,v)$. Note that it is necessary but not sufficient that $N_{\rm exp}\ge {\mathcal N}(u,v)$. 

Particular dithering patterns have been proposed for the various sampling cases, For example, a diagonal dither (Fig.~2.2 of \citealt{2012drzp.book.....G}) with $N_{\rm exp}=2$ positions at ${\boldsymbol s} = (0,0), (\frac12,\frac12)$ works for case (b), because in the ${\mathcal N}(u,v)=2$ regions (${\mathcal N}=2$ in Fig.~\ref{fig:alias}b) the lattice points contributing to the sum are $(\Delta u,\Delta v)=(0,0)$ and one of $(1,0)$, $(-1,0)$, $(0,1)$, or $(0,-1)$. Then the matrix ${\mathbfss L}$ is
\begin{equation}
{\mathbfss L} = \left( \begin{array}{rr}
1 & 1 \\ 1 & -1
\end{array}\right),
\end{equation}
which has rank 2.
Similarly, a ``$2\times 2$'' dither with $N_{\rm exp}=4$ positions (Fig.~2.3 of \citealt{2012drzp.book.....G}) at ${\boldsymbol s} = (0,0), (\frac12,0), (0,\frac12), (\frac12,\frac12)$ works for case (c). For example, in the lower-left corner region of Fig.~\ref{fig:alias}c, we have ${\mathcal N}(u,v)=4$ and the lattice points contributing to the sum are $(\Delta u,\Delta v)=(0,0)$, $(1,0)$, $(0,1)$, and $(1,1)$. Then the matrix ${\mathbfss L}$ is
\begin{equation}
{\mathbfss L} = \left( \begin{array}{rrrr}
1 & 1 & 1 & 1 \\ 1 & -1 & 1 & -1 \\
1 & 1 & -1 & -1 \\ 1 & -1 & -1 & 1
\end{array}\right),
\end{equation}
which has rank 4.

Cases with more small dither steps for other sampling cases could be discussed: for example, the $N_{\rm exp}=5$ ``knight's move'' pattern ${\boldsymbol s}_j = (\frac25j,\frac15j)$ works for case (d), and \citet{2012drzp.book.....G} discuss the $N_{\rm exp}=8$ point dither (for cases e--g) and $N_{\rm exp}=9$ point dither (for case h). [The $N_{\rm exp}=6$ pattern considered in Fig.~2.4 of \citet{2012drzp.book.....G} does not lead to an ${\mathbfss L}$ matrix of full rank for cases (e) and (f).] However such large numbers of small deterministic dithers --- which would have to be supplemented by large dithers over chip gaps --- do not fit naturally into a fast wide-angle survey and thus were not options for the {\slshape Roman} HLIS.

This approach does not apply to cases with rolls, since in an exposure rolled by angle $\alpha$ there are aliased Fourier modes with $(\Delta u,\Delta v) = (\cos\alpha,\sin\alpha)$, and the ``network'' of coupled modes (Eq.~\ref{eq:network}) does not close with a finite number of constraints and unknowns. The one straightforward analytic result in this case comes from mode counting: the number of ``measurements'' (input pixels) is $N_{\rm exp}{\cal A}$ where ${\cal A}$ is the survey area (again in units where $s_{\rm in}=1$), whereas the number of ``unknowns'' (output Fourier modes) is $\pi{\cal A}/Q^2$ (recall the number of modes is area in real space times area in Fourier space). Therefore to disambiguate all the modes, it is necessary (but not sufficient) that
\begin{equation}
N_{\rm exp}\ge \pi/Q^2.
\label{eq:KQ}
\end{equation}
The {\sc Imcom} formalism \citep{2011ApJ...741...46R} was developed to handle this additional case numerically, which is needed for the proposed {\slshape Roman} strategy since it includes multiple roll angles.

In general, we will denote by ${\mathcal E}$ the region where $\tilde S(u,v)$ can be reconstructed, with the understanding that for $N_{\rm exp}=1$ dither, ${\mathcal E} = {\mathcal N}^{-1}(1)$.

\subsection{The operations in shear calibration techniques}

We now recall the operations used in the various shear calibration techniques. There is a class of techniques that works directly with Fourier-domain quantities \citep{2014MNRAS.438.1880B, 2016MNRAS.459.4467B}, and in this case one wants to work in a region of the Fourier plane that is contained within ${\mathcal E}$, and actually contains a ``buffer'' region before one reaches the edge of ${\mathcal E}$; see, e.g., \citet[\S4]{2010MNRAS.406.2793B}. We will see that this behavior is generic.

We start with the ``standard'' Metacalibration \citep{2017arXiv170202600H, 2017ApJ...841...24S}, and then investigate ``deep'' Metacalibration \citep{2022arXiv220607683Z}, with a particular emphasis on the operations that are potentially problematic on undersampled data \citep[e.g.][]{2021A&A...646A.124H}. Finally, we consider the use of pixel responses as in \citet{2022arXiv220810522L}.

\subsubsection{Standard Metacalibration}

The basic operation in standard Metacalibration is to take an image; infer the sky scene $S$ (e.g., by de-convolving the PSF); apply a magnification $\kappa$ and shear ${\boldsymbol\gamma}$; and re-convolve it with a new PSF $G_{\rm r}$. The resulting image is
\begin{equation}
f_{\rm out} = G_{\rm r}\ast {\mathcal Y}_{0,{\boldsymbol\gamma},0}S,
\label{eq:fout}
\end{equation}
so that
\begin{equation}
\check f_{\rm out}(u,v) =
\!\!\!\sum_{(\Delta u,\Delta v)\in{\mathbb Z}^2} \!\!\! \frac{\tilde G_{\rm r}(u+\Delta u,v+\Delta v)}{\det{\mathbfss M}_{0,{\boldsymbol\gamma},0}} \tilde S({\mathbfss M}_{0,{\boldsymbol\gamma},0}^{-1\,\rm T}(u+\Delta u,v+\Delta v)).
\label{eq:fout-meta}
\end{equation}
We suppose that the re-convolution PSF has a Fourier transform $\tilde G_{\rm r}(u+\Delta u,v+\Delta v)$ with support in some circular region out to radius $R_{\rm r}$.

For non-zero shear, the Fourier modes ${\mathbfss M}_{0,{\boldsymbol\gamma},0}^{-1\,\rm T}(u+\Delta u,v+\Delta v)$ do not alias to each other. In this case, we requires knowledge of $\tilde S$ at all the points that contribute to the sum. That means that for every point $(u',v')$ in the support of $\tilde G_{\rm r}$, ${\mathbfss M}_{0,{\boldsymbol\gamma},0}^{-1\,\rm T}(u',v')$ must be in the reconstructible region ${\mathcal E}$. This leads to the following cases:
\begin{list}{$\bullet$}{}
\item For oversampled data ($Q>2$), ${\mathcal E} = {\mathcal D}_{1/Q}$. This means that the radius of support of $\tilde G_{\rm r}$ must have $R_{\rm r}\le (1-|{\boldsymbol\gamma}|)/Q$
(see Eq.~\ref{eq:Lambda}). Since the original PSF $\tilde G$ has support out to a radius $1/Q$, this encapsulates the usual notion that the re-convolution PSF must be at least a little bit bigger than the original PSF, with the definition of ``a little bit'' being determined by the applied shear ${\boldsymbol\gamma}$.
\item For weakly undersampled data ($1<Q<2$) without dithering, the largest disc contained within ${\mathcal E}$ has radius $1-Q^{-1}$. Thus we can carry out the standard metacalibration operation if the re-convolution PSF satisfies $R_{\rm r}\le (1-|{\boldsymbol\gamma}|)(1-Q^{-1})$. Note in this case that as the native PSF gets smaller at fixed pixel scale ($Q$ gets smaller), the Fourier-space cutoff $R_{\rm r}$ must get smaller and hence the re-convolution PSF has to get bigger in real space.
\item For strongly undersampled data ($Q<1$) without dithering, ${\mathcal E}$ is the null set and standard metacalibration cannot be implemented.
\item For data with enough dithers to de-alias all of the Fourier modes, the radius $R_{\rm r}$ is again limited by $(1-|{\boldsymbol\gamma}|)/Q$.
\end{list}
For single-epoch {\slshape Roman} images, ``standard'' Metacalibration does not apply in Y106. It theoretically can be applied in J129 ($Q=1.021$), however $1-Q^{-1}$ is so small that it would not be practical (the re-convolution PSF would have to be more compact in Fourier space, and thus larger in real space, than even a typical ground-based PSF). The algorithm could be of interest in the redder filters.

\subsubsection{Deep Metacalibration}

We now turn to ``deep'' Metacalibration \citep{2022arXiv220607683Z}. Here there is both a deep scene $S_{\rm deep}$ and a wide scene $S_{\rm wide}$ that have to be rendered with a reconvolution PSF $G_{\rm r}$. The shear operation, Eq.~(\ref{eq:fout-meta}), is applied only to the deep data, whereas no shear is applied to the wide data.
In the context of both ground- and space-based surveys, deep fields usually have a much larger number of observations than the wide survey; but for a diffraction-limited space telescope the band limit and sampling parameter $Q$ are the same in the deep and wide surveys. (This may or may not be the case on the ground.) The deep survey recovers information over some region of the $(u,v)$-plane ${\mathcal E}_{\rm deep}$. Because of the large number of dithers, this region may be larger than ${\mathcal E}_{\rm wide}$, but it is still bounded by the radius $1/Q$.

In order to recover all of the modes in the deep data that we need for Eq.~(\ref{eq:fout-meta}), we must have
\begin{equation}
{\mathbfss M}_{0,{\boldsymbol\gamma},0}^{-1\,\rm T}(u',v') \in{\mathcal E}_{\rm deep} ~~~{\rm if}~~~
\tilde G_{\rm r}(u',v')\neq 0.
\end{equation}
However, if we are working with an undersampled single-epoch image in the wide survey (with PSF $G_{\rm wide}$), we do not have access to all Fourier modes independently, but only the linear combinations:
\begin{equation}
\check f_{\rm wide}(u,v) = \sum_{(\Delta u,\Delta v)\in{\mathbb Z}^2} \tilde G_{\rm wide}(u+\Delta u,v+\Delta v) \tilde S_{\rm wide}(u+\Delta u,v+\Delta v).
\label{eq:check-fuv}
\end{equation}
This means that for a given $(u,v)$, if there is a $(\Delta u,\Delta v)\in{\mathbb Z}^2$ satisfying
\begin{equation}
\tilde G_{\rm wide}(u+\Delta u,v+\Delta v)\neq 0
~~~{\rm and}~~~
{\mathbfss M}_{0,{\boldsymbol\gamma},0}^{-1\,\rm T}(u+\Delta u,v+\Delta v) \notin{\mathcal E}_{\rm deep},
\label{eq:gw1}
\end{equation}
then the linear combination appearing in Eq.~(\ref{eq:check-fuv}) must simply be zero, and $\tilde G_{\rm r}$ must be zero for all modes that alias to $(u,v)$. And in particular, if there is a lattice point $(u+\Delta u,v+\Delta v)$ in the troublesome annulus:
\begin{equation}
\frac{1-|{\boldsymbol\gamma}|}Q < \sqrt{(u+\Delta u)^2 + (v+\Delta v)^2} < \frac1Q,
\label{eq:annulus}
\end{equation}
then there is an orientation for the shear such that Eq.~(\ref{eq:gw1}) is satisfied, and that Fourier mode must not be present in the reconvolved image. So if we were to implement Deep Metacalibration on an undersampled image, we must avoid all modes that alias to this annulus, i.e., set $\tilde G_{\rm r}(u,v)=0$ in these regions. In practice, since one uses Deep Metacalibration with small values of $|{\boldsymbol\gamma}|$, the troublesome regions of $(u,v)$-space are those near the circle $\sqrt{(u+\Delta u)^2 + (v+\Delta v)^2} = 1/Q$, i.e., the ``edges'' seen in Fig.~\ref{fig:alias}.

The aforementioned procedure is in principle possible, and the conditions required for it to work are slightly less restrictive than for standard Metacalibration --- indeed, these conditions could be met even in the strongly undersampled case. However, in practice the re-convolution PSF must be very large --- for example, if $Q=0.834$ (Y106 band) and $|{\boldsymbol\gamma}|=0.01$, then $\tilde G_{\rm r}$ must go to zero at a radius of 0.187 cycles pixel$^{-1}$, which is equivalent to an Airy disc with a full width at half maximum of 0.61 arcsec. This means that the re-convolution PSF effectively degrades the sky image to ground-based resolution, so even though the mathematics works this is not an attractive option for analyzing {\slshape Roman} data.

\subsubsection{Pixel basis function technique}

\citet{2022arXiv220810522L} present another approach to shear calibration, using many of the same ideas as Metacalibration but using analytic differentiation to compute the shear response. The idea is that the flux in a given pixel $(x,y)\in{\mathbb Z}^2$ can be written in terms of the sky scene $S$ via
\begin{equation}
f_{x,y} = \int \tilde S(u,v) \Phi^\ast_{x,y}(u,v) {\rm d}u\,{\rm d}v,
\end{equation}
where
\begin{equation}
\Phi_{x,y}(u,v) =  {\rm e}^{-2\pi{\rm i}(ux+vy)}\tilde G^\ast(u,v) .
\label{eq:pxyuv}
\end{equation}
is the {\em pixel basis function}.
[\citet{2022arXiv220810522L} define $\phi_{x,y}(u,v)$, which is our $\Phi_{x,y}(u,v)/(2\pi)^2$; the difference results from the choice of cycles per pixel versus radians per pixel as the unit of spatial frequency.] One may then consider a partial derivative of the pixel with respect to a shear component ($a=1,2$):
\begin{eqnarray}
f_{x,y;a}
&\equiv&
\left.\frac{\partial}{\partial \gamma_a} [G \star {\mathcal Y}_{0,{\boldsymbol\gamma},0}S](x,y)\right|_{{\boldsymbol\gamma}=0}
\nonumber \\
&=& \int \tilde S(u,v) \Phi^\ast_{x,y;a}(u,v) {\rm d}u\,{\rm d}v,
\end{eqnarray}
where
\begin{eqnarray}
\!\!\!\!
\Phi_{x,y;1}(u,v) \!\!\!\!&=&\!\!\!\! {\rm e}^{-2\pi{\rm i}(ux+vy)} 
\nonumber \\ && \times \!\!
\left[ -u \frac{\partial}{\partial u} + v\frac{\partial}{\partial v} + 2\pi{\rm i}(xu-yv) \right] \tilde G^\ast(u,v)
~~~~~~~~
\end{eqnarray}
and
\begin{eqnarray}
\!\!\!\!
\Phi_{x,y;2}(u,v) \!\!\!\!&=&\!\!\!\! {\rm e}^{-2\pi{\rm i}(ux+vy)} 
\nonumber \\ && \times \!\!
 \left[ -v \frac{\partial}{\partial u} - u\frac{\partial}{\partial v} + 2\pi{\rm i}(yu+xv) \right] \tilde G^\ast(u,v)
~~~~~~~~
\label{eq:P2}
\end{eqnarray}
are the shear responses of the pixel basis functions.
(These are Eq.~19 of \citealt{2022arXiv220810522L}, but written for a non-Gaussian PSF.)

We are interested in the conditions under which $f_{x,y;a}$ can be determined from the data, i.e., when the shear response functions can be written in terms of a linear combination of the pixel response functions themselves:
\begin{equation}
\Phi_{x,y;a}(u,v) \stackrel?= \sum_{(\Delta x,\Delta y)\in{\mathbb Z}^2} W^{(a)}_{x,y,\Delta x,\Delta y} \Phi_{x+\Delta x,y+\Delta y}(u,v).
\label{eq:phitest}
\end{equation}
We write the Fourier transform of the weights on the $(\Delta x,\Delta y)$ indices,
\begin{equation}
W^{(a)}_{x,y,\Delta x,\Delta y} = \int_{\mathcal B} \check W^{(a)}_{x,y}(u',v') {\rm e}^{2\pi{\rm i}(u'\Delta x+v'\Delta y)} \,{\rm d}u'\,{\rm d}v'.
\end{equation}
Substituting this and the expressions for $\Phi_{x,y}$ and $\Phi_{x,y;a}$ into Eq.~(\ref{eq:phitest}), we find that in the summation over $\Delta x$ and $\Delta y$, only the cases where $(u,v)$ and $(u',v')$ alias to each other survive. Then if we cancel the common factors of ${\rm e}^{-2\pi{\rm i}(ux+vy)}$ in Eq.~(\ref{eq:phitest}), we arrive at
\begin{equation}
\left[ -u \frac{\partial}{\partial u} + v\frac{\partial}{\partial v} + 2\pi{\rm i}(xu-yv) \right]
\tilde G^\ast(u,v) \stackrel?= 
\check W^{(1)}_{x,y}({\mathbb F}u,{\mathbb F}v)
\tilde G^\ast(u,v)
\label{eq:phitest2-1}
\end{equation}
and
\begin{equation}
\left[ -v \frac{\partial}{\partial u} -u\frac{\partial}{\partial v} + 2\pi{\rm i}(yu+xv) \right]
\tilde G^\ast(u,v) \stackrel?= 
\check W^{(2)}_{x,y}({\mathbb F}u,{\mathbb F}v)
\tilde G^\ast(u,v).
\label{eq:phitest2-2}
\end{equation}

For oversampled data, this is only non-trivial for modes $(u,v)$ that are within the band limit and hence in ${\mathcal B}$, and we can solve the problem by dividing by $\tilde G^\ast(u,v)$. A subtlety occurs at the band limit itself, $\sqrt{u^2+v^2} =1/Q$, where both $G(u,v)$ and its derivatives approach zero. \citet{2022arXiv220810522L} considered the case where the Fourier transform of the PSF declines smoothly toward zero as $(u,v)$ increases, in which case beyond some point both $\tilde G$ and its derivatives can be treated as negligible to the accuracy required for a given project. With oversampled data, one may always accomplish this by a convolution in pre-processing even if the native PSF of the telescope has a sharper band limit.

For undersampled data, however, Eqs.~(\ref{eq:phitest2-1}) and (\ref{eq:phitest2-2}) are overdetermined: we have ${\mathcal N}(u,v)$ constraints (each $u,v$ contributes a constraint) for the same 1 unknown $\check W^{(a)}_{x,y}({\mathbb F}u,{\mathbb F}v)$. With the exception of some special unrealistic cases, there is no solution. If the data are weakly undersampled, then a filter in pre-processing can remove all of the aliased modes and we can work in the region where ${\mathcal N}(u,v)=1$. In the strongly undersampled case, however, there is no such simple fix for the problem.

\bsp	
\label{lastpage}
\end{document}